\documentclass[structabstract]{aa} % printer format !!! THIS IS THE USUAL 2-COLUMN FORMAT
\usepackage{natbib}
\usepackage{array}
\bibpunct{(}{)}{;}{a}{}{,} 
\usepackage{graphicx}
\usepackage{txfonts}
\usepackage{lscape}
\usepackage{caption}
\usepackage{color}
\usepackage{amsmath}
\usepackage{textgreek}
\usepackage{float}
\usepackage{subfig}
\usepackage[titletoc]{appendix}
 \usepackage{booktabs,multirow,tabularx}
 \usepackage{natbib,twoopt}
 \bibpunct{(}{)}{;}{a}{}{,} %% natbib format like A&A and ApJ
 \newcommandtwoopt{\citeads}[3][][]{\href{http://adsabs.harvard.edu/abs/#3}{\citealp[#1][#2]{#3}}}
 \newcommandtwoopt{\citepads}[3][][]{\href{http://adsabs.harvard.edu/abs/#3}{\citep[#1][#2]{#3}}}
 \newcommandtwoopt{\citetads}[3][][]{\href{http://adsabs.harvard.edu/abs/#3}{\citet[#1][#2]{#3}}}
 \newcommandtwoopt{\citeyearads}[3][][]{\href{http://adsabs.harvard.edu/abs/#3}{\citeyear[#1][#2]{#3}}}
\usepackage[ % Farben fuer die Links
  breaklinks=true,
  colorlinks=true,         % Links erhalten Farben statt Kaeten
   urlcolor=pdfurlcolor,    % \href{...}{...} external (URL)
   filecolor=pdffilecolor,  % \href{...} local file
   linkcolor=pdflinkcolor,  %\ref{...} and \pageref{...}
   citecolor=pdfcitecolor,  %
   raiselinks=true,                     % calculate real height of the link
   breaklinks,              % Links berstehen Zeilenumbruch
   backref=page,            % Backlinks im Literaturverzeichnis (section, slide, page, none)
   pagebackref=false,       % Backlinks im Literaturverzeichnis mit Seitenangabe
   verbose,                                    % extra diagnostic messages are printed in the log file
   hyperindex=true,         % backlinkex index
   linktocpage=false,        % Inhaltsverzeichnis verlinkt Seiten
   hyperfootnotes=false,     % Keine Links auf Fussnoten
   bookmarks=true,          % Erzeugung von Bookmarks fuer PDF-Viewer
   bookmarksopenlevel=1,    % Gliederungstiefe der Bookmarks
   bookmarksopen=true,      % Expandierte Untermenues in Bookmarks
   bookmarksnumbered=true,  % Nummerierung der Bookmarks
   bookmarkstype=toc,       % Art der Verzeichnisses
   plainpages=false,        % Anchors even on plain pages ?
   pageanchor=true,         % Pages are linkable
   pdfauthor={Sara Saeedi},      % Autor
   pdfdisplaydoctitle=true, % Dokumententitel statt Dateiname im Fenstertitel
   pdfstartview=FitV,       % Dokument wird Fit height geoeffnet
   pdfpagemode=UseOutlines, % Bookmarks im Viewer anzeigen
   pdfpagelabels=true,           % set PDF page labels
   pdfpagelayout=TwoPageRight, % zweiseitige Darstellung: ungerade Seiten
]{hyperref} 
  
\definecolor{pdfurlcolor}{rgb}{0,0,0.6}
\definecolor{pdffilecolor}{rgb}{0.7,0,0}
\definecolor{pdflinkcolor}{rgb}{0,0,0.6}
\definecolor{pdfcitecolor}{rgb}{0,0,0.6}

\newcommand{\xmm}{{\it XMM-Newton}}

\begin{document}

  \title{Classification of low-luminosity stellar X-ray sources in the field of the Draco dwarf spheroidal galaxy  \thanks{Based on observations obtained with \xmm, an ESA science mission with instruments and contributions directly funded by ESA Member States and NASA.}
}
   \subtitle{}
    \titlerunning{Classification of low luminosity stellar X-ray sources of the Draco dSph}
   \authorrunning{S.~Saeedi et al}
   \author{Sara\,Saeedi\inst{1}, Manami\,Sasaki\inst{2}, Beate\,Stelzer\inst{1, 3}, \and Lorenzo\,Ducci\inst{1}
          %\fnmsep\thanks{}
   }

\institute{\\ \inst{1} Institut f\"ur Astronomie und Astrophysik, Kepler Center for Astro and Particle Physics, Eberhard Karls Universit\"at T\"ubingen, Sand~1, D-72076 T\"ubingen, Germany\\
    \email{saeedi@astro.uni-tuebingen.de}\\
    \inst{2} Dr. Karl Remeis Observatory and ECAP, Universit\"at Erlangen-N\"urnberg, Sternwartstr. 7, 96049 Bamberg, Germany\\
    \inst{3} INAF - Osservatorio Astronomico di Palermo, Piazza del Parlamento 1, I-90134 Palermo, Italy\\
         %\and
             %\thanks{}
             }
  \date{Received DATE; accepted DATE}

\abstract{}{}{}{}{} 
% 5 {} token are mandatory
 
  \abstract
   {}
   {A previous study of the X-ray luminosity function of the X-ray sources in the field of the Draco dwarf spheroidal (dSph) galaxy indicated the presence of a population of unknown X-ray sources in the soft energy range of 0.5-2 keV. In 2015, Draco dSph was observed again in twenty-six deep \xmm\, observations providing an opportunity for a new study of the yet unclassified sources.}
   {We apply the classification criteria presented in our previous multi-wavelength study of the X-ray sources of the Draco dSph to the sources detected in the combined 2009 and 2015 \xmm\, data set. These criteria are based on X-ray studies and properties of the optical, near-infrared, and mid-infrared counterparts and  allows us to distinguish background active galactic nuclei~(AGNs) and galaxies from other types of X-ray sources. In this work we perform X-ray spectral and timing analyses for fifteen sources in the field of Draco dSph with stellar counterparts.}
   {We present the classification of X-ray sources, for which the counterpart is identified as a stellar object based on our criteria from multi-wavelength data. We identify three new symbiotic stars in the Draco dSph with X-ray luminosities between $\sim$3.5$\times10^{34}$\,erg\,s$^{-1}$ and 5.5$\times10^{34}$\,erg\,s$^{-1}$. The X-ray spectral analysis shows that two of the classified symbiotic stars are $\beta$-type. This is the first identification of this class of symbiotic stars in a nearby galaxy. Eight sources are classified as Galactic M\,dwarfs in the field of the Draco dSph. The distances of these M\,dwarfs are between$\sim$140--800 pc, their X-ray luminosities are between $10^{28}-10^{29}$\,erg\,s$^{-1}$ and the logarithmic ratio of  X-ray to bolometric luminosity, log$(\frac{L_\text{X}}{L_\text{bol}})$, is between $-3.4$ to $-2.1$. The multiple observations allowed us to investigate flare activity of the M~dwarfs. For 5 M~dwarfs flare(s) are observed with a significance of >$3 \sigma$ level of confidence. Moreover, we classified three foreground sources, located at distances of the order of $\sim$1--3 kpc in the field of the Draco dSph. Based on both the X-ray luminosities of these foreground sources\,(>$10^{30}$\,erg\,s$^{-1}$) and their optical counterparts\,(late type G or K stars), these X-ray sources were classified as candidates of contact binary systems.}
     %conclusions heading (optional), leave it empty if necessary 
   {Our study of X-ray sources of the Draco dSph shows that accreting white dwarfs are the most promising X-ray population of dSphs, which is in line with theoretical expectations. The number of Galactic M~dwarfs detected at our X-ray sensitivity limit is consistent with the expectation based on the space density of M~dwarfs.}

    % SEARCH THEM ON THE A&A PAGE

   \maketitle
%
%________________________________________________________________

\section{Introduction}\label{intro}
Most of the satellite galaxies of our Milky Way are dwarf spheroidal galaxies\,(dSph), which are faint with luminosities of 10$^{5-7}$\,L$_{\sun}$ and have an approximately spherical shape. Observations of nearby dwarf galaxies show that most of them are metal-poor systems with metallicities as low as [Fe/H]$<-3$ \citep[e.g,][]{2008ApJ...685L..43K, 2009A&A...502..569A, 2010Natur.464...72F}. As many dSphs show no recent star formation, they are ideal targets to study the old stellar populations in galaxies formed in the early stages of galaxy evolution. However, in comparison to other nearby galaxies, the X-ray population of low-mass satellite galaxies of the Milky Way (e.g, dSphs) is poorly studied \citep[e.g,][]{2006ARA&A..44..323F}.

Theoretically, the old stellar population of the dSphs makes the presence of X-ray binaries\,(XRBs) very unlikely. Because these compact objects form at the end of the life of massive stars, XRBs with high-mass companions (high-mass X-ray binaries, HMXBs) are often found in stellar populations with a very young age. However, also the presence of X-ray binaries with low-mass companion stars (low-mass X-ray binaries, LMXBs) in a dSph is a theoretical challenge. LMXBs are supposed to form a few Gyr after the general star formation bursts in galaxies. Since the compact object consumes the mass of the less-massive donor star in a few hundred million years, any presence of persistently bright LMXBs in dSphs, which are dominated by very old stellar populations\,(> 10 Gyr), is not consistent with stellar evolution models \citep{2005MNRAS.364L..61M}. Recent observational studies suggest the presence of candidates for LMXBs in some dwarf galaxies \citep[e.g,][]{2005MNRAS.364L..61M, 2016A&A...586A..64S}. However, these few candidates cannot be considered as the main population of the X-ray sources that belong to the dSphs. In the study of the XLF of the Draco dSph, \citet{2016A&A...586A..64S} showed that the population of hard X-ray sources (2.0-10.0 keV) in the field of the galaxy does not exceed the population of background sources (background galaxies, AGNs). However, the XLF at 0.5-2.0 keV showed that there might be a population of soft X-ray sources with luminosities $\lesssim10^{34}$~erg\,s$^{-1}$. This result was consistent with the findings of \citet{2006A&A...459..777R}, who reported that there is a fraction of unclassified soft X-ray sources in Carina and Sagittarius dSphs. An investigation of the nature of the soft low-luminosity X-ray sources in the field of the Draco dSph is the aim of this work.

Low-luminosity X-ray point sources can either be transient LMXBs observed during the low-luminosity state or accreting white dwarfs (AWDs) \citep{2006csxs.book..623T}. Since late-type stars are the main population of dSphs \citep[e.g;][]{2009ARA&A..47..371T}, white dwarfs are expected to be the main population of compact objects in dSphs and a higher number of AWDs is expected than XRBs. AWDs can be observed in  X-rays, UV, optical, and infrared. White dwarfs with a red giant star form binary systems called symbiotic stars, which are considered to be likely candidates for type Ia supernova progenitors \citep[e.g,][]{1993ApJ...407L..81K}. Steady or quasi-steady burning of accreted matter on the surface of the white dwarf makes the system a super-soft X-ray source (SSS), whereas in wider binary systems, the X-ray emission can have different origins \citep[e.g;][]{2013A&A...559A...6L}. In soft X-ray bands, AWDs have typical luminosities of $~10^{31}-10^{34}$\,erg\,s$^{-1}$, but in the case of super-soft sources the luminosity range increases to $~10^{36}-10^{38}$\,erg\,s$^{-1}$  \citep[e.g;][]{2006csxs.book.....L}.

The Draco dSph hosts a known super-soft X-ray source, Draco\,C\,1, which was studied by \citet{2018MNRAS.473..440S}. In this work we focus on finding more candidates of AWDs in the Draco dSph based on the multi-wavelength study of all unclassified soft X-ray sources.

The first population study of the X-ray sources in the field of Draco dSph was performed based on five \xmm\,observations obtained in 2009 \citep{2016A&A...586A..64S, 2015MNRAS.451.2735M, 2016ApJ...821...54S}. Draco dSph was observed again in 26 deep \xmm\, observations in 2015. The new observations not only made an extended study of the unclassified sources of the previous catalogues possible, but also revealed many new sources in the field of Draco dSph owing to the increased total exposure time\,(see Table~\ref{obs-data}). \citet{2016A&A...586A..64S} presented  multi-wavelength criteria to distinguish background sources from  foreground sources and the members of the Draco dSph. The properties of the optical and infrared counterparts of the X-ray sources were used to identify background sources. This paper presents the classification of fourteen sources detected in the new combined analysis of 2006 and 2015 data, for which a stellar counterpart is confirmed in multi-wavelength studies. An updated catalogue of all X-ray sources in the field of Draco dSph is in preparation (Saeedi et al. in prep).

\section{\xmm\, data analysis}
\label{data-ana}
\begin{figure}[!htb]
\includegraphics[clip, trim={1.0cm 3.5cm 1.0cm 1.5cm}, width=0.50\textwidth]{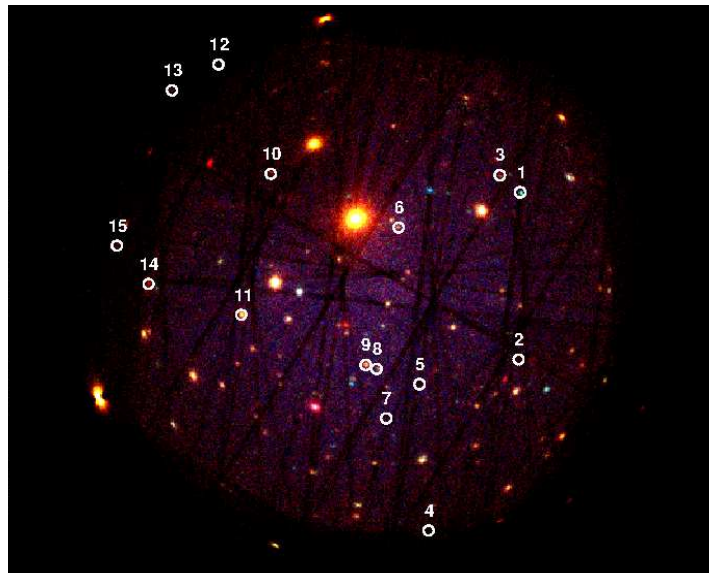}
\caption{The combined X-ray image of thirty-one \xmm\, observations in the field of Draco dSph in the energy band of 0.2--4.5 keV. Sources studied in this work are marked with source numbers from the catalogue in Table~2. \label{image-draco}}
\end{figure}

The thirty-one \xmm\, observations used for our studies are listed in Table~\ref{obs-data}. The cameras EPIC-pn \citep{2001A&A...365L..18S} and EPIC-MOS1,\,2 \citep{2001A&A...365L..27T} were in full frame mode and the thin filter was used in all observations. Data reduction and source detection were performed using the \xmm\, Science Analysis System\,(SAS,\,V.16.0.0).  The event files were screened for time intervals with high background caused by soft proton flares. We extracted single events with PI>10000, and PATTERN=0 for EPIC-MOS, and PI>10000 and PI<12000, as well as PATTERN=0 for EPIC-pn, and created light curves with time bins of 100 s in order to identify background flare intervals. We used a threshold rate of $\leq$ 0.35 count\,s$^{-1}$ for EPIC-MOS and rate $\leq$ 0.4 count\,s$^{-1}$ for EPIC-pn as criteria for the good time intervals. Also, the light curves were checked for good time intervals by eye and possible background flares have been removed. The final good time interval files were used to filter the event lists. Table\,\ref{obs-data} lists the final exposure time for each observation and EPIC camera. In this paper we use the number\,(OBS-No)  in Table\,\ref{obs-data} to specify an \xmm\, observation instead of the observation ID. Source detection for each observation has been performed using the \texttt{SAS} task \texttt{edetect-chain} in the five standard energy-bands of \xmm\, B1\,(0.2--0.5\,keV), B2\,(0.5--1.0\,keV), B3\,(1.0--2.0\,keV), B4\,(2.0--4.5\,keV), B5\,(4.5--12.0\,keV) with minimum detection likelihood of 7 \footnote{The detection likelihood is calculated by the probability of Poisson random fluctuations of the counts, $L$=--ln$(p)$, where $p$ is the probability, which is calculated on the basis of the raw counts of the source and the raw counts of the background maps.}. Table~\ref{source-list} lists the \xmm\, ID, position, and position uncertainty for each source. The source numbers\,(Src-No) given to each X-ray source in Table~\ref{source-list} are used in the entire paper. The coordinate and the positional error of each source in Table\,\ref{source-list} were taken from the observation in which the source was detected with the highest maximum likelihood. The Right Ascension\,(RA) and Declination\,(DEC) for each X-ray source were corrected by the offset of the X-ray to the optical position of the known symbiotic star Draco\,C1 (source No.~5 in this paper) in each observation. We produced a mosaic image using the \texttt{SAS} task \texttt{edetect-stack} from a combination of EPIC images of all observations. Figure\,\ref{image-draco} shows the three-colour combined image of all observations together in the energy range of 0.2--4.5 keV.

\begin{table*}[!htb]
\caption{
\label{obs-data}
\xmm\, observations of Draco dSph}
%\small
\centering
\begin{tabular}{cccccc}
\hline\hline\
OBS-No & OBS-ID & OBS-Date &  EPIC-pn & EPIC-MOS1 & EPIC-MOS2\\
 & & &   T.exp$^{\ast}$&T.exp$^{\ast}$ &T.exp$^{\ast}$\\
 & & &    (ks)& (ks) & (ks)\\
\hline
       1&    0603190101&    2009-08-04 &      16.2   &    18.8    &       18.8     \\
       2&    0603190201&    2009-08-06 &      16.9   &    19.7     &      19.7      \\
       3&    0603190301&    2009-08-08 &      9.7    &    13.5     &      13.8      \\
       4&    0603190401&    2009-08-20 &      5.8    &    15.8     &      15.9     \\
       5&    0603190501&    2009-08-28 &      16.8   &    19.7     &      19.7       \\
       6&    0764800101&    2015-03-18 &      27.2   &    49.3     &      50.1      \\
       7&    0764800301&    2015-03-26 &      25.5   &    56.5     &      57.1      \\
       8&    0764800401&    2015-03-28 &      55.5   &    60.6     &      60.2      \\
       9&    0764800201&    2015-04-05 &      26.3   &    41.4     &      40.2      \\
      10&    0764800501&    2015-04-07 &      48.2   &    61.7     &      61.7      \\
      11&    0764800601&    2015-04-09 &      53.7   &    56.5     &      56.6      \\
      12&    0764800801&    2015-04-19 &      28.5   &    48.7     &      50.2      \\
      13&    0764800901&    2015-04-25 &      37.2   &    51.7     &      51.7      \\
      14&    0770180101&    2015-04-27 &      34.1   &   55.2      &      55.2     \\
      15&    0770180201&    2015-05-25 &      54.7   &    58.7     &      58.7      \\
      16&    0764800701&    2015-06-15 &      56.1   &    57.2     &      57.2      \\
      17&    0770180401&    2015-06-18 &      50.9   &    52.7     &      52.7      \\
      18&    0770180301&    2015-07-01 &      56.8   &   56.1      &      58.7     \\
      19&    0770180501&    2015-07-31 &      51.1   &    54.7     &      54.5      \\
      20&    0770180701&    2015-08-22 &      38.5   &    50.7     &      50.3      \\
      21&    0770180601&    2015-09-01 &      38.8   &    67.0     &      66.6     \\
      22&    0770180801&    2015-09-03 &      66.6   &    77.4     &      77.8     \\
      23&    0770190401&    2015-09-11 &      56.2   &    70.3     &      69.7      \\
      24&    0770190301&    2015-09-21 &      20.7   &    43.2     &      40.4     \\
      25&    0770190101&    2015-09-23 &      18.3   &    42.8     &      41.9      \\
      26&    0770190201&    2015-09-25 &      18.2   &    40.6     &      37.8     \\
      27&    0770190501&    2015-10-11 &      32.3   &    44.6     &      44.6      \\
      28&    0770180901&    2015-10-13 &      20.7   &    35.5     &      33.5      \\
      29&    0770190601&    2015-10-15 &      7.9    &    23.1     &      24.9      \\
      30&    0770190701&    2015-10-17 &      45.3   &    52.0     &      51.6     \\
      31&    0770190801&    2015-10-19 &      32.6   &   48.8      &      48.7     \\
\hline
\end{tabular}
%\vspace{-2mm}
\centering
\tablefoot{$\ast$: Exposure time after screening for high background.}
\end{table*}

\begin{table}[!htb]
\caption{Catalogue of X-ray sources}
\label{source-list}
\small
\centering
\addtolength{\tabcolsep}{-0.1cm}    
\begin{tabular}{llccc}
\hline \hline
Src-No & ID & RA& DEC &r1$\sigma$\\
&&(J2000)&(J2000)&($''$)\\
\hline
1  & XMMUJ171919.8+575943$^{\ast}$  &   17 19 19.80      & +57 59 43.8    & 0.65 \\
2 &  XMMUJ171920.6+575120           &   17 19 20.61	 & +57 51 20.2   & 0.95 \\
3 &  XMMUJ171927.1+580035$^{\ast}$   &   17 19 27.24	 & +58 00 36.2	 & 1.30\\
4 &  XMMUJ171954.1+574244           &   17 19 54.18	 & +57 42 44.0   & 1.21 \\
5 &  XMMUJ171957.6+575005$^{\ast}$   &   17 19 57.65	 & +57 50 05.5	 & 0.42 \\
6 &  XMMUJ172005.6+575759$^{\ast}$   &   17 20 05.62 	 & +57 57 59.2	 & 1.45 \\      
7 &  XMMUJ172010.2+574823           &   17 20 10.27 	 & +57 48 23.1   & 1.20 \\
8 &  XMMUJ172013.3+575051$^{\ast}$   &   17 20 13.39 	 & +57 50 51.6	 &  0.99 \\       
9 &  XMMUJ172017.9+575105$^{\ast}$   &   17 20 17.99 	 & +57 51 05.7	 & 0.48 \\
10&  XMMUJ172053.8+580044$^{\ast}$   &   17 20 53.89 	 & +58 00 44.6	 & 0.80 \\
11&  XMMUJ172104.8+575333$^{\ast}$   &   17 21 04.79 	 & +57 53 33.5	 & 0.44 \\
12&  XMMUJ172113.6+580610           &   17 21 13.67 	 & +58 06 10.1   &  1.26\\
13&  XMMUJ172131.2+580451           &   17 21 31.20     & +58 04 51.6   &  1.46\\
14&  XMMUJ172139.6+575506           &   17 21 39.67 	 & +57 55 06.9   & 1.54 \\
15&  XMMUJ172151.4+575700$^{\ast}$   &   17 21 51.48 	 & +57 57 00.9	 & 1.76 \\
\hline
\end{tabular}
\vspace{-2mm}
\tablefoot{$\ast$ These sources are also listed in the enhanced 3XMM catalogue \citep[3XMMe,][]{2016yCat.9047....0R}.}
 \end{table}

\subsection{X-ray timing analysis}
\label{X-time}
We studied the short and long-term variability of the detected sources of Table\,\ref{source-list}. To study variability  on long time-scales, we calculated the weighted flux of the source  in each observation from the  EPIC cameras, in which the source was detected in the field of view\,(FOV) in the energy range of 0.2--4.5 keV. Band\,5\,(4.5--12 keV) is excluded because the EPIC cameras have a high background contamination and low sensitivity in this energy band. Moreover, except for source No.~1, all the sources studied in this paper emit predominantly X-rays at energies $\lesssim$2~keV (see Figs.~\ref{srcno.1.spec.sym}, \ref{src.spec.mdwarf}, and \ref{src.spec.forgrounds}). For each source, the count rates have been converted to flux using the energy conversion factor obtained based on the spectral model fitted to the source (see Sect.~\ref{sa}). If the spectrum of a source was not available, the energy conversion factor was calculated based on typical models and parameter values for the given source type (see Sect.\,\ref{x-ana-m-dwarf} and Sect.\,\ref{diss-foreground}). For observations, in which the position of an X-ray source was in the FOV of the EPIC cameras but the source was not detected, an upper limit was calculated at the position of the source using the sensitivity map in the energy range of 0.2-4.5~keV, created using the \texttt{SAS} task \texttt{esensmap}. The long-term X-ray light curves of the sources are shown in Appendix~\ref{appen-light-curve}.

We checked the variability of each source by calculating the ratio of maximum to minimum flux from all observations. The variability\,($Var$) and its significance\,($S$) were calculated using the following formulae:
\begin{equation}
\label{var-eq}
Var=\frac{F_{\rm max}}{F_{\rm min}},\\
  \vspace{1.cm}
S=\frac{F_{\rm max}-F_{\rm min}}{\sqrt{EF_{\rm max}^{2}+EF_{\rm min}^{2} }},
\end{equation}\\
where $F_{\rm max}$ and $F_{\rm min}$ are the maximum and minimum X-ray fluxes, and $EF_{\rm max}$ and $EF_{\rm min}$ are the errors of the maximum and minimum flux of the source, respectively \citep{1993ApJ...410..615P}. In sources with upper limits lower than the minimum flux, the lowest upper limit was taken to calculate the variability.  A source was considered variable if the significance~$S$ of the variability factor was > 3.

We also searched for short-term variability in two steps as follows: First, for each source we identified those observations where the flux was $>3 \sigma$ above the minimum flux measurement for this source. Here, $\sigma$ is the standard deviation of the flux built from all observations, and the minimum flux is either the lowest detected flux or the lowest upper limit. For all individual observations fullfilling this $3 \sigma$ criterion we examined the background-subtracted lightcurve using the same approach, i.e. we measured the standard deviation of the count rate taking into account all bins in the short-term light curve and we identified the bin with the lowest count rate. All bins with count rate $> 3\sigma$ above this minimum were flagged as variable. The results of this analysis are presented in Sect.~\ref{diss}.

To search for possible periodicity of the sources, we extracted the barycentric corrected event files of each source from the observation, where it had the highest number of counts.  We applied the $Z^2_1$ analysis to the arrival photons (0.2$-$12\,keV) of each source \citep[]{1983A&A...128..245B, 1988A&A...201..194B}. We searched for periodic signals in the range of $0.146-10^4$\,s (0.146 is the Nyquist limit based on the time resolution of EPIC-pn in full-frame mode, while $10^4$\,s is of the order of the duration of the observation). No periodic variability was found for any of the sources. Following the method described in \citet{10.1093/mnras/268.3.709}, we found a 90\% upper limit of the pulsed fraction (the pulsed fraction is defined as the semi-amplitude of the modulation divided by the mean source magnitude) of $\approx 85$\% in the period range of $0.146-10^2$\,s and $\approx 60$\% in the period range of $100-1000$\,s. The pulsed fractions of accreting pulsars and magnetic white dwarfs can be lower than the aforementioned upper limits \citep[see e.g,][and references therein]{1997ApJS..113..367B, 2009ApJ...701.1992K}, suggesting that these observations are not sensitive enough to set a stringent upper limit on the pulsed fractions of the sources studied in this work.

%\textbf{Following the method described in \citet{10.1093/mnras/268.3.709}, we found for the studied sources a 90\% upper-limit of the pulsed fraction (the pulsed fraction is defined as the semiamplitude of the modulation divided by the mean source magnitude) of $\approx 85$\% in the period range $0.146-10^2$\,s and $\approx 60$\% in the period range $100-1000$\,s. The pulsed fractions of accreting pulsars and magnetic white dwarfs can be lower than the aformentioned upper-limits \citep[see e.g,][and references therein]{1997ApJS..113..367B, 2009ApJ...701.1992K}, suggesting that these observations are not sensitive enough to set a stringent upper-limit on the pulsed fractions of the studied sources.}

\subsection {Optical and $UV$ timing analysis}
We reduced the data from the Optical Monitor\,(OM) onboard \xmm\,\citep{2001A&A...365L..36M} using the SAS (V.16.0.0) task \texttt{omchain}. The sources were observed using the $U$~band optical filter (300–390 nm) and the $UV$ filter, $UVW1$\,(245–320 nm), of the OM telescope in both series of observations in 2009 and 2015. The \textit{XMM}-OM telescope provides coverage of the central 17$\arcmin$ square region of the EPIC FOV. Therefore, optical light curves are only available for a fraction of the X-ray sources.
 For observations, in which the optical counterpart of the source was located in the FOV of the OM camera, but not detected, the magnitude of the upper limit was calculated using the following formula considering the zero-points of 18.24 mag, and 17.37 mag for the $U$ and the $UVW1$ band, respectively \citep{2001A&A...365L..36M}.
\begin{equation}
m= -2.5\times {\rm log}_{10}(DN / EXPTIME) + ZEROPOINT, 
\end{equation}
where $DN$ is the background number of counts and $EXPTIME$ is the exposure time of the observation \citep{1856MNRAS..17...12P}. The long-term optical/UV light curves of all sources are shown in Appendix~\ref{appen-light-curve}. In addition, We applied the Lomb-Scargle technique for the unevenly sampled time series \citep{1982ApJ...263..835S} to search for the possible periodicity in optcial/UV data of the sources . No significant periodicity was found.

\subsection{Spectral analysis}
\label{sa}

For sources, which have been frequently detected over 31 observations, we obtain a single spectrum with high statistics by merging the spectra of all observations, in which the source was detected.  We excluded the spectrum of observations, in which significant flare activity  was observed in the light curve of the source according to the analysis described in Sect.~\ref{X-time}.
The source spectrum, background spectrum, ancillary response, and response matrix files of all available observations were combined separately for each EPIC camera, using the \texttt{SAS} task \texttt{epicspeccombine}. In most of the cases, we only used the EPIC-pn data because of their higher statistics compared to EPIC-MOS data. We used EPIC-MOS data in addition to EPIC-pn for sources that have a number of detections in EPIC-MOS comparable with the EPIC-pn. In fact, except for source No.\,9, the detection in the FOV of EPIC-pn was more frequent than for EPIC-MOS. The data of the combined spectrum has been grouped to a minimum of 20 counts per bin. We used XSPEC\,(Ver. 12.10.0) to fit the spectra. The models are fitted using the $\chi^2$ statistics. 

\section{Multi wavelength studies of counterparts}

In the following, we discuss the multi-wavelength photometry used to uncover the stellar nature of our sources. 
\subsection{Optical counterparts of the sources}
\label{exp-opt}
\begin{figure}[!htb]
\centering
\includegraphics[trim={3.2cm 0.0cm 0.0cm 0.cm},width=0.45\textwidth]{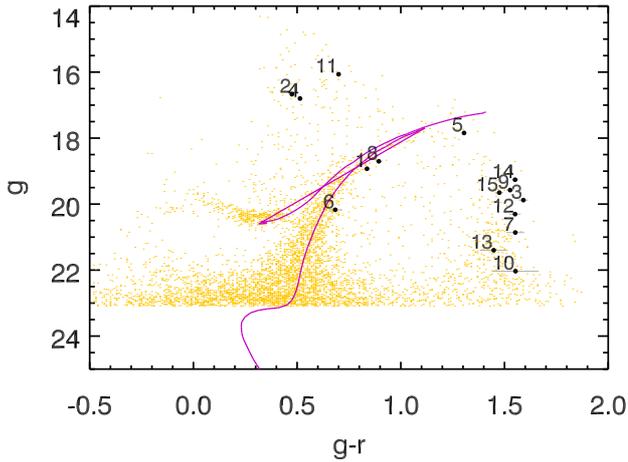}
\caption{Colour-magnitude diagram ($g$ versus $g-r$) for the SDSS9 optical counterparts of the X-ray sources (black data points with source No.~as labels). Orange dots represent the SDSS7 members\,($g<23$ mag) of the Draco dSph as classified by \citet{2003ApJS..145..245R}. The red line is the stellar isochrone for the age of 10~Gyrs and metallicity of 0.0004 solar metallicity of the Draco dSph according to \citet{2004A&A...422..205G}. \label{opt-counterpart}}
\end{figure}

\begin{figure}[!htb]
\centering
\includegraphics[trim={3.2cm 0.0cm 0.0cm 0.cm},width=0.45\textwidth]{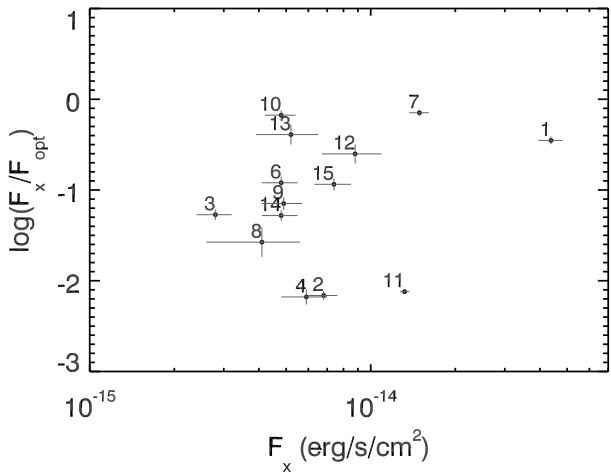}
\caption{Logarithmic X-ray to optical flux ratio log$(\frac{F_\text{X}}{F_\text{opt}})$ versus $F_{\rm x}$.  log$(\frac{F_\text{X}}{F_\text{opt}})$ is calculated using Equation~\ref{fx-fopt}. \label{opt-x}}
\end{figure}

The most recent optical survey in the field of Draco dSph is the 9th released data of the Sloan Digital Sky Survey \citep[SDSS9,][]{2012ApJS..203...21A}. The SDSS9 catalogue contains the magnitudes of the objects in five different energy bands from the near ultraviolet\,(UV) to the near infrared ($u=3551\AA$, $g=4686\AA$, $r=6165\AA$, $i=7481\AA$, $z=8931\AA$). Table\,\ref{opt-count-table} presents the magnitudes in different bands of SDSS9 for the optical counterparts of the X-ray sources. Appendix\,\ref{SDSS-image} shows the images of the optical counterparts of the X-ray sources taken from SDSS9\footnote{\url{https://dr9.sdss.org/fields}}. The colour-magnitude diagram of the optical counterparts was plotted using the magnitudes of the $g$ and $r$ bands (Fig.~\ref{opt-counterpart}). We also plotted the logarithmic X-ray to optical flux ratio log$(\frac{F_\text{X}}{F_\text{opt}})$, versus the X-ray flux (Fig.\,\ref{opt-x}). The flux ratio log$(\frac{F_\text{X}}{F_\text{opt}})$ was calculated using the modified equation of \citet{1988ApJ...326..680M} in SDSS optical bands \citep{2016A&A...586A..64S}:
\begin{equation}
  \label{fx-fopt}
{\rm log}\bigg(\frac{F_ \text{X}}{F_\text{opt}}\bigg)={\rm log_{10}(F_\text{X})}+\frac{g+r}{2\times2.5}+5.37,
\end{equation}
where $F_\text{X}$ is the X-ray flux and $g$ and $r$ are the SDSS magnitudes of the optical counterpart associated with the X-ray source. Fig.~\ref{opt-x} shows that all sources have log$(\frac{F_\text{X}}{F_\text{opt}})<0$, typical for stars.

For the $u$, $g$, $r$, $i$, and $z$ bands, the Galactic extinction of 0.12, 0.09, 0.06, 0.04, and 0.03 mag is used, respectively, in the direction of the Draco dSph \citep{2011ApJ...737..103S}.

\subsection{The infrared counterparts of the sources}
\label{exp-infra}
We also searched for mid-infrared counterparts in the WISE All-Sky Data in four energy bands \citep[3.4, 4.6, 12, and 22 $\mathrm \mu$m, called \textit{W1, W2, W3, W4}, respectively;][]{2012yCat.2311....0C} and near-infrared counterparts in the 2MASS All-Sky Catalogue of Point Sources in the $J$, $H$, $K$ bands \citep{2003yCat.2246....0C}. Table\,\ref{inf-count} lists the WISE and 2MASS magnitudes of counterparts of the X-ray sources. For the $J$, $H$, and $K$ bands the Galactic extinction in the direction of the Draco dSph of 0.02, 0.01, 0.01 mag,  respectively, are taken into account \citep{1998ApJ...500..525S}.

\begin{figure}[!htb]
  \centering
  \includegraphics[trim={3.2cm 0.0cm 0.cm 0.cm},width=0.45\textwidth]{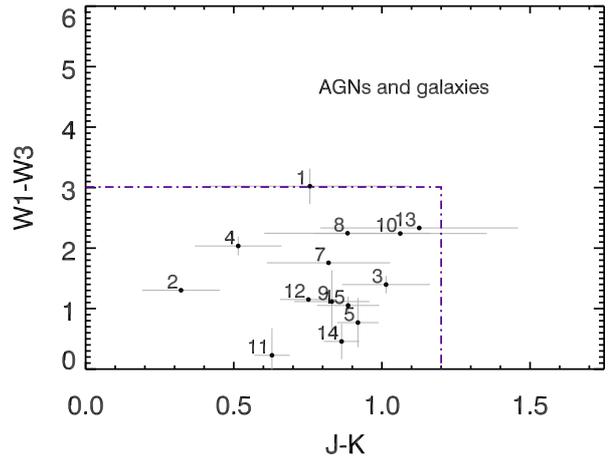}
\caption{Colour-colour diagram of mid-infrared WISE $W1\,(3.4\, \mathrm{\mu m})- W3 \,(12\, \mathrm{\mu m})$ colour index versus near-infrared 2MASS $J-K$ colour index for the counterparts of X-ray sources in the field of the Draco dSph detected by \xmm. Dash-dotted lines separate the regions, which are occupied by the counterparts of background sources from that of the others \citep[Fig.\,3,][]{2016A&A...586A..64S}. \label{infra}}
\end{figure}

\subsection{Results of multi-wavelength studies}
\label{multi-diss}

Photometric studies of the Draco dSph showed that the main population of stars in this dwarf galaxy has already left the main sequence and is on the red giant branch \citep[e.g,][]{2001AJ....121..841P, 2002AJ....124.3222B, 2003ApJS..145..245R, 2007MNRAS.375..831S}. The orange dots in Fig.~\ref{opt-counterpart} show the main population of stars in the Draco dSph classified by \citet{2003ApJS..145..245R}.

In colour-magnitude diagrams, the optical and infrared counterparts of X-ray sources, which are background objects (galaxies and AGNs) are located in different regions compared to the foreground objects and members of Draco dSph. According to \citet{2016A&A...586A..64S} the optical and infrared counterparts of non-background sources satisfy the conditions of log$(\frac{F_\text{X}}{F_\text{opt}})$\,<0.0, $J-K$\,< 1.2, and $W1-W3$\,< 3.0. All the optical and infrared counterparts associated with the X-ray sources, which are studied in this work fulfill these criteria for non-background sources. The colour-colour diagrams of WISE \,($W1-W3,\,W3$) and 2MASS\,($J-K, K$) show that none of the counterparts is found in the region, where background sources are located \citep[see our Fig.~\ref{infra}, and Fig.\,3 in][]{2016A&A...586A..64S}. Also, all these X-ray sources have log$(\frac{F_\text{X}}{F_\text{opt}})$\,<0.0.

These criteria, which separate the X-ray sources associated with stars from those associated with galaxies and AGNs, are consistent with other photometric classification criteria. \citet{2012MNRAS.419...80A} performed a photometric study to distinguish the sources detected in the SDSS into stars, galaxies, and quasars. They showed that stars have very different $u-g$ and $g-r$ colours than quasars. The optical counterparts of all X-ray sources of this work have $u-g$>1.5 and $g-r$>1.0. According to \citet[][]{2012MNRAS.419...80A}, they can be classified as stars.

The optical colour-magnitude diagram (Fig.~\ref{opt-counterpart}) shows that the optical counterparts of some X-ray sources are located at the end of the tail of the red-giant branch of the Draco dSph \,(sources No.\,1, 5, 6, 8) and can be identified as members of the Draco dSph (Sect.~\ref{diss-symbio}). Optical counterparts associated with sources No.\,3, 7, 9, 10, 12, 13, 14, 15 appear redder and are Galactic M\,dwarfs (see Sect.~\ref{diss-m-dwarfs}). The optical counterparts of sources No.\,2, 4, 11 are located above the red giant branch and seem to be hotter Galactic stars than M\,dwarfs (Sect.~\ref{diss-foreground}). 
 
\begin{figure}[!htb]
\includegraphics[ width=0.51\textwidth, trim=1.0cm 2.cm 1.0cm 0.5cm]{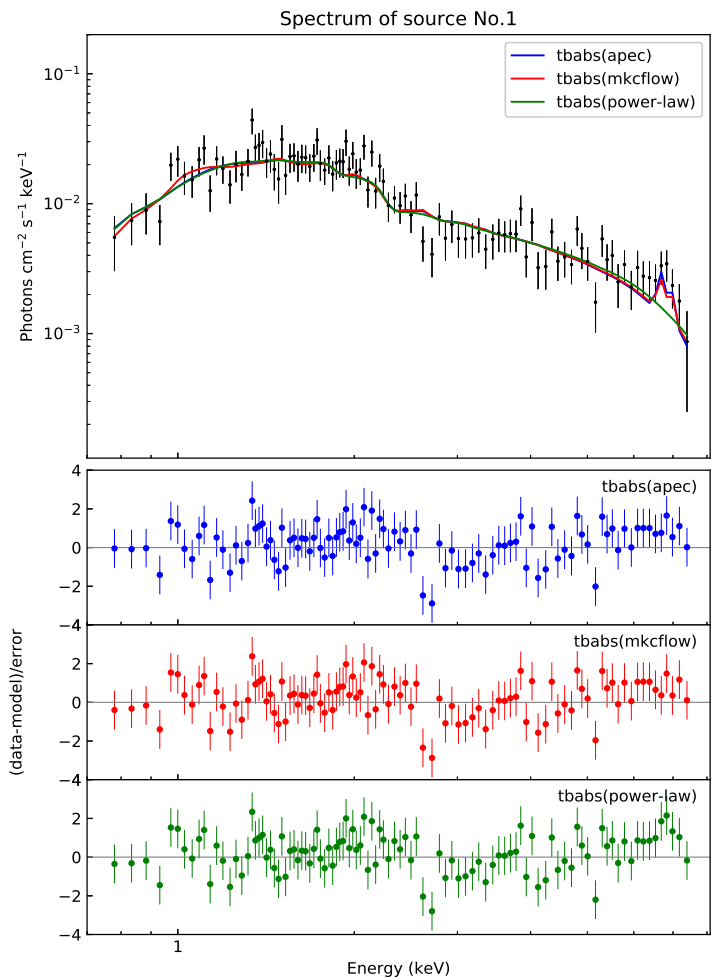}
\caption{Combined spectrum of all \xmm\, observations and best-fit models assuming different types of emission for the source No.\,1, together with the residuals of the fitted models in units of the standard deviation. \label{srcno.1.spec.sym}}
\end{figure}

%\begin{figure*}[!htb]
%\includegraphics[angle=270, width=0.51\textwidth, trim=1.5cm 0.cm 0.cm 2.0cm]{src-6-1.ps}
%\includegraphics[angle=270, width=0.51\textwidth, trim=1.5cm 2.cm 0.cm 0.0cm]{src-27-5.ps}
%\includegraphics[angle=270, width=0.51\textwidth, trim=1.5cm 0.cm 0.cm 2.0cm]{src-25-6.ps}\\
%\includegraphics[angle=270, width=0.51\textwidth, trim=1.5cm 2.cm 0.cm 0.0cm]{src-31-8.ps}
%\caption{Combined spectrum of all \xmm\,observations for the symbiotic systems in the Draco dSph studied in this work, together with the residuals in units of the standard deviation. \label{src.spec.sym}} 
%\end{figure*}

\begin{figure*}
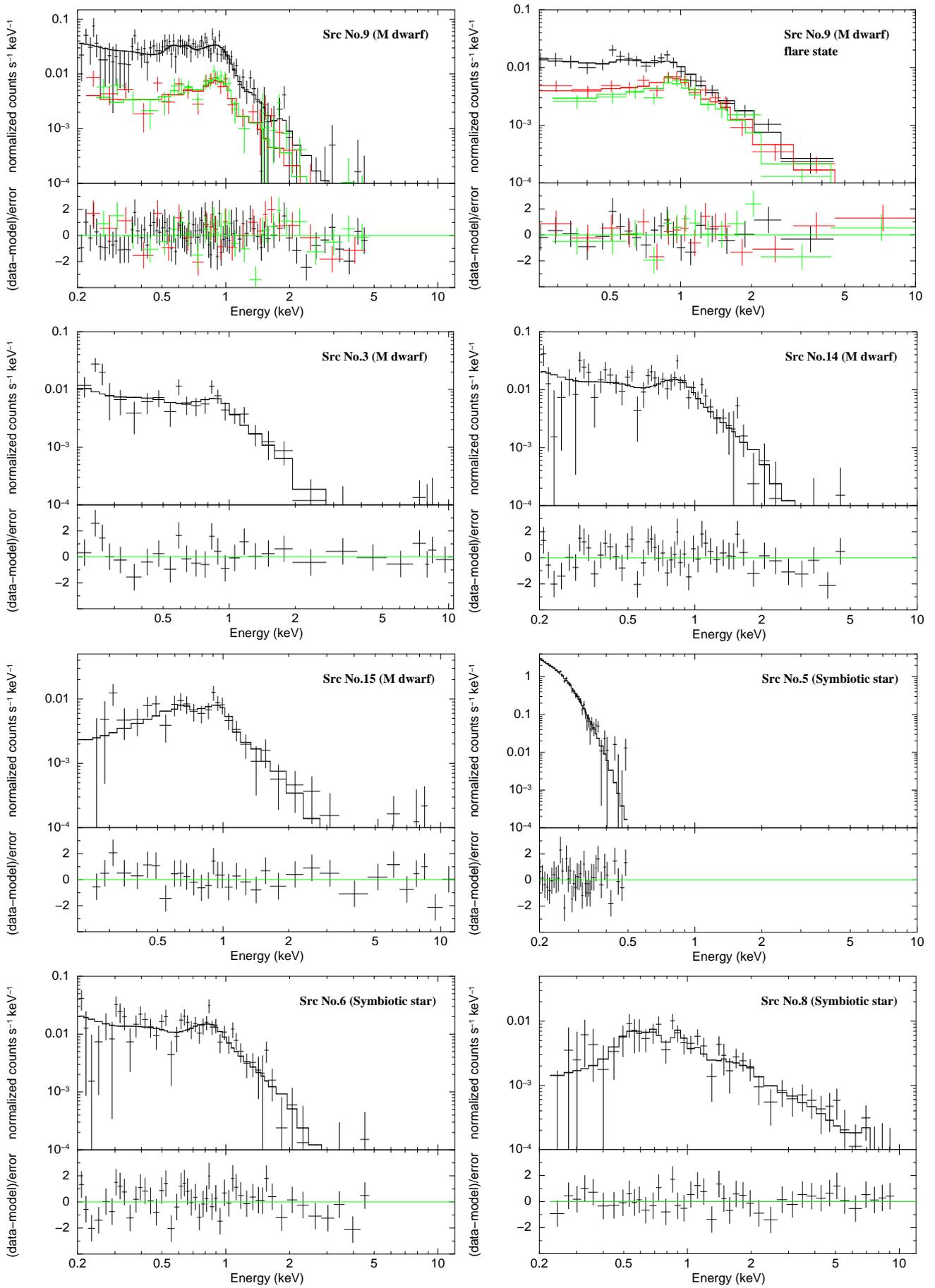

\includegraphics[angle=270, width=0.5\textwidth, trim=1.5cm 0.cm 0.cm 2.0cm]{src-33-9.ps}
\includegraphics[angle=270, width=0.5\textwidth, trim=1.5cm 2.cm 0.cm 0.0cm]{src-33-flare-state.ps}\\
\includegraphics[angle=270, width=0.5\textwidth, trim=1.5cm 0.cm 0.cm 2.0cm]{spec-3-5.ps}
\includegraphics[angle=270, width=0.5\textwidth, trim=1.5cm 2.cm 0.cm 0.0cm]{src-13-2.ps}\\
\includegraphics[angle=270, width=0.5\textwidth, trim=1.5cm 0.cm 0.cm 2.0cm]{src-14-4.ps}
\includegraphics[angle=270, width=0.5\textwidth, trim=1.5cm 2.cm 0.cm 0.0cm]{src-27-5.ps}\\
\includegraphics[angle=270, width=0.5\textwidth, trim=1.5cm 0.cm 0.cm 2.0cm]{src-25-6.ps}
\includegraphics[angle=270, width=0.5\textwidth, trim=1.5cm 2.cm 0.cm 0.0cm]{src-31-8.ps}
\caption{Combined spectrum of all \xmm\, observations of sources\,(M\,dwarfs and symbiotic stars) in the field of Draco dSph, which have enough statistics for spectral analysis: EPIC-pn\,(black), EPIC-MOS1 (red), and EPIC-MOS2 (green), together with the residuals in units of the standard deviation. \label{src.spec.mdwarf}} 
\end{figure*}
\section{Discussion}
\label{diss}
\subsection{Symbiotic stars in the Draco dSph}
\label{diss-symbio}

\begin{figure}[!htb]
\includegraphics[clip, trim={0.cm 7.5cm 0.cm 2.cm},width=0.50\textwidth]{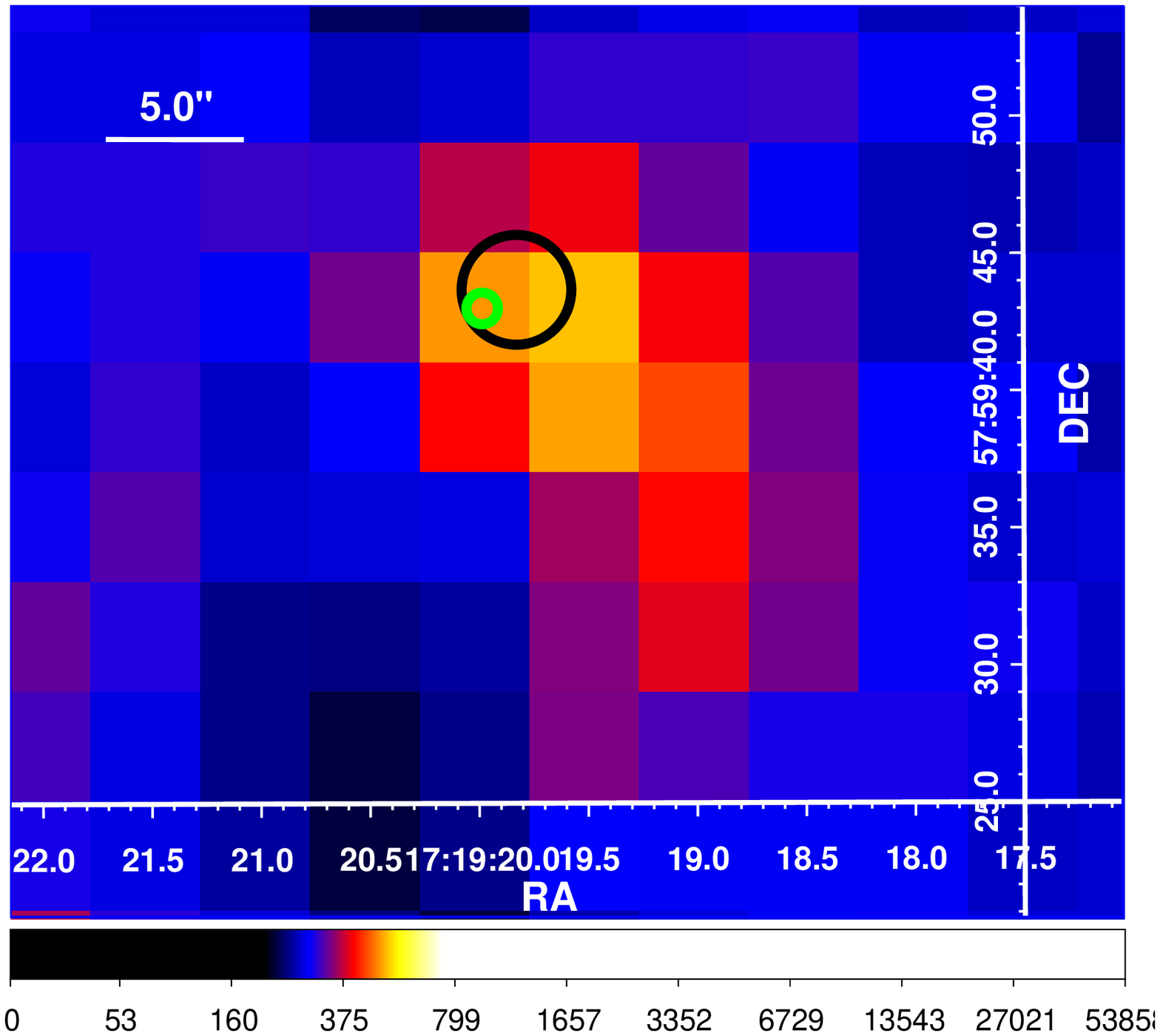}
\caption{The X-ray mosiac image of Source No.~1 in the energy band 0.2-12.0 keV. The black circle shows the X-ray position of the source. The green circle shows the position of the optical counterpart.\label{DS9-src1}}
\end{figure}

\begin{table}[!htb]
\caption{Best-fit parameters of the X-ray spectrum of source No.\,1 for three different models. Errors are at the $90\%$ confidence level.}  \label{spectral-Source-No.1}
\small
\centering
\addtolength{\tabcolsep}{7pt}
\begin{tabular}{lcc}
\hline\hline
\multicolumn{3}{c}{\texttt{tbabs$\times$(apec)$^{\ast}$}} \\
\hline
%\multicolumn{3}{c}{------------------------------------------} \\
%\multicolumn{1}{c}{Multi-column}\\
\vspace{1mm}
$N_{\rm H}$& $10^{22}$ cm$^{-2}$ & 0.83$^{+0.12}_{-0.11}$\\
\vspace{1mm}
$kT$&keV&9.57$^{+4.28}_{-2.22}$\\
\vspace{1mm}
Norm.&&(2.66$^{+0.18}_{-0.17}$)$\times10^{-5}$\\
\vspace{1mm}
$\chi^2$ (d.o.f)& & 1.05\,(99) \\
\vspace{1mm}
$F_{\rm X}^{\star}$ &erg\,s$^{-1}$\,cm$^{-2}$& (4.20$^{+0.26}_{-0.25}$)$\times10^{-14}$ \\
\hline
\multicolumn{3}{c}{\texttt{tbabs$\times$(mkcflow)$^{\ast}$}}\\
\hline
\vspace{1mm}
$N_{\rm H}$& $10^{22}$ cm$^{-2}$ &0.98$^{+0.16}_{-0.16}$\\
\vspace{1mm}
$kT_{\rm min}$&keV& 0.08 frozen\\
\vspace{1mm}
$kT_{\rm max}$&keV&31.19$^{+36.38}_{-13.42}$\\
\vspace{1mm}
Norm.&& (8.67$^{+5.15}_{-3.83}$)$\times10^{-9}$\\
\vspace{1mm}
$\chi^2$ (d.o.f)&& 1.05\,(99)\\
\vspace{1mm}
$F_{\rm X}^{\star}$&erg\,s$^{-1}$\,cm$^{-2}$&(4.46$^{+0.30}_{-0.29}$)$\times10^{-14}$\\
\hline
\multicolumn{3}{c}{\texttt{tbabs$\times$(power-law)}}\\
\hline
%\vspace{1mm}
$N_{\rm H}$&$10^{22}$ cm$^{-2}$&0.92$^{+0.17}_{-0.15}$\\
\vspace{1mm}
PhoIndex&&1.62$^{+0.17}_{-0.16}$\\
\vspace{1mm}
Norm.&& (7.73$^{+1.80}_{-1.43}$)$\times10^{-6}$\\
\vspace{1mm}
$\chi^2$ (d.o.f)&& 1.10\,(99)\\
\vspace{1mm}
$F_{\rm X}^{\star}$&erg\,s$^{-1}$\,cm$^{-2}$& (4.33$^{+0.29}_{-0.27}$)$\times10^{-14}$\\
\hline
%\hline
%\vspace{1mm}
%$N_{\rm H}$&$10^{22}$ cm$^{-2}$&1.06$^{+0.33}_{-0.19}$\\
%\vspace{1mm}
%PhoIndex&&1.83$^{+0.52}_{-0.22}$\\
%\vspace{1mm}
%Norm.&& (9.53$^{+5.98}_{-2.10}$)$\times10^{-6}$\\
%\vspace{1mm}
%LineE&keV&6.71$^{2.05}_{-0.48}$\\
%\vspace{1mm}
%Sigma&keV&0.50$^{+18.59}_{-0.34}$\\
%\vspace{1mm}
%Norm.&& (4.55$^{+1.08}_{-2.51}$)$\times10^{-7}$\\
%\vspace{1mm}
%$\chi^2$ (d.o.f)&& 0.97\,(96)\\
%\vspace{1mm}
%$F_{\rm X}^{\star}$&erg\,s$^{-1}$\,cm$^{-2}$& (4.81$^{+0.47}_{-0.37}$)$\times10^{-14}$\\
%\hline
\end{tabular}
\tablefoot{$\star$: The absorbed flux is calculated in the energy range of 0.7--8.0~keV. $\ast$: The abundances of the \texttt{apec} and \texttt{mkcflow} models are fixed to the Solar abundances of 1.0.}
 \end{table}

\begin{table*}
\caption{Best-fit parameters of the X-ray spectra. Errors are at the 90$\%$ confidence level.
\label{spectral-Table-sym}
}
\small
\centering
\addtolength{\tabcolsep}{-0.1cm}   
\begin{tabular}{lllcccccc}
\hline\hline\
Src-No &Type$^{\dagger}$& Model & $N_{\rm H}$& $kT$  & Abundance&$\chi^2$ (d.o.f) &  Absorbed $F_{\rm X}^{\ast}$&$L_{\rm X}^{\ast}$\\
&&&$10^{22}$ cm$^{-2}$&keV&&&$10^{-15}$erg\,s$^{-1}$\,cm$^{-2}$&erg\,s $^{-1}$\\
\hline
\vspace{1mm}
%1&SS&\texttt{tbabs$\times$(apec)}&0.62$^{+0.07}_{-0.07}$  &8.8$^{+3.6}_{-1.8}$ &0.7$^{+0.6}_{-0.3}$& 1.14\,(129)& 5.84$^{+0.32}_{-0.28}$& 4.7$\times 10^{34}$\\
%\vspace{2mm}
3&MD&\texttt{tbabs$\times$(apec)}&<0.02&0.89$^{+0.20}_{-0.15}$& <0.1 &0.94(26) & 3.28$^{+0.34}_{-0.29}$&1.0$\times$10$^{29}$\\
\vspace{2mm}
5$^{\star}$&SS &\texttt{tbabs$\times$bb} & 0.04$^{+0.02}_{-0.02}$ & 0.015$^{+0.003}_{-0.003}$ &&1.35\,(33)&68.22$^{+3.75}_{-3.35}$&5.5$\times 10^{34}$\\
\vspace{2mm}
6&SS&\texttt{tbabs$\times$apec}&<0.03&4.06$^{+2.62}_{-1.29}$&&0.82\,(28)&4.75$^{+0.73}_{-0.73}$&3.9$\times 10^{33}$\\
\vspace{2mm}
8&SS& \texttt{tbabs$\times$apec}&<0.08&6.2$^{+19.4}_{-2.3}$&&0.76\,(40)&4.08$^{+1.53}_{-1.14}$&3.8$\times 10^{33}$\\
\vspace{2mm}
9&MD& \texttt{tbabs$\times$(apec+apec)}& <0.03 &0.21$^{+0.03}_{-0.03}$ & 0.33$^{+0.45}_{-0.15}$ & 1.03\,(128) & 4.87$^{+0.78}_{-0.38}$ &1.4$\times10^{29}$\\
&&&&0.92$^{0.07}_{-0.08}$&&&\\
\vspace{2mm}
9 &MD (flare-state)& \texttt{tbabs$\times$(apec+apec+apec)}& <0.03 &frozen to 0.21 & frozen to 0.33 & 0.95\,(46) & 36.80$^{+3.13}_{-3.56}$&1.1$\times10^{30}$ \\
\vspace{2mm}
&&&&frozen to 0.92 &&&\\
\vspace{2mm}
&&& &2.84$^{+1.61}_{0.72}$&&&\\
\vspace{2mm}
%10&MD&\texttt{tbabs$\times$(apec)}&<1.1& 0.22$^{+0.58}_{-0.18}$&&0.60\,(11)&1.83$^{+8.09}_{-9.27}$&7.8$\times10^{28}$\\
%vspace{2mm}
11&FG&\texttt{tbabs$\times$(apec+apec)}&0.05$^{+0.02}_{-0.01}$&0.27$^{+0.06}_{-0.04}$&0.17$^{+0.06}_{-0.05}$&1.00\,(174)& 13.2$^{+0.05}_{-0.05}$ & 1.2$\times10^{30}$ \\
&&&& 1.07$^{+0.06}_{-0.06}$ &&&\\
\vspace{2mm}
14&MD&\texttt{tbads$\times$(apec)}&<0.07&0.78$^{+0.13}_{-0.10}$&<0.1&1.18\,(46)&5.36$^{+0.65}_{-0.66}$&1.2$\times10^{28}$\\
\vspace{2mm}
15&MD&\texttt{tbabs$\times$(apec+apec)}&frozen to 0.02&0.21$^{+0.11}_{-0.06}$&0.14$^{+0.41}_{-0.11}$&1.66\,(29)& 7.38$^{1.14}_{-1.12}$&2.3$\times10^{29}$\\
&&&0.12$^{+1.05}_{-0.08}\times 10^{22}$&1.02$^{+0.25}_{-0.19}$&&&\\
\hline
\end{tabular}
\vspace{-2mm}
\tablefoot{$\dagger$: The types of sources are symbiotic stars (SS), Galactic M~dwarfs (MD), and other types of foreground star (FG). $\ast$:~Flux and luminosity of  all sources are calculated in the energy range of 0.2--5.0 keV,  except sources  No.\,1 and No.\,8, which are in the energy range of 0.2--10 keV.  $\star$:~Source No.\,5 is the super-soft symbiotic white dwarf system Draco C1. The fitted parameters for this source are taken from \citet{2018MNRAS.473..440S}.}
 \end{table*}

The colour-magnitude diagram in Fig.~\ref{opt-counterpart} shows that the optical counterparts of the four sources, No.\,1, 5, 6, and 8 are located on the red-giant branch of Draco dSph. These four sources have also been classified as members of the Draco dSph in other surveys. According to the infrared photometric analysis, symbiotic stars are distinguished into two main groups of either cool star (S-type) with a typical infrared colour of the red giant branch, or as a star with a significant contribution of warm dust (D-type) typical for the asymptotic giant branch \citep[][]{2008A&A...480..409C}. The infrared counterparts of symbiotic stars of the Draco dSph satisfy the conditions of $J-H$~<1.0, $H- K_{\rm s}$~<0.5, $H-W2$<1.0, and $J-W1$<2 (see Table.\,\ref{inf-count}), which make them candidates for S-type symbiotic stars \citep[][]{2019MNRAS.483.5077A, 2008A&A...480..409C}. The properties of these systems, which we identify as symbiotic systems in Draco dSph are discussed in the following. 

\textbf{Source No.\,1 (XMMUJ171919.8+575943):} In the previous work, based on the five \xmm\, observations of 2009, no optical/infrared counterpart was found for the source. Only a radio counterpart was found and the source was classified as an AGN candidate \citep[Source No.\,12,][]{2016A&A...586A..64S}. The longer exposure times of the \xmm\, observations in 2015 yielded a more accurate position of the X-ray source and a bright optical/infrared counterpart was found. Figure~\ref{DS9-src1} shows the $3\sigma$ error circle of the X-ray source position in observation 17,  where source No.\,1 had the highest detection maximum likelihood. The optical counterpart of the source is a red giant in Draco dSph. The radial velocity of the source is --306.59\,km\,s$^{-1}$, which is consistent with that of the Draco dSph members \citep{2002MNRAS.330..792K}. \citet{2010ApJS..191..352K} measured an effective temperature of $T_{\rm eff}=4660$\,K and a metallicity of [Fe/H]=--2.12 for the red giant. %The time-averaged X-ray spectrum was fitted very well with an absorbed collisionally-ionized thermal gas model\,(\texttt{apec})\textbf{, An isobaric cooling flow model\,(\texttt{mkcflow}) and a power law model (see Table\,\ref{spectral-Table-sym}} and Fig.\,\ref{src.spec.sym}). 
%The spectrum of source No.\,1 shows an emission line at $\sim$6.7\,keV, which can be ascribed to Fe\,XXV transitions in a thermal plasma, as observed in other known AWDs e.g, 4\,Dra \citep[][]{2016ApJ...824...23N}.  %The radio counterpart of source is observed with a flux density of $56.8\pm6.6$\,mJy  at $150\,MHZ$ \citep[GMRT survey,][]{2017A&A...598A..78I} and has a flux of $8.7\pm0.5$\,mJy at $1.4GHZ$ \citep[NVSS survey,][]{1998AJ....115.1693C}, similar flux of $10.37\pm0.14$ at $1.4GHZ$ \citep[FIRST survey,][]{2011ApJ...737...45O}. The radio counterparts are very common in the symbiotic satrs \citep[e.g,][]{1984ApJ...284..202S, 1991MNRAS.249..374I}. The radio emission of the symbiotic systems can have two origins: first, the radio jets of the collision of the winds white dwarf and the giant companion star, which can be seen particularly in the outbursts of the system, and second, more permeant jets, which originate from the relativistic ejections of the accretion disk, as it have seen in some cases of symbiotic stars \citep[e.g, Z\,And][]{2004MNRAS.347..430B}. The accretion disk of source No.1 suggests that the radio emission of this source can be from the relativistic ejection. The radio analysis has to be investigated for this system.
The X-ray flux of the source steadily decreased and again increased within about one month in the last observations of 2015 (see Appendix\,\ref{appen-light-curve}), which can be a sign of eclipse in the system. The typical orbital period of symbiotic stars is of the order of hundred days \citep{2000A&AS..146..407B}. If the observed minimum in the X-ray light curve is related to an eclipse of the white dwarf, its inclination must therefore be very low to explain the short duration of eclipse. On the other hand, UV and optical light curves do not show any evidence for a drop in flux. Therefore, the eclipse scenario remains controversial.

The time-averaged X-ray spectrum is fitted very well with an absorbed collisionally-ionized thermal gas model\,(\texttt{apec}), an absorbed isobaric cooling flow model\,(\texttt{mkcflow}), and also an absorbed power-law model (see Table\,\ref{spectral-Source-No.1} and Fig.\,\ref{srcno.1.spec.sym}). The parameters of the spectrum of the \texttt{apec} model and \texttt{mkcflow} are very similar to those of symbiotic stars, which show hard X-ray emission and are classified as $\delta$-type symbiotic stars \citep[][]{1997A&A...319..201M, 2013A&A...559A...6L}. These symbiotic stars are highly absorbed hard X-ray sources with thermal emission above 2.4 keV, which is assumed to originate from the boundary layer of an accretion disk \citep{2013A&A...559A...6L} like in e.g, Hen~3-461 \citep{2013A&A...559A...6L}, CD-28~3719, 4~Dra \citep{2016ApJ...824...23N}, T\,CrB \citep{2008ASPC..401..342L}, and  RT\,Cru \citep{2007ApJ...671..741L, 2016A&A...592A..58D}. The spectrum is also well fitted with an absorbed power law model. Residuals obtained by subtracting the continuum (Fig.\,\ref{srcno.1.spec.sym}, lower panel) show a possible emission feature within the energy range 6.3$-$7\,keV, that could be interpreted as noise fluctuation or one (or more) iron line emission. When a Gaussian component is added to the model, the fit is improved by $\Delta \chi^2 \approx 15.6$. To determine whether the data require this additional component, we used the XSPEC script lrt to perform the likelihood ratio test on $10^4$ simulated datasets. We find that the probability that the observed spectrum can be described with an absorbed power law without a Gaussian component is $\sim 0.23$\%. Therefore, we conclude that the line significance is marginally larger than $3\sigma$. The 90\% uncertainties on the line energy parameter ($E_{\rm line}=6.7^{+2.0}_{-0.5}$\,keV) does not allow us  to determine which emission lines (Fe XXV, Fe XXVI, and Fe XXIII) are responsible for the observed feature.
%An isobaric cooling flow model\,(\texttt{mkcflow}) has been used to fit the X-ray spectrum of $\delta$-type symbiotic stars \citep[e.g,][]{2013A&A...559A...6L,2007ApJ...671..741L,2008ASPC..401..342L}. 
%The low statistics and the lack of spectral coverage at energy greater than 10\,keV did not allow us to obtain a good fit of the spectrum with \texttt{mkcflow}.

Assuming a distance of $\sim$82\,kpc, the X-ray luminosity of the source is $>10^{34}$\,erg\,s$^{-1}$ in the energy range of 0.7--8.~keV. Such a high $L_\text{X}$, together with the relatively hard X-ray emission of the source is compatible with the emission coming from the boundary layer around a non-magnetic white dwarf \citep[see][]{2016A&A...592A..58D, 2008ASPC..401..342L}. It is also reminiscent of the typical X-ray emission from magnetic white dwarfs \citep[e.g,][]{2005A&A...435..191S}. Also, the good fit of power-law model emphasises that the compact object can be a neutron star too. Therefore, the nature of the compact object remains unclear. %To distinguish the neutron star from, e.g., intermediate polars or optically-thin boundary layers around non-magnetic white dwarfs deeper observations at energies above 10 keV are necessary.
%The power-law model for the spectrum, which also yielded a good fit, was used assuming that the compact object can be a neutron star. Therefore, the nature of the compact object remains unclear.

\textbf{Source No.\,5 (Draco\,C1, XMMUJ171957.6+575005):} This source is a known super-soft symbiotic star in the Draco dSph detected by ROSAT in X-rays \citep{1984PASAu...5..369A, 1997A&A...319..201M}. We recently studied the properties of this source using all available \xmm\, observations \citep{2018MNRAS.473..440S}. 

\textbf{Source No.\,6 (XMMUJ172005.6+575759):} The optical counterpart of this source was classified as a red giant in the Draco dSph with an effective temperature of $T_{\rm eff}=4973$~K and a metallicity of [Fe/H]=--2.28 \citep{2015ApJ...801..125K}. The long-term light curve of the source (Appendix\,\ref{appen-light-curve}) shows higher X-ray flux in observations 23 and 31. We checked the short-term light curves of the source in these two observations but the count rate was too low to significantly observe flares or outburst activity. These high X-ray fluxes are not correlated with changes in the UV and optical $U$ magnitudes. The source showed variations in UV and optical $U$-band light curves, but no evidence of periodicity was seen in the light curves. We could fit the X-ray spectrum with  an absorbed \texttt{apec} model (see Table~\ref{spectral-Table-sym} and Fig.~\ref{srcno.1.spec.sym}). The system has an $N_{\rm H}$ similar to the  $N_{\rm H}$ in the direction of the Draco dSph. The statistics of the spectrum are too poor to determine the element abundance of the emitting gas. Therefore, the abundance is left at the default abundance of the \texttt{apec} model (1.0 times the solar values).  The soft X-ray spectrum of the source makes it a candidate for a $\beta$-type symbiotic-star.  $\beta$-type symbiotic stars have the most part of emission at energies $\lesssim$2~keV \citep{2013A&A...559A...6L}. The X-ray emission is most likely caused by the collision of the wind of the white dwarf with the wind of the red giant \citep{1997A&A...319..201M}. However, in this case the system must have been observed during the outburst \citep{2013A&A...559A...6L}. Another model assumes that if the system is observed almost edge-on, the scattering of hard photons in a ionized medium around the white dwarf causes $\beta$-type emission \citep{2006MNRAS.372.1602W}. However, \citet{2013A&A...559A...6L} argued that this model makes it unlikely to detect many symbiotic systems with soft X-ray emission, whereas there are plenty of these systems. As an alternative explanation, \citet{2013A&A...559A...6L} suggested that the collision of the wind from the accretion disk with the red-giant wind can cause soft X-ray emission. The high temperature of the source suggests that the X-ray emission comes from the colliding region of fast winds. This source has no counterpart in WISE and 2MASS catalogues.

\textbf{Source No.\,8 (XMMUJ172013.3+575051):} The optical counterpart of this source was classified as a member of the Draco dSph with a radial velocity of --291.3 km\,s$^{-1}$ \citep{1995AJ....110.2131A}. The long-term light curve of the source shows that the X-ray fluxes in the observations of 2009 were higher than during the observations in 2015. The observations of 2015 show no significant variation over the 26 observations ($Var$=1.8, see Sect.~\ref{X-time}). The X-ray spectrum of the source is soft, which makes it a candidate for a $\beta$-type symbiotic-star. The spectrum is fitted well with an absorbed \texttt{apec} model (see Table~\ref{spectral-Table-sym} and Fig.~\ref{srcno.1.spec.sym}).

Based on available catalogues of symbiotic stars \citep[e.g,][]{2000A&AS..146..407B, 2013A&A...559A...6L}, Sources No. 6 and 8 in the Draco dSph are the first identified extragalactic $\beta$-type symbiotic stars.

\subsection{Galactic M~dwarfs in the field of Draco dSph}
\label{diss-m-dwarfs}
\subsubsection{Optical properties of M dwarfs}
\label{diss-m-dwarfs-opt}

The counterparts of sources No.\,3, 7, 9, 10, 12, 13, 14, and 15 have infrared  magnitudes and colours consistent with those expected from symbiotic systems in the Draco dSph (see Fig.~\ref{infra}), while in the optical colour-magnitude diagram (Fig.~\ref{opt-counterpart}) they are separated from the red giant branch of the Draco dSph. All of them were classified as stars (see Sect.~\ref{multi-diss}). However, their peculiar position in the colour-magnitude diagram of Fig.~\ref{opt-counterpart} required further studies to determine their spectral type. We found that all these sources satisfy the conditions for M~dwarfs for the SDSS colours $r-i<0.42$ and $i-z<0.24$ \citep{2005PASP..117..706W}.  We also estimated the spectral type of the counterparts using the characterization of M~dwarfs given by \citet[][]{2011AJ....141...97W} through the observed colours $r-i$, $i-z$, $z-J$, and $J-H$. Moreover, based on the optical SDSS colours of the M dwarfs the absolute magnitude of the M dwarfs in SDSS $i$ band using $i-z$ colours are estimated (see Table\,\ref{info-Mdwarf-Table}). The photometric distances derived this way show that all sources are Galactic M dwarfs located at distances of few hundred parsecs.

The 2nd release of Gaia \citep[DR2,][]{2018arXiv180409365G} provides distances based on parallax measurements. We examined whether the differences between the photometric and astrometric distances (of up to a factor of two; see Table\,\ref{info-Mdwarf-Table}) might be due to problematic solutions in Gaia DR\,2. To this end, we used the filters defined by \citet[]{2018A&A...616A..17A} and 
\citet[appendix C, equations C-1 and C-2]{2018A&A...616A...2L} and additional quality indicators of the solutions ({\sc astrometric\_excess\_noise}, {\sc astrometric\_gof\_al}).  
The only criterion of this quality check that is not fullfilled by all stars is the {\sc astrometric\_excess\_noise}. However, the values of the {\sc astrometric\_excess\_noise} are small for all objects and similar values have been deemed acceptable in other studies \citep{2018A&A...616A...2L}. Therefore, we decided to consider the Gaia DR\,2 distances as validated and we used them for the calculation of the luminosities.  The bolometric luminosity of each M dwarf was calculated using the bolometric correction of the $i$ band \citep{2015ApJ...804...64M}. 
We also estimated the logarithmic ratio of X-ray to bolometric luminosity log$(\frac{L_\text{X}}{L_\text{bol}})$ for each source. X-ray luminosity and X-ray to bolometric luminosity ratio are listed in Table\,\ref{info-Mdwarf-Table}.

\begin{table*}[!htb]
\caption{Optical and X-ray properties of classified Galactic M dwarfs in the filed of the Draco dSph}
\label{info-Mdwarf-Table}
\centering
\addtolength{\tabcolsep}{-0.1cm}   
\begin{tabular}{lcccccccccc}
  \hline\hline\
  \small
Src-No & $r-i$ &$i-z$& $z-J$&Spectral-type&$M_{i}$  & $d_{\rm phot}$\,(pc)& $d_{\rm Gaia}$\,(pc)&$M_{\rm bol}$&L$_{\rm x}$\,(erg\,s$^{-1}$)&log$(\frac{L_\text{X}}{L_\text{bol}})$\\
\hline
\vspace{0.5mm}
3   &   0.92$\pm$0.01   &     0.52$\pm$0.02      & 1.21$\pm$0.08 &            M2  & 8.8$\pm$0.1   &    506     &--      & 8.6 &8.5$\times$10$^{28}$ &--2.9\\
\vspace{1mm}
7   &    1.33$\pm$0.02    &     0.70$\pm$0.02     &   1.38$\pm$0.10    &        M4& 9.34$\pm$0.3   & 520 &  366$\pm 23$&  9.5 &  3.2$\times$10$^{29}$&--2.1\\
\vspace{1mm}
9   &    1.15$\pm$0.01    &     0.64$\pm$0.01     &    1.28$\pm$0.06    &       M3 & 8.8$\pm$0.3    & 413  &   497$\pm 20$&8.2&1.4$\times$10$^{29}$&--3.0\\
\vspace{1mm}
10  &     1.74$\pm$0.04   &      0.98$\pm$0.03    &    1.59$\pm$0.13   &      M5  & 11.7$\pm$0.4 &   242 &             657$\pm 160$&9.1&2.5$\times$10$^{29}$&--2.5\\
\vspace{1mm}
12  &     1.40$\pm$0.02   &      0.78$\pm$0.02     &   1.51$\pm$0.05   &          M4   & 10.0$\pm$0.3  &    351 &    446$\pm 28$&8.7&2.0$\times$10$^{29}$&--2.7\\
\vspace{1mm}
13  &   1.14$\pm$0.04     &      0.64$\pm$0.04   &   1.32$\pm$0.19      &         M3 & 9.2$\pm$0.5    &  820 & --&  9.7&3.0$\times$10$^{29}$& --2.4\\
\vspace{1mm}
14  &     1.45$\pm$0.01   &      0.80$\pm$0.01   &     1.44$\pm$0.04    &          M4-M5  & 10.2$\pm$0.3   &    158 &      140$\pm 2$&9.7&1.2$\times10^{28}$&--3.3\\
\vspace{1mm}
15  &     1.15$\pm$0.01  &      0.62$\pm$0.01    &     1.37$\pm$0.06    &       M3  & 9.1$\pm$0.1   &   375 & 510$\pm70$&8.3&2.3$\times10^{29}$&--2.8\\
\hline
\end{tabular}
\vspace{-2mm}
\tablefoot{The spectral types of the M dwarfs are identified based on the optical/ infrared colours \citep[][]{2011AJ....141...97W}. The table also presents the absolute magnitude ($M_{i}$), bolometric magnitude ($M_{bol}$), photometric distance ($d_{\rm phot}$), Gaia 2DR distance ($d_{\rm Gaia}$), X-ray luminosity ($L_\text{X}$) in the energy range of 0.2--5.0~keV, and X-ray to bolometric luminosity, log$(\frac{L_\text{X}}{L_\text{bol}})$, of M~dwarfs.
}
\end{table*}

\subsubsection{X-ray analysis of M~dwarfs}
\label{x-ana-m-dwarf}
               M dwarfs are also known as `flare stars' due to their frequent brightness increases resulting from magnetic reconnection events. The timescale of stellar X-ray flares is hours to days \citep[e.g,][]{2004A&ARv..12...71G}. As the typical duration of the observations analysed here is $\sim30-50$\,ks, we expect to identify the typical variability of flares -- if present -- in the lightcurve of individual observations. In fact, with the procedure described in Sect.~\ref{X-time} we have identified flares for sources No.\,3, 9, 10, 14 and 15. The poor count statistics imply that most of these events are charaterized by a single bin in the short-term lightcurve. Therefore, we cannot examine the morphology of the flares. The exception is a long and bright flare of source No.\,9 that we describe in the following. For the other events we list the minimum duration and minimum flux increase in Table~\ref{flare-M-dwarf}. Highlight results of timing and spectral analyses are explained in the following:

\begin{table}[!htb]
  \begin{center}
    \caption{Flux \text{bf variation} of M dwarfs in flaring states}
    \label{flare-M-dwarf}
\begin{tabular}{llll}\hline
  Source No. & OBS-No. & ${F}_{\rm min}$  & $\Delta F_{\rm min}$ \\ 
   & & (erg\,s$^{-1}$\,cm$^{-2}$)  &  (erg\,s$^{-1}$\,cm$^{-2}$)\\
  \hline
3 & 11 & (1.5$\pm0.4$)$\times$10$^{-15}$&  3.2$\times$10$^{-15}$\\
9 & 3  & (5.8$\pm1.5$)$\times$10$^{-15}$& 3.8$\times$10$^{-14}$\\
9 & 4  & (5.8$\pm1.5$)$\times$10$^{-15}$& 1.8$\times$10$^{-13}$ \\
9 & 5  & (5.8$\pm1.5$)$\times$10$^{-15}$& 1.1$\times$10$^{-14}$\\
9 & 20  & (5.8$\pm1.5$)$\times$10$^{-15}$& 7.0$\times$10$^{-14}$\\
9 & 22  & (5.8$\pm1.5$)$\times$10$^{-15}$& 9.4$\times$10$^{-15}$\\
10 & 17  & (1.5$\pm0.4$)$\times$10$^{-15}$& 3.7$\times$10$^{-15}$\\
14 & 19 &(2.2$\pm0.6$)$\times$10$^{-15}$& 4.0$\times$10$^{-15}$\\
15 & 14  & (2.0$\pm0.7$)$\times$10$^{-15}$& 1.3$\times$10$^{-14}$\\
15 & 30  & (2.0$\pm0.7$)$\times$10$^{-15}$& 1.0$\times$10$^{-14}$\\
\hline
\end{tabular}
\end{center}
\end{table}

              \begin{figure}[!ht]
%\centring
\includegraphics[clip, trim={2.0cm 1.0cm 0.0cm 0.cm},width=0.5\textwidth]{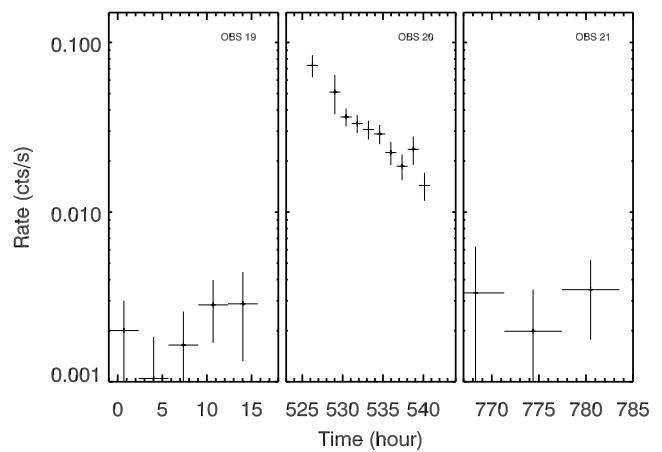}
\caption{The X-ray light curves of source No.9 in the observations 19, 20 (during the strong flare), and 21. The data is taken from the EPIC-pn camera. \label{short-lightcurve-src9}}
\end{figure}

\textbf{Source No.\,9:} This source has been classified as the most variable source in the field of the Draco dSph due to its very high flux in observation 4 \citep[source No.\,33,][]{2016A&A...586A..64S}. In observation~20 the source again showed strong variability. The long exposure time of observation~20 allowed us to study the short-term X-ray light curve and the X-ray spectrum of the star during this flare. Fig.\,\ref{short-lightcurve-src9} shows the light curve of the source in the observations 19, 20, and 21, which represent the times before, during, and after the strong flare. It seems that the flare started before the beginning of observation 20 and therefore, we can only give a lower limit to the flare duration of >15 hours. The X-ray spectrum of the source in the quiescent state was fitted with two absorbed \texttt{apec} models (see Table~\ref{spectral-Table-sym} and Fig.~\ref{src.spec.mdwarf}).  The spectrum of the source in observation 20, when the source was flaring, required an additional third thermal component. The first two components were fixed to the values previously obtained during the quiescent state. The third \texttt{apec} component is significantly hotter than the temperatures representing the quiescent spectrum. The spectral analysis yielded a temperature of $\sim$32~MK and an average X-ray luminosity of $10^{30}$\,erg\,s$^{-1}$ (see Table\,\ref{spectral-Table-sym}), which are typical for X-ray flares of cool stars \citep[e.g,][]{2004A&ARv..12...71G}.  The log$(\frac{L_\text{X}}{L_\text{bol}})$ in the quiescent state was --3.0 and increased to --2.1 in the strong flare. The long-term light curve in the $U$ band shows evidence for enhanced emission during the X-ray flare (observations~20, Appendix\,\ref{appen-light-curve}).

For the \textbf{sources No.\,3, 14, and 15} we fitted absorbed one or two-temperature \texttt{apec} model to their spectra (Table~\ref{spectral-Table-sym}). The best fit temperatures of the sources and their abundance (< with respect to the solar abundance) are very similar to each other. The X-ray luminosity of all these sources, based on the Gaia distances, are presented in Table~\ref{spectral-Table-sym}.

For the \textbf{Sources No.\,7, 10, 12, and 13} we could not fit the spectrum due to the very low statistics of their combined spectra. Therefore, we estimated the flux of these sources  assuming a typical model fitted to the spectrum of a M\,dwarf using the parameters of Table\,\ref{spectral-Table-sym}, which means an absorbed \texttt{apec} model with a Galactic $N_{\rm H}$ of $2.5\times10^{20}$ cm$^{-2}$, a temperature of  $kT$=0.7\,keV, and a sub-solar abundance  of 0.2. The X-ray luminosity of these sources are presented in Table~\ref{info-Mdwarf-Table}. Source No.\,7 had log$(\frac{L_\text{X}}{L_\text{bol}})$ of $-2.1$, which is higher than the typical log$(\frac{L_\text{X}}{L_\text{bol}})$ of M~dwarfs. The source is detected only twice over all observations. All the upper limit fluxes at the position of the source were lower than the flux of the source measured in observations 6 and 18 (Appendix\,\ref{appen-light-curve}). This suggests that source No.\,7 was detected only during flare activity. However, the count rates were too low to identify a flare signature in the short-term light curves of observations 6 and 18.  Sources No.\,10, 12, and 13 have typical X-ray to bolometric luminosity for M\,dwarfs in the quiescent state. %These three sources were outside the FOV of the OM camera.

\subsubsection{X-ray population}
M dwarfs are the most numerous stars in the Galaxy \citep{1994ASPC...64..520L}. Their X-ray emission originates from  magnetic activity caused by a  stellar dynamo \citep{1975ApJ...198..205P}. Stars with spectral types later than $\sim$ M3 are fully convective, therefore the field generating mechanism is expected to be different from the solar-type $\alpha\Omega$\,dynamo. The spectral types we derived for M~dwarfs from \xmm\, data in the Draco dSph field suggest that they belong to this category. All of them have high log$(\frac{L_\text{X}}{L_\text{bol}})$ ratios typical for fast-rotating M~dwarfs in the saturation regime, while the full population of M~dwarfs is characterized by a spread of more than three dex reaching down as low as log$(\frac{L_\text{X}}{L_\text{bol}})\sim-6$  \citep{2013MNRAS.431.2063S}. Evidently, only the most active M dwarfs in the Draco field were detected with \xmm.

From the space density of M dwarfs presented by Bochanski et al. (2010) based on the SDSS we estimate a total number of $\sim480$ M dwarfs in the $0.25$\,deg$^{-2}$ FOV of {\em XMM-Newton} for a volume reaching to a distance of $500$\,pc (i.e, the average distance of the M dwarfs detected in this work). This shows that these
exceptionally deep X-ray observations have allowed us to detect $\sim2$\,\% of the Galactic M~dwarfs in the surveyed area. This is consistent with the results derived from the X-ray luminosity function of the volume-limited sample of M dwarfs within $10$\,pc of the Sun, where only $3$ out of $159$ stars (i.e. $2$\,\%) have $L_\text{X}$ values above the ones observed in our sample, i.e, $> 10^{29}$\,erg\,s$^{-1}$\citep{2013MNRAS.431.2063S}.

\subsection{Other classified Galactic sources in the field of Draco dSph}
\label{diss-foreground}
Sources No.\,2, 4, and 11 have infrared colours and magnitudes similar to the symbiotic stars in the Draco dSph (see Fig.\,\ref{infra}). However, their optical apparent magnitude is higher than those of  members of the Draco dSph (Fig.\,\ref{opt-counterpart}). The parallax measurements of Gaia DR2 show that the counterparts of these X-ray sources are Galactic objects located at distances of $\sim$1--3~kpc. Table\,\ref{info-foreground-Table} shows the distance presented for the classified foreground X-ray sources in the field of Draco dSph in the work of \citet{2016A&A...586A..64S} and the three new classified foreground X-ray sources of this work. The filtering conditions of Gaia parallax (see Sect.\,\ref{diss-m-dwarfs-opt}) confirmed that the parallax measurement of all them are acceptable. The distances of these three new foreground sources are larger than the distances of the previous six foreground X-ray sources in the work of \citet{2016A&A...586A..64S}, which were located at $\sim$100-600 pc, consistent with the higher sensitivity for the detection of faint X-ray sources thanks to the increased exposure time available now.

\begin{table}[!htb]
\caption{Distance of additional classified foreground stars in the field of the Draco dSph}
\label{info-foreground-Table}
\centering
\begin{tabular}{lcc}
  \hline\hline\
  \small
  Src-ID &distance(pc)&\\
  \hline
  XMMUJ171925.97+575020.1$\ast$&224$\pm$36&  \\
  XMMUJ172021.83+575827.3$\ast$&95.1$\pm$0.6& \\
  XMMUJ172025.65+575304.4$\ast$&566$\pm$5&     \\
  XMMUJ172037.64+580211.9$\ast$&515$\pm$7&    \\
  XMMUJ172116.97+580113.6$\ast$&299$\pm$2&     \\
  XMMUJ172158.29+574922.5$\ast$&431$\pm$6&   \\
  XMMUJ172104.8+575333.5\,(No.\,11) &860$\pm$22&        \\
  XMMUJ171920.6+575120.2\,(No.\,2) & 2094$\pm$168&      \\
  XMMUJ171954.2+574244.0\,(No.\,4) &2665$\pm$300&   \\
\hline
\end{tabular}
\vspace{-2mm}
\tablefoot{$\ast$: Foreground sources, which were classified in \citet{2016A&A...586A..64S}.}
\end{table}

\begin{figure}[!htb]
\includegraphics[angle=270, width=0.48\textwidth, trim=1.5cm 0.cm 0.cm 0.0cm]{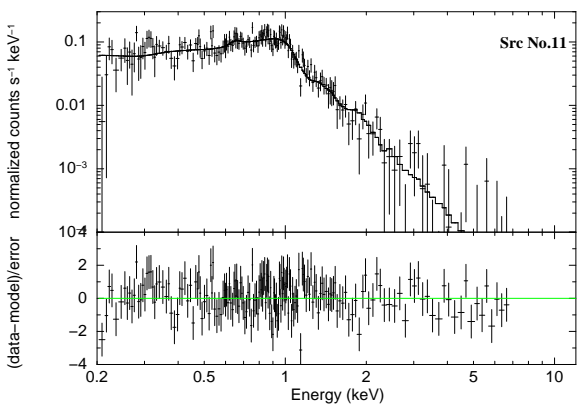}
\caption{Combined spectrum of \xmm\,observations (EPIC-pn) of the  foreground star source No.\,11, together with the residuals in units of the standard deviation. \label{src.spec.forgrounds}} 
\end{figure}

\textbf{Source No.\,2} was detected in X-rays only three times over the observations of 2015. The statistics are too poor to analyse the spectrum of the source. Most source counts were received in the soft energy range of 0.2--2.0\,keV, therefore we assumed an absorbed \texttt{apec} model with the Galactic $N_\mathrm{H}$ of $2.5\times10^{22}$\,cm$^{-2}$ and a temperature of $kT=1.0$\,keV.  The minimum and maximum flux of the source was between (2--7)$\times10^{-15}$~erg\,s$^{-1}$\,cm $^{-2}$. Thus, assuming a distance of $\sim$2~kpc  (see Table~\ref{info-foreground-Table}) the X-ray luminosity is $\sim$(1--4)$\times$10$^{30}$ erg\,s$^{-1}$. The flux of the source in observation 30 is significantly higher than in observations 8 and 10, but no flare is observed in the short-term light curve of observation~30. The $g-r$ colour of $\sim$0.45\,mag suggests a spectral type of F/G for the optical counterpart \citep{2009AJ....137.4377Y}. The UV/optical counterpart of the source was always detected by OM if it was inside its FOV. 

  \textbf{Source No.\,4} has an X-ray detection in four observations during the 2015 campaign (see Appendix\,\ref{appen-light-curve}). The statistics were too low for a spectral fit. Therefore, we assumed the same model as for source No.\,2 to estimate the flux of the source. Its average is between $(2-6)\times10^{-15}$~erg\,s$^{-1}$\,cm$^{-2}$, i.e, a luminosity of $\sim$(2--5)$\times10^{30}$\,erg\,s$^{-1}$ at the distance of 2.6\,kpc. No variability is detected for this source. The $g-r$ colour of $\sim$0.5\,mag suggests a spectral type of G for the optical counterpart \citep{2009AJ....137.4377Y}.  The source was outside the FOV of the OM telescope.

  \textbf{Source No.\,11} was already detected during five observations of 2009. However, it remained unclassified due to the lack of information on the counterpart. The extended X-ray light curve including the data from year 2015 shows that it is a variable  X-ray source. We identified a flare in observation~25. 
  The  time-averaged X-ray spectrum of the source was well fitted with an absorbed two-temperature \texttt{apec} model (see Table\,\ref{spectral-Table-sym} and Fig.\ref{src.spec.forgrounds}). The $g-r$ colour of $\sim$0.6\,mag suggests a spectral type of K for the optical counterpart \citep{2009AJ....137.4377Y}. 
  
Both the luminosity of the three classified foreground sources in this work (>10$^{30}$ erg\,s$^{-1}$) and the spectral types of their optical counterparts (G or K type) make them candidates for contact binary systems \citep[e.g,][]{1996ApJ...461..951S, 2006AJ....131..633G}. However, more observations in the optical band are necessary to study the light curves and emission lines of these sources.

\section{Summary}
This study presents multi-wavelength criteria to distinguish X-ray sources in the Draco dSph from Galactic stars using deep \xmm\, observations. We showed that all sources with a stellar counterpart have $J-K<1.2$ and $W1-W3<3.0$ colours in the infrared. All classified accreting white dwarfs and foreground stars have log$(\frac{F_\text{X}}{F_\text{opt}})<0.0$.  Also, the colour magnitude diagram of optical counterparts of the X-ray sources is a useful key to identify the spectral type of the stellar counterparts. These criteria are specially helpful in the classification of accreting white dwarfs, which have bright optical/infrared counterparts in nearby galaxies.

Based on the above criteria, we classified fifteen X-ray sources with stellar counterparts in the field of the Draco dSph. We provided X-ray timing (periodicity and other variability) and spectral analyses for these sources. 

We classified three new symbiotic stars in the Draco dSph, in addition to the known super-soft symbiotic star Draco\,C1 \citep{ 2018MNRAS.473..440S}. The X-ray spectra of the three symbiotic stars revealed one $\delta$-type symbiotic star~(source No.\,1) and two $\beta$-type symbiotic stars~(sources No.\,6 and No.\,8), which are the first classified extragalactic $\beta$-type symbiotic stars. This study showed that white dwarf binaries are the most prominent class of X-ray sources in the field of dSphs as expected from theoretical studies (see Sect.\,\ref{intro}). 

 Eight sources are classified as Galactic M\,dwarfs in the field of the Draco dSph with spectral types between M2—M5. The X-ray luminosity of these  M\,dwarfs range between $10^{28}- 3 \times10^{29}$~erg\,s$^{-1}$. We calculate a detection fraction of $\sim 2$\,\% based on estimates of the space density of M dwarfs. The observed X-ray luminosities are consistent with those of the upper $2$\,\% of the M dwarf X-ray luminosity function of the volume-limited sample of M~dwarfs within 10~pc distance from the Sun. For five M dwarfs we observed flaring activity. The X-ray spectrum and light curve of source No.\,9 was studied during both the quiescent and the flaring states showing the typical temporal and spectral signatures of flares, i.e. exponentially decaying brightness and increased X-ray temperature during the bright state.  

Finally, we  classified three sources as more distant Galactic stars, which are candidates for contact binaries. 

 This study is part of a new X-ray study of the Draco dSph. Here, we have presented the study of sources, which have a stellar counterpart.  The new catalogue, which will contain the classification of all other X-ray sources (i.e, mainly classified background objects, X-ray sources without any counterpart, and X-ray sources with unclassified counterparts) is in preparation (Saeedi et al, in prep).

\begin{acknowledgements}
We thank the anonymous referee for the useful comments that helped us to improve the paper. This research was funded by the Deutsche Forschungsgemeinschaft\,(DFG) through the Heisenberg research grant SA 2131/4-1. M.S. acknowledges support by the DFG through the Heisenberg professor grants SA 2131/5-1, 12-1. L.D. acknowledges support by the Bundesministerium f\"ur Wirtschaft und Technologie and the Deutsches Zentrum f\"ur Luft und Raumfahrt through the grant FKZ 50 OG 1602. This study is based on observations obtained with XMM–Newton, an ESA science mission with instruments and contributions directly funded by ESA Member States and NASA. This research has made use of the SIMBAD and VIZIER database, operated at CDS, Strasbourg, France, and of the NASA/IPAC Extragalactic Database (NED), which is operated by the Jet Propulsion Laboratory, California Institute of Technology, under contract with the National Aeronautics and Space Administration. This publication makes use of data products from the Wide-field Infrared Survey Explorer, which is a joint project of the University of California, Los Angeles, and the Jet Propulsion Laboratory/California Institute of Technology, funded by the National Aeronautics and Space Administration. This publication has made use of data products from the Two Micron All Sky Survey, which is a joint project of the University of Massachusetts and the Infrared Processing and Analysis Center, funded by the National Aeronautics and Space Administration and the National Science Foundation. Funding for SDSS and SDSS-III has been provided by the Alfred P. Sloan Foundation, the Participating Institutions, the National Science Foundation, and the U.S. Department of Energy Office of Science. The SDSS-III web site is http://www.sdss3.org/. SDSS-III is managed by the Astrophysical Research Consortium for the Participating Institutions of the SDSS-III Collaboration including the University of Arizona, the Brazilian Participation Group, Brookhaven National Laboratory, University of Cambridge, University of Florida, the French Participation Group, the German Participation Group, the Instituto de Astrofisica de Canarias, the Michigan State/Notre Dame/JINA Participation Group, Johns Hopkins University, Lawrence Berkeley National Laboratory, Max Planck Institute for Astrophysics, New Mexico State University, New York University, Ohio State University, Pennsylvania State University, University of Portsmouth, Princeton University, the Spanish Participation Group, University of Tokyo, University of Utah, Vanderbilt University, University of Virginia, University of Washington, and Yale University. This research has made use of SAOImage DS9, developed by Smithsonian Astrophysical Observatory.
\end{acknowledgements}

\bibliographystyle{aa} % style aa.bst
\bibliography{bibtex.bib}

\begin{thebibliography}{67}
\expandafter\ifx\csname natexlab\endcsname\relax\def\natexlab#1{#1}\fi

\bibitem[{{Abraham} {et~al.}(2012){Abraham}, {Philip}, {Kembhavi}, {Wadadekar},
  \& {Sinha}}]{2012MNRAS.419...80A}
{Abraham}, S., {Philip}, N.~S., {Kembhavi}, A., {Wadadekar}, Y.~G., \& {Sinha},
  R. 2012, \mnras, 419, 80

\bibitem[{{Ahn} {et~al.}(2012){Ahn}, {Alexandroff}, {Allende Prieto},
  {Anderson}, {Anderton}, {Andrews}, {Aubourg}, {Bailey}, {Balbinot}, {Barnes},
  \& et~al.}]{2012ApJS..203...21A}
{Ahn}, C.~P., {Alexandroff}, R., {Allende Prieto}, C., {et~al.} 2012, \apjs,
  203, 21

\bibitem[{{Akras} {et~al.}(2019){Akras}, {Leal-Ferreira}, {Guzman-Ramirez}, \&
  {Ramos-Larios}}]{2019MNRAS.483.5077A}
{Akras}, S., {Leal-Ferreira}, M.~L., {Guzman-Ramirez}, L., \& {Ramos-Larios},
  G. 2019, \mnras, 483, 5077

\bibitem[{{Allen}(1984)}]{1984PASAu...5..369A}
{Allen}, D.~A. 1984, Proceedings of the Astronomical Society of Australia, 5,
  369

\bibitem[{{Aoki} {et~al.}(2009){Aoki}, {Arimoto}, {Sadakane}, {Tolstoy},
  {Battaglia}, {Jablonka}, {Shetrone}, {Letarte}, {Irwin}, {Hill}, {Francois},
  {Venn}, {Primas}, {Helmi}, {Kaufer}, {Tafelmeyer}, {Szeifert}, \&
  {Babusiaux}}]{2009A&A...502..569A}
{Aoki}, W., {Arimoto}, N., {Sadakane}, K., {et~al.} 2009, \aap, 502, 569

\bibitem[{{Arenou} {et~al.}(2018){Arenou}, {Luri}, {Babusiaux}, {Fabricius},
  {Helmi}, {Muraveva}, {Robin}, {Spoto}, {Vallenari}, {Antoja},
  {Cantat-Gaudin}, {Jordi}, {Leclerc}, {Reyl{\'e}}, {Romero-G{\'o}mez}, {Shih},
  {Soria}, {Barache}, {Bossini}, {Bragaglia}, {Breddels}, {Fabrizio},
  {Lambert}, {Marrese}, {Massari}, {Moitinho}, {Robichon}, {Ruiz-Dern},
  {Sordo}, {Veljanoski}, {Eyer}, {Jasniewicz}, {Pancino}, {Soubiran}, {Spagna},
  {Tanga}, {Turon}, \& {Zurbach}}]{2018A&A...616A..17A}
{Arenou}, F., {Luri}, X., {Babusiaux}, C., {et~al.} 2018, \aap, 616, A17

\bibitem[{{Armandroff} {et~al.}(1995){Armandroff}, {Olszewski}, \&
  {Pryor}}]{1995AJ....110.2131A}
{Armandroff}, T.~E., {Olszewski}, E.~W., \& {Pryor}, C. 1995, \aj, 110, 2131

\bibitem[{{Belczy{\'n}ski} {et~al.}(2000){Belczy{\'n}ski}, {Miko{\l}ajewska},
  {Munari}, {Ivison}, \& {Friedjung}}]{2000A&AS..146..407B}
{Belczy{\'n}ski}, K., {Miko{\l}ajewska}, J., {Munari}, U., {Ivison}, R.~J., \&
  {Friedjung}, M. 2000, \aaps, 146, 407

\bibitem[{{Bellazzini} {et~al.}(2002){Bellazzini}, {Ferraro}, {Origlia},
  {Pancino}, {Monaco}, \& {Oliva}}]{2002AJ....124.3222B}
{Bellazzini}, M., {Ferraro}, F.~R., {Origlia}, L., {et~al.} 2002, \aj, 124,
  3222

\bibitem[{{Bildsten} {et~al.}(1997){Bildsten}, {Chakrabarty}, {Chiu}, {Finger},
  {Koh}, {Nelson}, {Prince}, {Rubin}, {Scott}, {Stollberg}, {Vaughan},
  {Wilson}, \& {Wilson}}]{1997ApJS..113..367B}
{Bildsten}, L., {Chakrabarty}, D., {Chiu}, J., {et~al.} 1997, \apjs, 113, 367

\bibitem[{Brazier(1994)}]{10.1093/mnras/268.3.709}
Brazier, K. T.~S. 1994, Monthly Notices of the Royal Astronomical Society, 268,
  709

\bibitem[{{Buccheri} {et~al.}(1983){Buccheri}, {Bennett}, {Bignami}, {Bloemen},
  {Boriakoff}, {Caraveo}, {Hermsen}, {Kanbach}, {Manchester}, {Masnou},
  {Mayer-Hasselwander}, {{\"O}zel}, {Paul}, {Sacco}, {Scarsi}, \&
  {Strong}}]{1983A&A...128..245B}
{Buccheri}, R., {Bennett}, K., {Bignami}, G.~F., {et~al.} 1983, \aap, 128, 245

\bibitem[{{Buccheri} {et~al.}(1988){Buccheri}, {di Gesu}, {Maccarone}, \&
  {Sacco}}]{1988A&A...201..194B}
{Buccheri}, R., {di Gesu}, V., {Maccarone}, M.~C., \& {Sacco}, B. 1988, \aap,
  201, 194

\bibitem[{{Corradi} {et~al.}(2008){Corradi}, {Rodr{\'{\i}}guez-Flores},
  {Mampaso}, {Greimel}, {Viironen}, {Drew}, {Lennon}, {Mikolajewska}, {Sabin},
  \& {Sokoloski}}]{2008A&A...480..409C}
{Corradi}, R.~L.~M., {Rodr{\'{\i}}guez-Flores}, E.~R., {Mampaso}, A., {et~al.}
  2008, \aap, 480, 409

\bibitem[{{Cutri} \& {et al.}(2012)}]{2012yCat.2311....0C}
{Cutri}, R.~M. \& {et al.} 2012, VizieR Online Data Catalog, 2311

\bibitem[{{Cutri} {et~al.}(2003){Cutri}, {Skrutskie}, {van Dyk}, {Beichman},
  {Carpenter}, {Chester}, {Cambresy}, {Evans}, {Fowler}, {Gizis}, {Howard},
  {Huchra}, {Jarrett}, {Kopan}, {Kirkpatrick}, {Light}, {Marsh}, {McCallon},
  {Schneider}, {Stiening}, {Sykes}, {Weinberg}, {Wheaton}, {Wheelock}, \&
  {Zacarias}}]{2003yCat.2246....0C}
{Cutri}, R.~M., {Skrutskie}, M.~F., {van Dyk}, S., {et~al.} 2003, VizieR Online
  Data Catalog, 2246

\bibitem[{{Ducci} {et~al.}(2016){Ducci}, {Doroshenko}, {Suleimanov},
  {Niko{\l}ajuk}, {Santangelo}, \& {Ferrigno}}]{2016A&A...592A..58D}
{Ducci}, L., {Doroshenko}, V., {Suleimanov}, V., {et~al.} 2016, \aap, 592, A58

\bibitem[{{Fabbiano}(2006)}]{2006ARA&A..44..323F}
{Fabbiano}, G. 2006, \araa, 44, 323

\bibitem[{{Frebel} {et~al.}(2010){Frebel}, {Kirby}, \&
  {Simon}}]{2010Natur.464...72F}
{Frebel}, A., {Kirby}, E.~N., \& {Simon}, J.~D. 2010, \nat, 464, 72

\bibitem[{{Gaia Collaboration} {et~al.}(2018){Gaia Collaboration}, {Brown},
  {Vallenari}, {Prusti}, {de Bruijne}, {Babusiaux}, \&
  {Bailer-Jones}}]{2018arXiv180409365G}
{Gaia Collaboration}, {Brown}, A.~G.~A., {Vallenari}, A., {et~al.} 2018, ArXiv
  e-prints [\eprint[arXiv]{1804.09365}]

\bibitem[{{Geske} {et~al.}(2006){Geske}, {Gettel}, \&
  {McKay}}]{2006AJ....131..633G}
{Geske}, M.~T., {Gettel}, S.~J., \& {McKay}, T.~A. 2006, \aj, 131, 633

\bibitem[{{Girardi} {et~al.}(2004){Girardi}, {Grebel}, {Odenkirchen}, \&
  {Chiosi}}]{2004A&A...422..205G}
{Girardi}, L., {Grebel}, E.~K., {Odenkirchen}, M., \& {Chiosi}, C. 2004, \aap,
  422, 205

\bibitem[{{G{\"u}del}(2004)}]{2004A&ARv..12...71G}
{G{\"u}del}, M. 2004, \aapr, 12, 71

\bibitem[{{Kennea} {et~al.}(2009){Kennea}, {Mukai}, {Sokoloski}, {Luna},
  {Tueller}, {Markwardt}, \& {Burrows}}]{2009ApJ...701.1992K}
{Kennea}, J.~A., {Mukai}, K., {Sokoloski}, J.~L., {et~al.} 2009, \apj, 701,
  1992

\bibitem[{{Kenyon} {et~al.}(1993){Kenyon}, {Livio}, {Mikolajewska}, \&
  {Tout}}]{1993ApJ...407L..81K}
{Kenyon}, S.~J., {Livio}, M., {Mikolajewska}, J., \& {Tout}, C.~A. 1993, \apjl,
  407, L81

\bibitem[{{Kirby} {et~al.}(2010){Kirby}, {Guhathakurta}, {Simon}, {Geha},
  {Rockosi}, {Sneden}, {Cohen}, {Sohn}, {Majewski}, \&
  {Siegel}}]{2010ApJS..191..352K}
{Kirby}, E.~N., {Guhathakurta}, P., {Simon}, J.~D., {et~al.} 2010, \apjs, 191,
  352

\bibitem[{{Kirby} {et~al.}(2015){Kirby}, {Guo}, {Zhang}, {Deng}, {Cohen},
  {Guhathakurta}, {Shetrone}, {Lee}, \& {Rizzi}}]{2015ApJ...801..125K}
{Kirby}, E.~N., {Guo}, M., {Zhang}, A.~J., {et~al.} 2015, \apj, 801, 125

\bibitem[{{Kirby} {et~al.}(2008){Kirby}, {Simon}, {Geha}, {Guhathakurta}, \&
  {Frebel}}]{2008ApJ...685L..43K}
{Kirby}, E.~N., {Simon}, J.~D., {Geha}, M., {Guhathakurta}, P., \& {Frebel}, A.
  2008, \apjl, 685, L43

\bibitem[{{Kleyna} {et~al.}(2002){Kleyna}, {Wilkinson}, {Evans}, {Gilmore}, \&
  {Frayn}}]{2002MNRAS.330..792K}
{Kleyna}, J., {Wilkinson}, M.~I., {Evans}, N.~W., {Gilmore}, G., \& {Frayn}, C.
  2002, \mnras, 330, 792

\bibitem[{{Lewin} \& {van der Klis}(2006)}]{2006csxs.book.....L}
{Lewin}, W.~H.~G. \& {van der Klis}, M. 2006, {Compact Stellar X-ray Sources}

\bibitem[{{Liebert}(1994)}]{1994ASPC...64..520L}
{Liebert}, J. 1994, in Astronomical Society of the Pacific Conference Series,
  Vol.~64, Cool Stars, Stellar Systems, and the Sun, ed. J.-P. {Caillault}, 520

\bibitem[{{Lindegren} {et~al.}(2018){Lindegren}, {Hern{\'a}ndez}, {Bombrun},
  {Klioner}, {Bastian}, {Ramos-Lerate}, {de Torres}, {Steidelm{\"u}ller},
  {Stephenson}, {Hobbs}, {Lammers}, {Biermann}, {Geyer}, {Hilger}, {Michalik},
  {Stampa}, {McMillan}, {Casta{\~n}eda}, {Clotet}, {Comoretto}, {Davidson},
  {Fabricius}, {Gracia}, {Hambly}, {Hutton}, {Mora}, {Portell}, {van Leeuwen},
  {Abbas}, {Abreu}, {Altmann}, {Andrei}, {Anglada}, {Balaguer-N{\'u}{\~n}ez},
  {Barache}, {Becciani}, {Bertone}, {Bianchi}, {Bouquillon}, {Bourda},
  {Br{\"u}semeister}, {Bucciarelli}, {Busonero}, {Buzzi}, {Cancelliere},
  {Carlucci}, {Charlot}, {Cheek}, {Crosta}, {Crowley}, {de Bruijne}, {de
  Felice}, {Drimmel}, {Esquej}, {Fienga}, {Fraile}, {Gai}, {Garralda},
  {Gonz{\'a}lez-Vidal}, {Guerra}, {Hauser}, {Hofmann}, {Holl}, {Jordan},
  {Lattanzi}, {Lenhardt}, {Liao}, {Licata}, {Lister}, {L{\"o}ffler},
  {Marchant}, {Martin-Fleitas}, {Messineo}, {Mignard}, {Morbidelli}, {Poggio},
  {Riva}, {Rowell}, {Salguero}, {Sarasso}, {Sciacca}, {Siddiqui}, {Smart},
  {Spagna}, {Steele}, {Taris}, {Torra}, {van Elteren}, {van Reeven}, \&
  {Vecchiato}}]{2018A&A...616A...2L}
{Lindegren}, L., {Hern{\'a}ndez}, J., {Bombrun}, A., {et~al.} 2018, \aap, 616,
  A2

\bibitem[{{Luna} \& {Sokoloski}(2007)}]{2007ApJ...671..741L}
{Luna}, G.~J.~M. \& {Sokoloski}, J.~L. 2007, \apj, 671, 741

\bibitem[{{Luna} {et~al.}(2008){Luna}, {Sokoloski}, \&
  {Mukai}}]{2008ASPC..401..342L}
{Luna}, G.~J.~M., {Sokoloski}, J.~L., \& {Mukai}, K. 2008, in Astronomical
  Society of the Pacific Conference Series, Vol. 401, RS Ophiuchi (2006) and
  the Recurrent Nova Phenomenon, ed. A.~{Evans}, M.~F. {Bode}, T.~J. {O'Brien},
  \& M.~J. {Darnley}, 342

\bibitem[{{Luna} {et~al.}(2013){Luna}, {Sokoloski}, {Mukai}, \&
  {Nelson}}]{2013A&A...559A...6L}
{Luna}, G.~J.~M., {Sokoloski}, J.~L., {Mukai}, K., \& {Nelson}, T. 2013, \aap,
  559, A6

\bibitem[{{Maccacaro} {et~al.}(1988){Maccacaro}, {Gioia}, {Wolter}, {Zamorani},
  \& {Stocke}}]{1988ApJ...326..680M}
{Maccacaro}, T., {Gioia}, I.~M., {Wolter}, A., {Zamorani}, G., \& {Stocke},
  J.~T. 1988, \apj, 326, 680

\bibitem[{{Maccarone} {et~al.}(2005){Maccarone}, {Kundu}, {Zepf}, {Piro}, \&
  {Bildsten}}]{2005MNRAS.364L..61M}
{Maccarone}, T.~J., {Kundu}, A., {Zepf}, S.~E., {Piro}, A.~L., \& {Bildsten},
  L. 2005, \mnras, 364, L61

\bibitem[{{Mann} {et~al.}(2015){Mann}, {Feiden}, {Gaidos}, {Boyajian}, \& {von
  Braun}}]{2015ApJ...804...64M}
{Mann}, A.~W., {Feiden}, G.~A., {Gaidos}, E., {Boyajian}, T., \& {von Braun},
  K. 2015, \apj, 804, 64

\bibitem[{{Manni} {et~al.}(2015){Manni}, {Nucita}, {De Paolis}, {Testa}, \&
  {Ingrosso}}]{2015MNRAS.451.2735M}
{Manni}, L., {Nucita}, A.~A., {De Paolis}, F., {Testa}, V., \& {Ingrosso}, G.
  2015, \mnras, 451, 2735

\bibitem[{{Mason} {et~al.}(2001){Mason}, {Breeveld}, {Much}, {Carter},
  {Cordova}, {Cropper}, {Fordham}, {Huckle}, {Ho}, {Kawakami}, {Kennea},
  {Kennedy}, {Mittaz}, {Pandel}, {Priedhorsky}, {Sasseen}, {Shirey}, {Smith},
  \& {Vreux}}]{2001A&A...365L..36M}
{Mason}, K.~O., {Breeveld}, A., {Much}, R., {et~al.} 2001, \aap, 365, L36

\bibitem[{{Muerset} {et~al.}(1997){Muerset}, {Wolff}, \&
  {Jordan}}]{1997A&A...319..201M}
{Muerset}, U., {Wolff}, B., \& {Jordan}, S. 1997, \aap, 319, 201

\bibitem[{{Nu{\~n}ez} {et~al.}(2016){Nu{\~n}ez}, {Nelson}, {Mukai},
  {Sokoloski}, \& {Luna}}]{2016ApJ...824...23N}
{Nu{\~n}ez}, N.~E., {Nelson}, T., {Mukai}, K., {Sokoloski}, J.~L., \& {Luna},
  G.~J.~M. 2016, \apj, 824, 23

\bibitem[{{Parker}(1975)}]{1975ApJ...198..205P}
{Parker}, E.~N. 1975, \apj, 198, 205

\bibitem[{{Piatek} {et~al.}(2001){Piatek}, {Pryor}, {Armandroff}, \&
  {Olszewski}}]{2001AJ....121..841P}
{Piatek}, S., {Pryor}, C., {Armandroff}, T.~E., \& {Olszewski}, E.~W. 2001,
  \aj, 121, 841

\bibitem[{{Pogson}(1856)}]{1856MNRAS..17...12P}
{Pogson}, N. 1856, \mnras, 17, 12

\bibitem[{{Primini} {et~al.}(1993){Primini}, {Forman}, \&
  {Jones}}]{1993ApJ...410..615P}
{Primini}, F.~A., {Forman}, W., \& {Jones}, C. 1993, \apj, 410, 615

\bibitem[{{Ramsay} \& {Wu}(2006)}]{2006A&A...459..777R}
{Ramsay}, G. \& {Wu}, K. 2006, \aap, 459, 777

\bibitem[{{Rave} {et~al.}(2003){Rave}, {Zhao}, {Newberg}, {Yanny}, {Schneider},
  {Brinkman}, \& {Lamb}}]{2003ApJS..145..245R}
{Rave}, H.~A., {Zhao}, C., {Newberg}, H.~J., {et~al.} 2003, \apjs, 145, 245

\bibitem[{{Rosen}(2016)}]{2016yCat.9047....0R}
{Rosen}, S. 2016, VizieR Online Data Catalog: Enhanced 3XMM catalogue (3XMMe),
  9047

\bibitem[{{Saeedi} {et~al.}(2016){Saeedi}, {Sasaki}, \&
  {Ducci}}]{2016A&A...586A..64S}
{Saeedi}, S., {Sasaki}, M., \& {Ducci}, L. 2016, \aap, 586, A64

\bibitem[{{Saeedi} {et~al.}(2018){Saeedi}, {Sasaki}, \&
  {Ducci}}]{2018MNRAS.473..440S}
{Saeedi}, S., {Sasaki}, M., \& {Ducci}, L. 2018, \mnras, 473, 440

\bibitem[{{Scargle}(1982)}]{1982ApJ...263..835S}
{Scargle}, J.~D. 1982, \apj, 263, 835

\bibitem[{{Schlafly} \& {Finkbeiner}(2011)}]{2011ApJ...737..103S}
{Schlafly}, E.~F. \& {Finkbeiner}, D.~P. 2011, \apj, 737, 103

\bibitem[{{Schlegel} {et~al.}(1998){Schlegel}, {Finkbeiner}, \&
  {Davis}}]{1998ApJ...500..525S}
{Schlegel}, D.~J., {Finkbeiner}, D.~P., \& {Davis}, M. 1998, \apj, 500, 525

\bibitem[{{S{\'e}gall} {et~al.}(2007){S{\'e}gall}, {Ibata}, {Irwin}, {Martin},
  \& {Chapman}}]{2007MNRAS.375..831S}
{S{\'e}gall}, M., {Ibata}, R.~A., {Irwin}, M.~J., {Martin}, N.~F., \&
  {Chapman}, S. 2007, \mnras, 375, 831

\bibitem[{{Shaw} {et~al.}(1996){Shaw}, {Caillault}, \&
  {Schmitt}}]{1996ApJ...461..951S}
{Shaw}, J.~S., {Caillault}, J.-P., \& {Schmitt}, J.~H.~M.~M. 1996, \apj, 461,
  951

\bibitem[{{Sonbas} {et~al.}(2016){Sonbas}, {Rangelov}, {Kargaltsev}, {Dhuga},
  {Hare}, \& {Volkov}}]{2016ApJ...821...54S}
{Sonbas}, E., {Rangelov}, B., {Kargaltsev}, O., {et~al.} 2016, \apj, 821, 54

\bibitem[{{Stelzer} {et~al.}(2013){Stelzer}, {Marino}, {Micela},
  {L{\'o}pez-Santiago}, \& {Liefke}}]{2013MNRAS.431.2063S}
{Stelzer}, B., {Marino}, A., {Micela}, G., {L{\'o}pez-Santiago}, J., \&
  {Liefke}, C. 2013, \mnras, 431, 2063

\bibitem[{{Str{\"u}der} {et~al.}(2001){Str{\"u}der}, {Briel}, {Dennerl},
  {Hartmann}, {Kendziorra}, {Meidinger}, {Pfeffermann}, {Reppin}, {Aschenbach},
  {Bornemann}, {Br{\"a}uninger}, {Burkert}, {Elender}, {Freyberg}, {Haberl},
  {Hartner}, {Heuschmann}, {Hippmann}, {Kastelic}, {Kemmer}, {Kettenring},
  {Kink}, {Krause}, {M{\"u}ller}, {Oppitz}, {Pietsch}, {Popp}, {Predehl},
  {Read}, {Stephan}, {St{\"o}tter}, {Tr{\"u}mper}, {Holl}, {Kemmer}, {Soltau},
  {St{\"o}tter}, {Weber}, {Weichert}, {von Zanthier}, {Carathanassis}, {Lutz},
  {Richter}, {Solc}, {B{\"o}ttcher}, {Kuster}, {Staubert}, {Abbey}, {Holland},
  {Turner}, {Balasini}, {Bignami}, {La Palombara}, {Villa}, {Buttler},
  {Gianini}, {Lain{\'e}}, {Lumb}, \& {Dhez}}]{2001A&A...365L..18S}
{Str{\"u}der}, L., {Briel}, U., {Dennerl}, K., {et~al.} 2001, \aap, 365, L18

\bibitem[{{Suleimanov} {et~al.}(2005){Suleimanov}, {Revnivtsev}, \&
  {Ritter}}]{2005A&A...435..191S}
{Suleimanov}, V., {Revnivtsev}, M., \& {Ritter}, H. 2005, \aap, 435, 191

\bibitem[{{Tauris} \& {van den Heuvel}(2006)}]{2006csxs.book..623T}
{Tauris}, T.~M. \& {van den Heuvel}, E.~P.~J. 2006, {Formation and evolution of
  compact stellar X-ray sources}, ed. W.~H.~G. {Lewin} \& M.~{van der Klis},
  623--665

\bibitem[{{Tolstoy} {et~al.}(2009){Tolstoy}, {Hill}, \&
  {Tosi}}]{2009ARA&A..47..371T}
{Tolstoy}, E., {Hill}, V., \& {Tosi}, M. 2009, \araa, 47, 371

\bibitem[{{Turner} {et~al.}(2001){Turner}, {Abbey}, {Arnaud}, {Balasini},
  {Barbera}, {Belsole}, {Bennie}, {Bernard}, {Bignami}, {Boer}, {Briel},
  {Butler}, {Cara}, {Chabaud}, {Cole}, {Collura}, {Conte}, {Cros}, {Denby},
  {Dhez}, {Di Coco}, {Dowson}, {Ferrando}, {Ghizzardi}, {Gianotti}, {Goodall},
  {Gretton}, {Griffiths}, {Hainaut}, {Hochedez}, {Holland}, {Jourdain},
  {Kendziorra}, {Lagostina}, {Laine}, {La Palombara}, {Lortholary}, {Lumb},
  {Marty}, {Molendi}, {Pigot}, {Poindron}, {Pounds}, {Reeves}, {Reppin},
  {Rothenflug}, {Salvetat}, {Sauvageot}, {Schmitt}, {Sembay}, {Short},
  {Spragg}, {Stephen}, {Str{\"u}der}, {Tiengo}, {Trifoglio}, {Tr{\"u}mper},
  {Vercellone}, {Vigroux}, {Villa}, {Ward}, {Whitehead}, \&
  {Zonca}}]{2001A&A...365L..27T}
{Turner}, M.~J.~L., {Abbey}, A., {Arnaud}, M., {et~al.} 2001, \aap, 365, L27

\bibitem[{{West} {et~al.}(2011){West}, {Morgan}, {Bochanski}, {Andersen},
  {Bell}, {Kowalski}, {Davenport}, {Hawley}, {Schmidt}, {Bernat}, {Hilton},
  {Muirhead}, {Covey}, {Rojas-Ayala}, {Schlawin}, {Gooding}, {Schluns},
  {Dhital}, {Pineda}, \& {Jones}}]{2011AJ....141...97W}
{West}, A.~A., {Morgan}, D.~P., {Bochanski}, J.~J., {et~al.} 2011, \aj, 141, 97

\bibitem[{{West} {et~al.}(2005){West}, {Walkowicz}, \&
  {Hawley}}]{2005PASP..117..706W}
{West}, A.~A., {Walkowicz}, L.~M., \& {Hawley}, S.~L. 2005, \pasp, 117, 706

\bibitem[{{Wheatley} \& {Kallman}(2006)}]{2006MNRAS.372.1602W}
{Wheatley}, P.~J. \& {Kallman}, T.~R. 2006, \mnras, 372, 1602

\bibitem[{{Yanny} {et~al.}(2009){Yanny}, {Rockosi}, {Newberg}, {Knapp},
  {Adelman-McCarthy}, {Alcorn}, {Allam}, {Allende Prieto}, {An}, {Anderson},
  {Anderson}, {Bailer-Jones}, {Bastian}, {Beers}, {Bell}, {Belokurov},
  {Bizyaev}, {Blythe}, {Bochanski}, {Boroski}, {Brinchmann}, {Brinkmann},
  {Brewington}, {Carey}, {Cudworth}, {Evans}, {Evans}, {Gates}, {G{\"a}nsicke},
  {Gillespie}, {Gilmore}, {Nebot Gomez-Moran}, {Grebel}, {Greenwell}, {Gunn},
  {Jordan}, {Jordan}, {Harding}, {Harris}, {Hendry}, {Holder}, {Ivans},
  {Ivezi{\v c}}, {Jester}, {Johnson}, {Kent}, {Kleinman}, {Kniazev},
  {Krzesinski}, {Kron}, {Kuropatkin}, {Lebedeva}, {Lee}, {French Leger},
  {L{\'e}pine}, {Levine}, {Lin}, {Long}, {Loomis}, {Lupton}, {Malanushenko},
  {Malanushenko}, {Margon}, {Martinez-Delgado}, {McGehee}, {Monet}, {Morrison},
  {Munn}, {Neilsen}, {Nitta}, {Norris}, {Oravetz}, {Owen}, {Padmanabhan},
  {Pan}, {Peterson}, {Pier}, {Platson}, {Re Fiorentin}, {Richards}, {Rix},
  {Schlegel}, {Schneider}, {Schreiber}, {Schwope}, {Sibley}, {Simmons},
  {Snedden}, {Allyn Smith}, {Stark}, {Stauffer}, {Steinmetz}, {Stoughton},
  {SubbaRao}, {Szalay}, {Szkody}, {Thakar}, {Sivarani}, {Tucker}, {Uomoto},
  {Vanden Berk}, {Vidrih}, {Wadadekar}, {Watters}, {Wilhelm}, {Wyse}, {Yarger},
  \& {Zucker}}]{2009AJ....137.4377Y}
{Yanny}, B., {Rockosi}, C., {Newberg}, H.~J., {et~al.} 2009, \aj, 137, 4377

\end{thebibliography}

\clearpage

\begin{appendices}
\appendix
\onecolumn

\section{Long-term light curves}
\label{appen-light-curve} 
The plots show the light curve of each source over thirty-one \xmm\, observations of EPIC cameras and OM cameras. The absorbed weighted flux ($0.2-4.5$ keV) of each source in the different observations is
plotted over time. If the source was not detected in an observation, the upper limit is shown. In X-ray light curves the red dots show the flux of each observation and the orange arrows show the upper limits. In optical and UV light curves, the black dots show the magnitude of the source and the gray arrows show the upper limits.
\begin{figure*}[!htb]
\centering 
\subfloat{\includegraphics[clip,trim={0.0cm 0.cm 0.cm 0.8cm},width=1.0\textwidth]{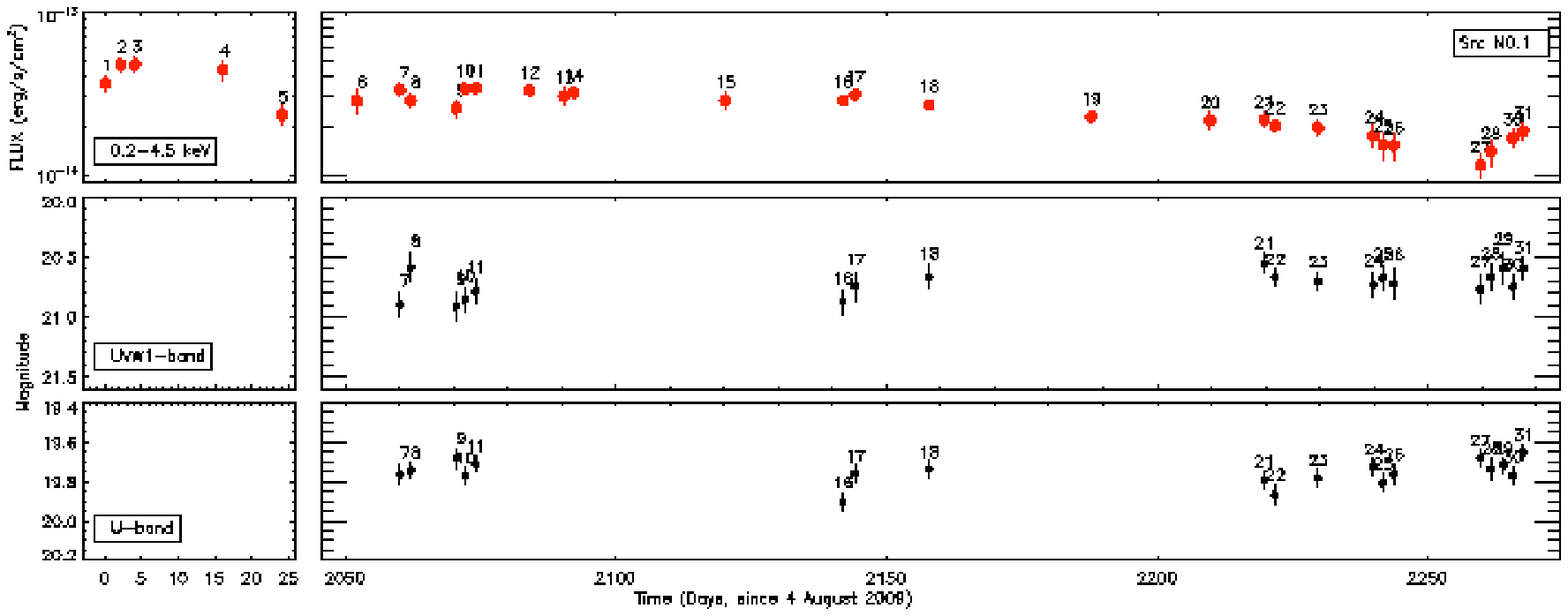}}\\
\subfloat{\includegraphics[clip, trim={0.0cm 0.cm 0.cm 0.8cm},width=1.0\textwidth]{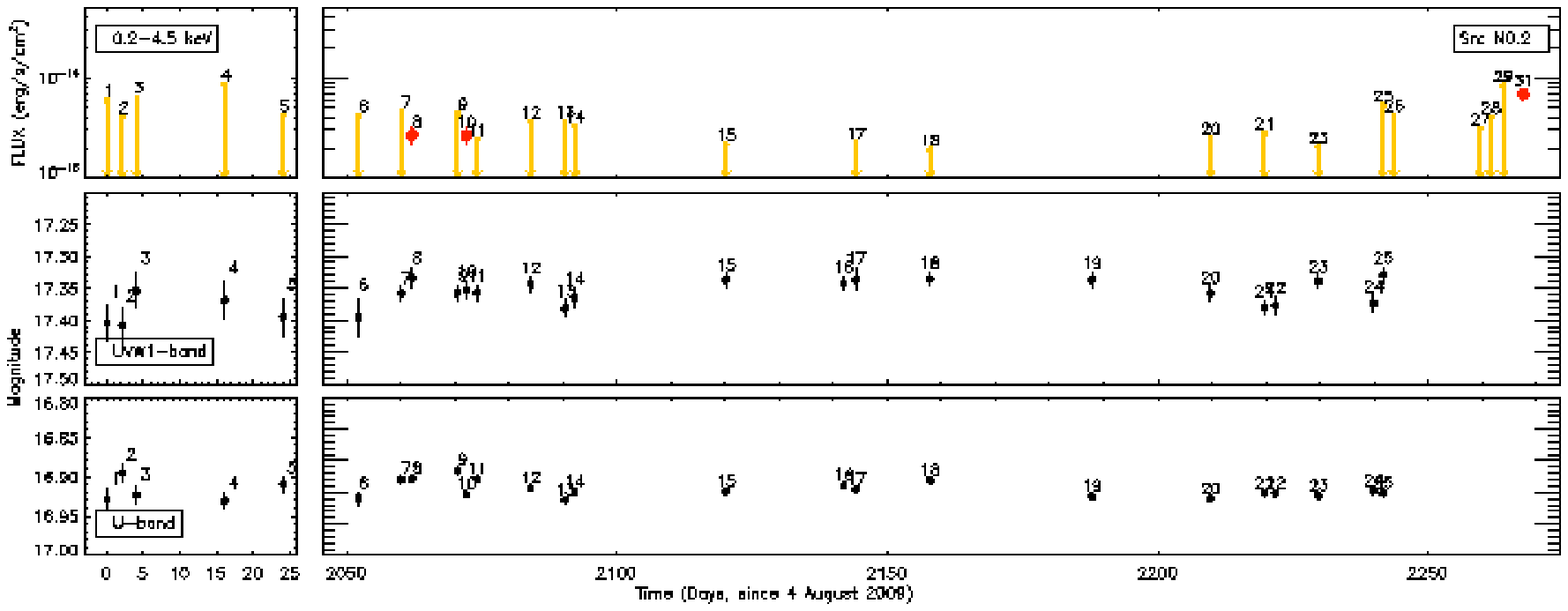}}\\
\subfloat{\includegraphics[clip, trim={0.0cm 0.cm 0.cm 0.8cm},width=1.0\textwidth]{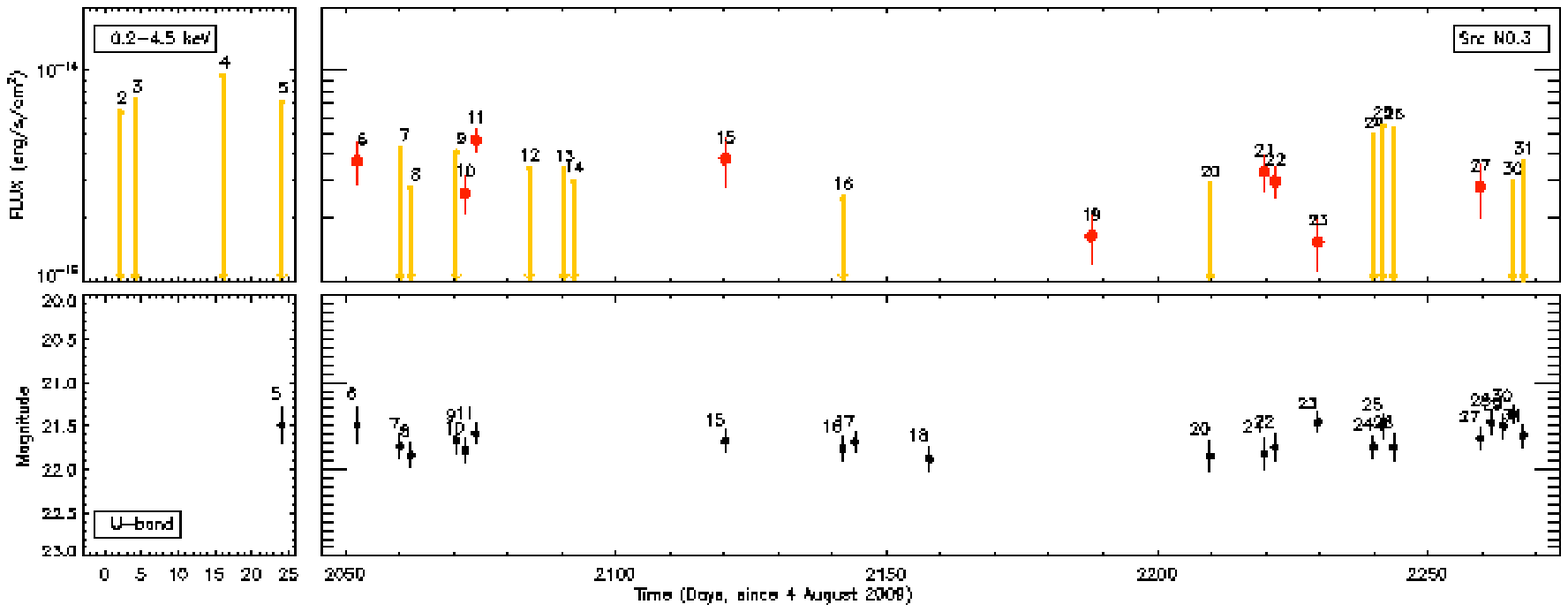}}\\
\end{figure*} 
\pagebreak
\clearpage
\hspace{0.3cm}Appendix A continued: long-term light curves
\vspace{-5cm}
\begin{figure*}[!htb]
\subfloat{\includegraphics[clip, trim={0.0cm 0.cm 0.cm 0.5cm},width=1.0\textwidth]{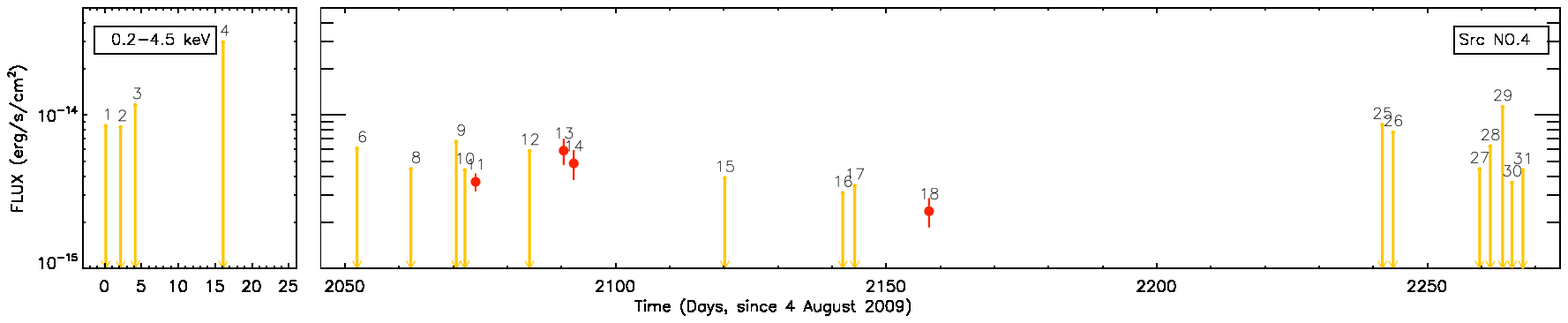}}\\
\subfloat{\includegraphics[clip, trim={0.0cm 0.cm 0.cm 0.5cm},width=1.0\textwidth]{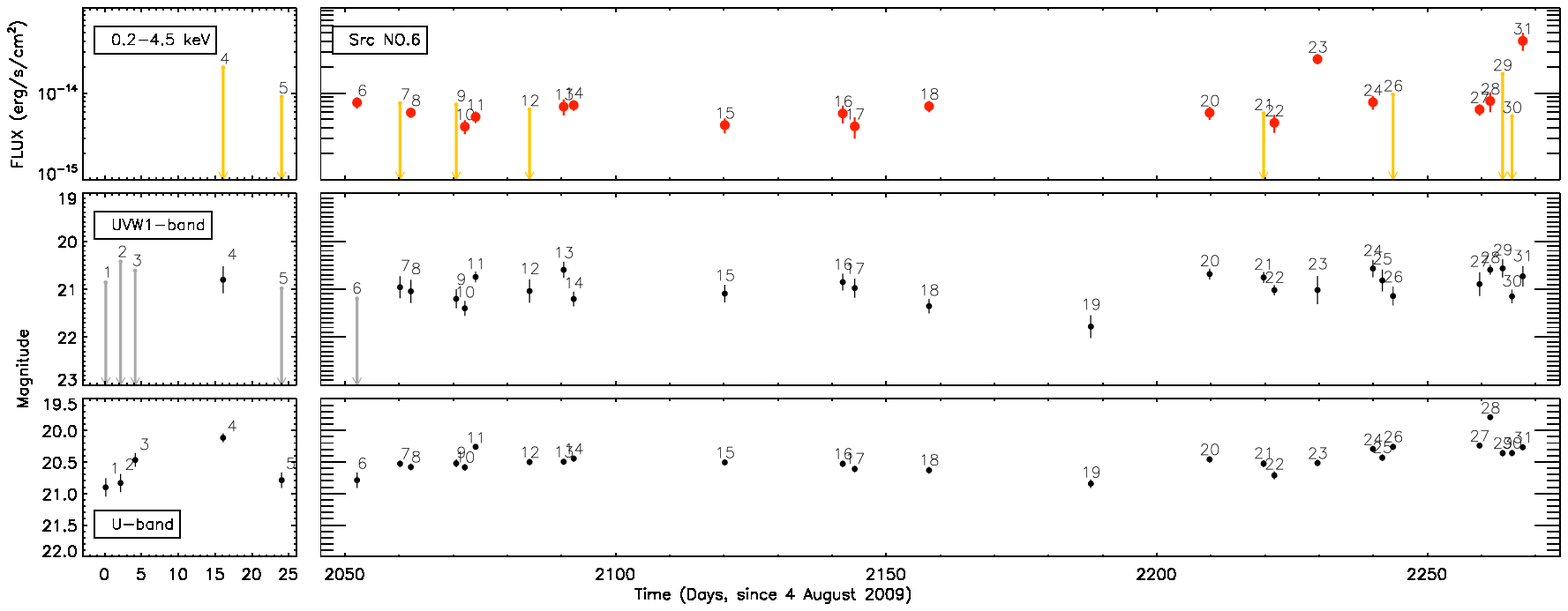}}\\
\subfloat{\includegraphics[clip, trim={0.0cm 0.cm 0.cm 0.5cm},width=1.0\textwidth]{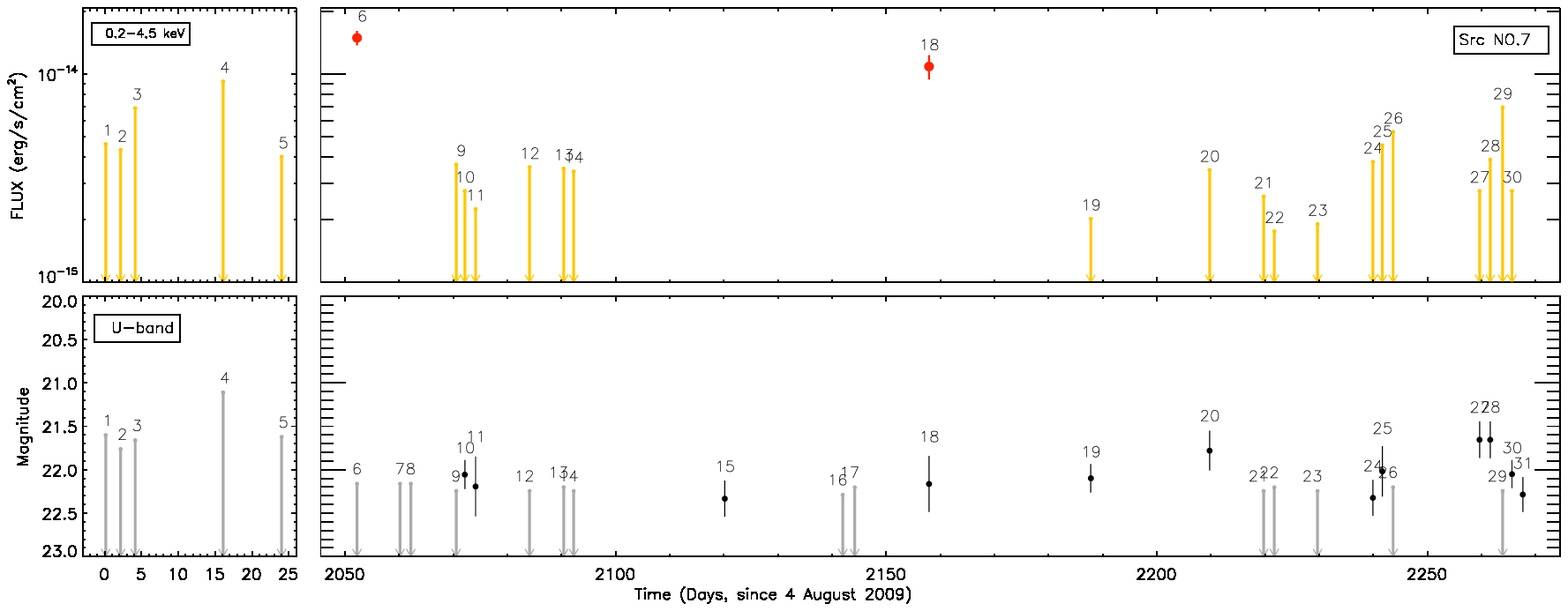}}\\
\end{figure*} 
\pagebreak
\clearpage
\hspace{0.3cm}Appendix A continued: long-term light curves
%\vspace{-5cm}
\begin{figure*}[!htb]
\subfloat{\includegraphics[clip, trim={0.0cm 0.cm 0.cm 0.5cm},width=1.0\textwidth]{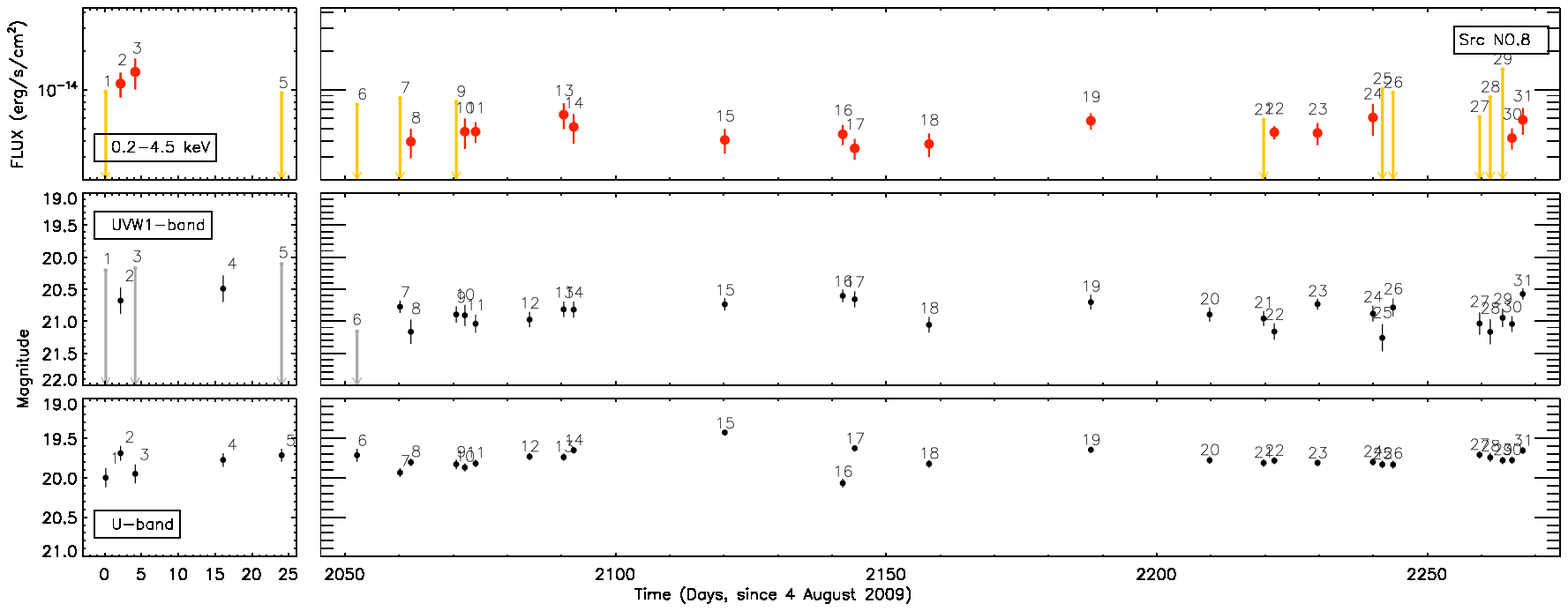}}\\
\subfloat{\includegraphics[clip, trim={0.0cm 0.cm 0.cm 0.5cm},width=1.0\textwidth]{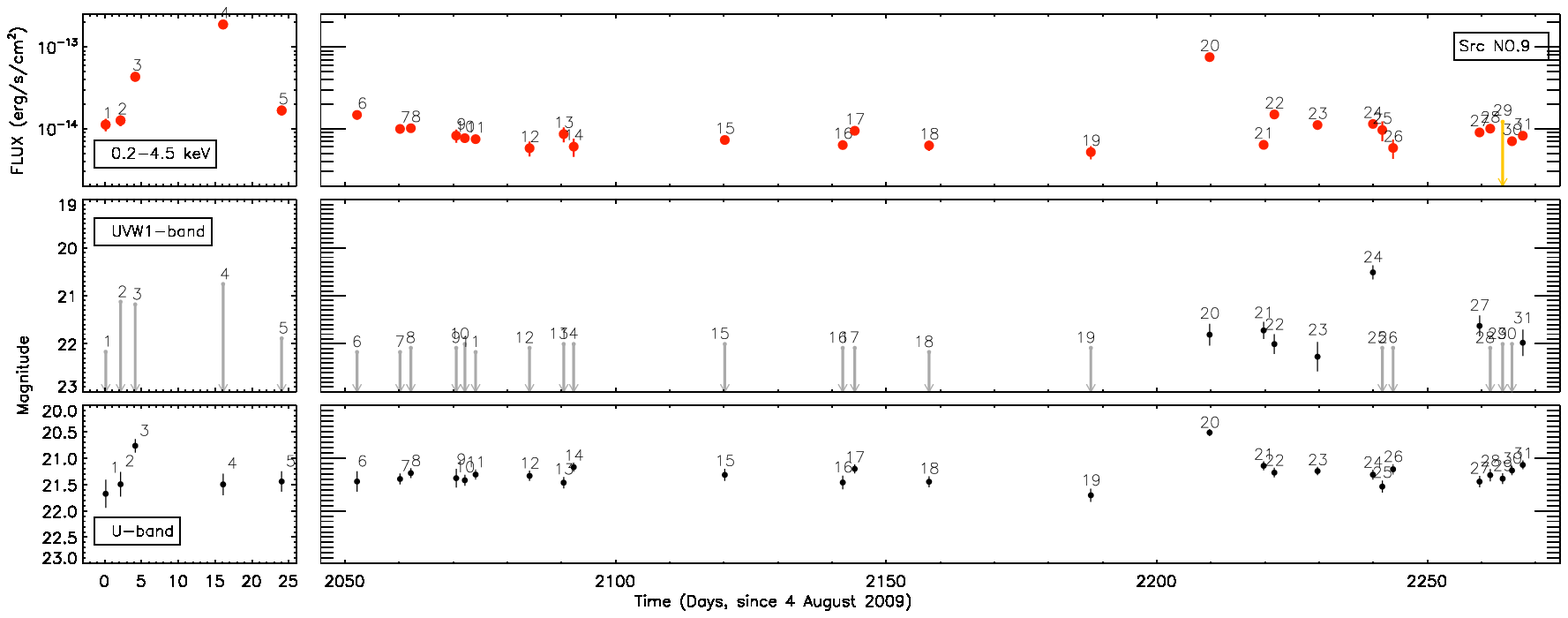}}\\
\subfloat{\includegraphics[clip, trim={0.0cm 0.cm 0.cm 1.0cm},width=1.0\textwidth]{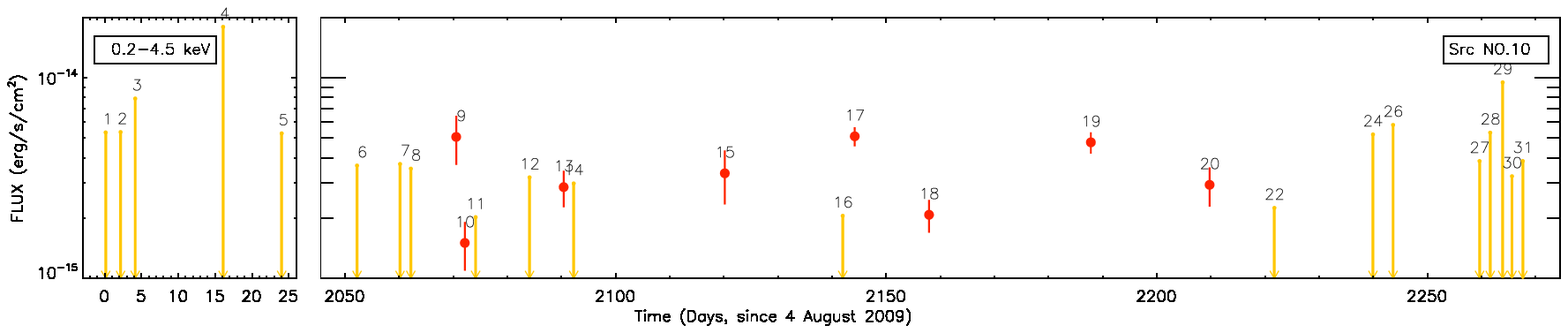}}\\
\end{figure*}
\pagebreak
\clearpage
\hspace{0.3cm}Appendix A continued: long-term light curves
\vspace{-0.4cm}
\begin{figure*}[!htb]
  \subfloat{\includegraphics[clip, trim={0.0cm 0.cm 0.cm 0.5cm},width=1.0\textwidth]{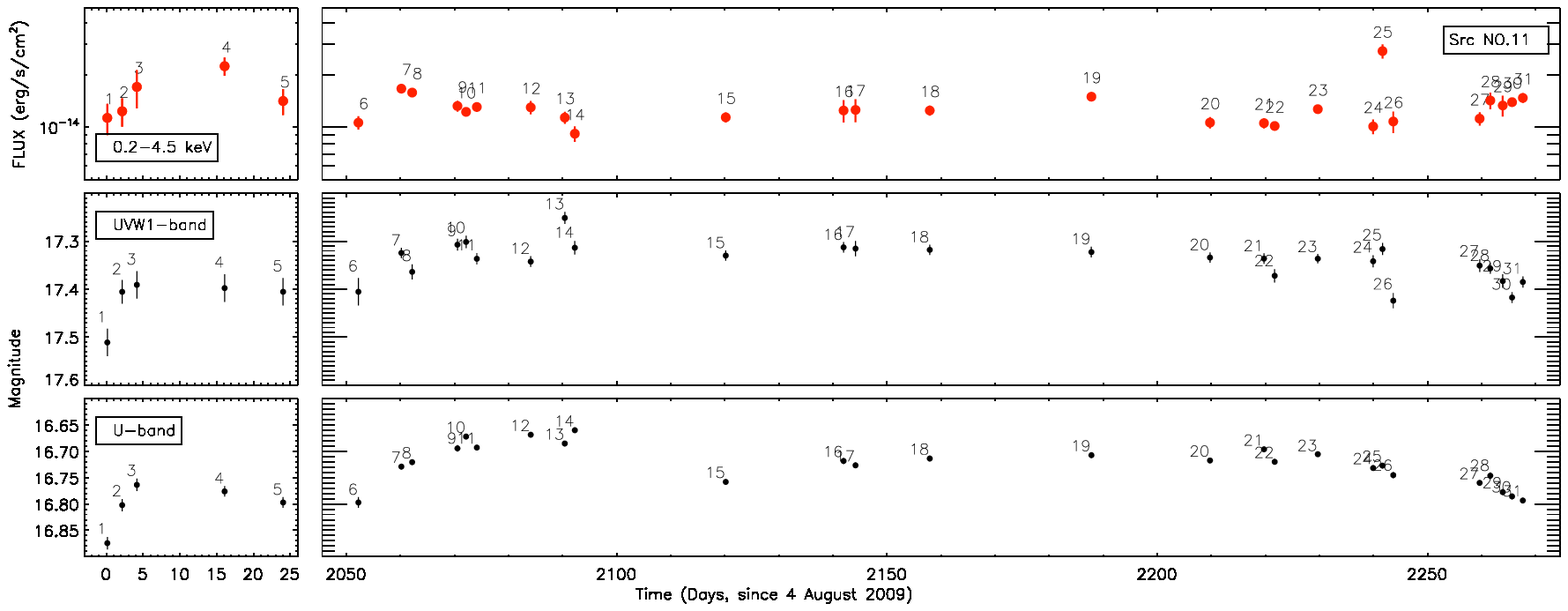}}\\
  \vspace{-0.5cm}
  \subfloat{\includegraphics[clip, trim={0.0cm 0.cm 0.cm 3.5cm},width=1.0\textwidth]{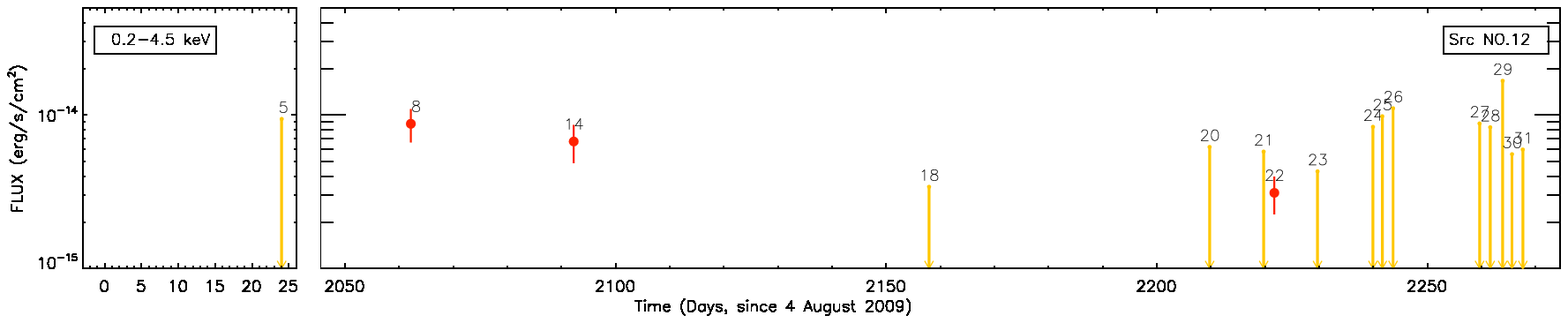}}\\
  \vspace{-0.5cm}
  \subfloat{\includegraphics[clip, trim={0.0cm 0.cm 0.cm 3.5cm},width=1.0\textwidth]{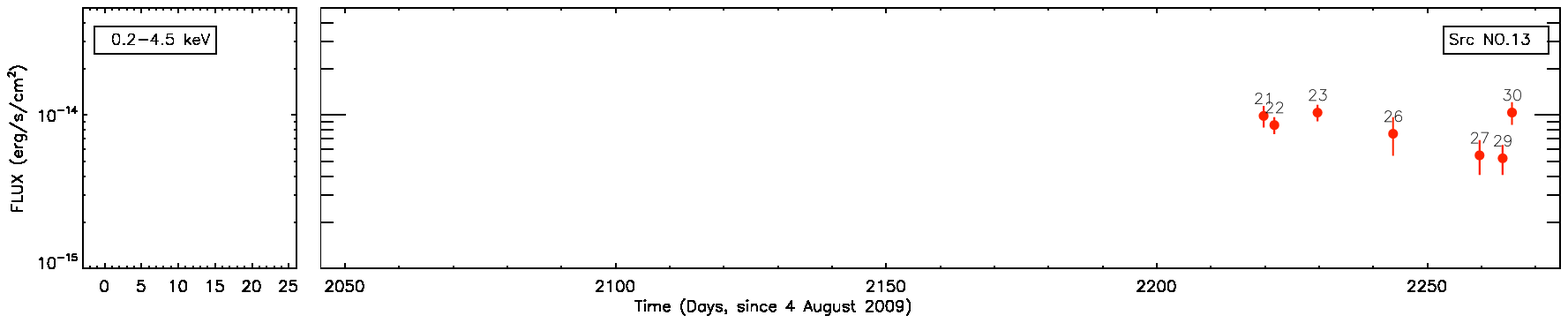}}\\
  \vspace{-0.5cm}
  \subfloat{\includegraphics[clip, trim={0.0cm 0.cm 0.cm 3.5cm},width=1.0\textwidth]{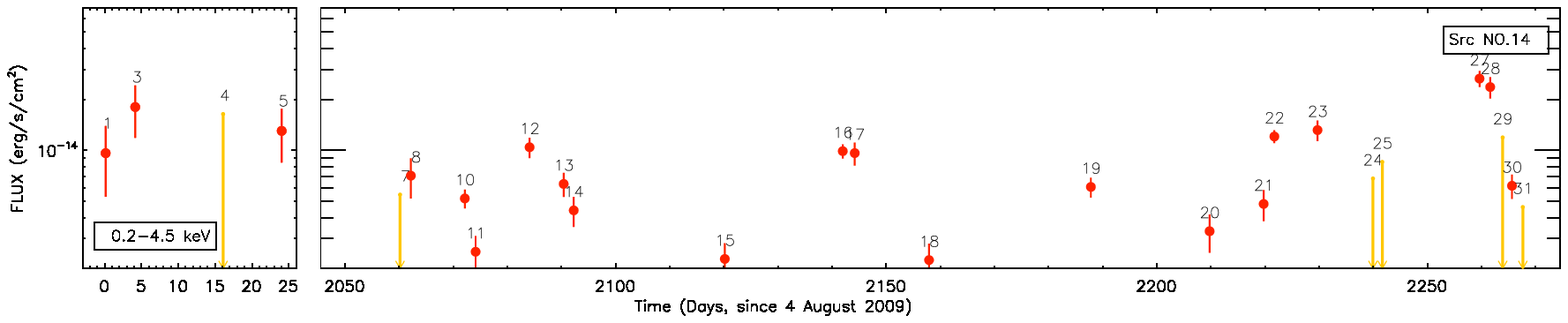}}\\
  \vspace{-0.5cm}
\subfloat{\includegraphics[clip, trim={0.0cm 0.cm 0.cm 3.5cm},width=1.0\textwidth]{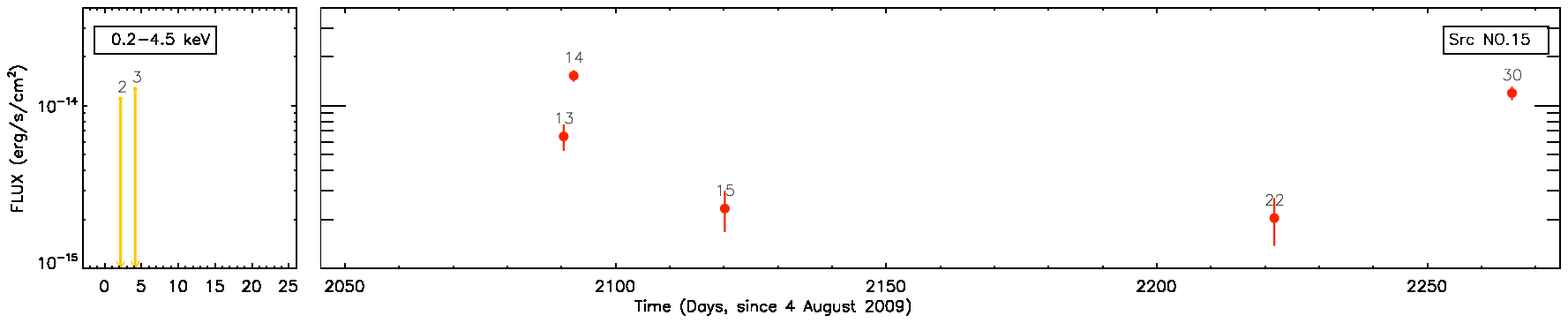}}\\
\end{figure*}
\pagebreak
\onecolumn

\section{Optical and infrared magnitudes of the counterparts of X-ray sources.}
\label{oni-tables}
\begin{table*}[!htb]
\caption{Optical  magnitudes of counterparts of X-ray sources in different energy filters of the SDSS9 survey.
\label{opt-count-table}
}
\centering
\begin{tabular}{lccccc}
\hline\hline\
Src-No&$u$ mag&$g$ mag&$r$ mag&$i$ mag&$z$ mag\\
\hline

 1    &   20.64 $\pm $   0.05   &      18.837  $\pm $      0.008 & 18.028 $\pm $ 0.006 &     17.682  $ \pm $ 0.007   & 17.49  $\pm $  0.02\\
 2    &   17.75 $\pm $   0.01   &      16.580  $\pm $      0.004 & 16.133 $\pm $ 0.004 &     15.971  $ \pm $ 0.004   & 15.91  $\pm $  0.01\\
 3    &   22.58 $\pm $   0.27   &      19.788  $\pm $      0.016 & 18.225 $\pm $ 0.007 &     17.322  $ \pm $ 0.006   & 16.82  $\pm $  0.01\\
 4    &   18.05 $\pm $   0.01   &      16.709  $\pm $      0.004 & 16.223 $\pm $ 0.004 &     16.084  $ \pm $ 0.004   & 16.05  $\pm $  0.01\\
 5    &   19.08 $\pm $   0.02   &      17.753  $\pm $      0.005 & 16.476 $\pm $ 0.004 &     16.191  $ \pm $ 0.004   & 15.77  $\pm $  0.01\\
 6    &   21.34 $\pm $   0.10   &      20.082  $\pm $      0.019 & 19.426 $\pm $ 0.014 &     19.148  $ \pm $ 0.017   & 19.06  $\pm $  0.05\\
 7    &   23.79 $\pm $   0.65   &      20.770  $\pm $      0.030 & 19.246 $\pm $ 0.013 &     17.931  $ \pm $ 0.008   & 17.24  $\pm $  0.01\\
 8    &   20.69 $\pm $   0.06   &      18.612  $\pm $      0.008 & 17.746 $\pm $ 0.006 &     17.387  $ \pm $ 0.006   & 17.17  $\pm $  0.01\\
 9    &   22.06 $\pm $   0.17   &      19.484  $\pm $      0.012 & 17.985 $\pm $ 0.006 &     16.845  $ \pm $ 0.005   & 16.21  $\pm $  0.01\\
 10   &   23.79 $\pm $   0.66   &      21.942  $\pm $      0.081 & 20.417 $\pm $ 0.030 &     18.693  $ \pm $ 0.012   & 17.73  $\pm $  0.02\\
 11   &   17.59 $\pm $   0.01   &      15.973  $\pm $      0.003 & 15.301 $\pm $ 0.003 &     15.033  $ \pm $ 0.004   & 14.88  $\pm $  0.00\\
 12   &   22.95 $\pm $   0.39   &      20.213  $\pm $      0.018 & 18.690 $\pm $ 0.009 &     17.317  $ \pm $ 0.006   & 16.55  $\pm $  0.01\\
 13   &   24.80 $\pm $ 	 1.07	&      21.307  $\pm $	   0.045 & 19.887 $\pm $ 0.022 &     18.761  $\pm $  0.013   & 18.13 $\pm $   0.02\\
 14   &   22.00 $\pm $   0.20   &      19.170  $\pm $      0.011 & 17.647 $\pm $ 0.006 &     16.210  $ \pm $ 0.005   & 15.42  $\pm $  0.01\\
 15   &   21.93 $\pm $   0.17   &      19.562  $\pm $      0.012 & 18.115 $\pm $ 0.007 &     16.974  $ \pm $ 0.005   & 16.36  $\pm $  0.01\\

\hline
\end{tabular}
%\vspace{-2mm}
%\tablefoot{}
\end{table*}

\begin{table*}[!htb]
\caption{Infrared magnitudes of counterparts of X-ray sources in different energy filters of 2MASS and WISE surveys.
\label{inf-count}
}
\centering
\begin{tabular}{lccccccc}
\hline\hline\
Src-No&$J$ mag&$H$ mag&$K$ mag&$W1$ mag&$W2$ mag&$W3$ mag&$W4$ mag\\
\hline
1    & 16.54   $\pm $     0.12  &  15.74   $\pm $ 0.14  &  15.78   $\pm $ 0.22  &  15.49  $\pm $  0.04   & 15.04   $\pm $ 0.06   & 12.46   $\pm $ 0.25  &   9.08   $\pm $ 0.3\\
2    & 15.07   $\pm $     0.04  &  14.64   $\pm $ 0.07  &  14.74   $\pm $ 0.09  &  14.57  $\pm $  0.03   & 14.66   $\pm $ 0.05   & <13.27$^{\ast}$       &   <9.61$^{\ast}$           \\
3    & 15.62   $\pm $     0.07  &  14.93   $\pm $ 0.08  &  14.60   $\pm $ 0.08  &  14.52  $\pm $  0.15   & 14.59   $\pm $ 0.04   & <13.12$^{\ast}$       &  <9.37$^{\ast}$           \\
5    & 14.38   $\pm $     0.03  &  13.71   $\pm $ 0.04  &  13.46   $\pm $ 0.04  &  13.25  $\pm $  0.11   & 13.27   $\pm $ 0.03   & 12.49   $\pm $ 0.29  &   <9.51$^{\ast}$   \\
6    & --              &     --         &       --         &     --       &     --           &    --            &    --           \\          
7    & 15.87   $\pm $     0.09  &  15.10   $\pm $ 0.09  &  15.05   $\pm $ 0.12  &  14.60  $\pm $  0.03   & 14.47   $\pm $ 0.04   & <12.84$^{\ast}$       &  <9.50$^{\ast}$             \\
8    & 16.31   $\pm $     0.10  &  15.68   $\pm $ 0.13  &  15.42   $\pm $ 0.18  &  15.35  $\pm $  0.04   & 15.55   $\pm $ 0.10   & <13.10$^{\ast}$       &   <9.52$^{\ast}$         \\
9    & 14.95   $\pm $     0.05  &  14.40   $\pm $ 0.06  &  14.12   $\pm $ 0.08  &  13.93  $\pm $  0.10   & 13.80   $\pm $ 0.03   & 12.81   $\pm $ 0.42  &   <9.32$^{\ast}$   \\
10   & 16.16   $\pm $     0.12  &  15.55   $\pm $ 0.14  &  15.09   $\pm $ 0.18  &  <15.06$^{\ast}$         & 15.05   $\pm $ 0.05   & <12.82$^{\ast}$       &  <9.47$^{\ast}$             \\
11   & 13.96   $\pm $     0.03  &  13.43   $\pm $ 0.03  &  13.33   $\pm $ 0.03  &  13.27  $\pm $  0.03   & 13.25   $\pm $ 0.08   & 13.05   $\pm $ 0.42  &   <9.51$^{\ast}$  \\
12   & 15.05   $\pm $     0.04  &  14.57   $\pm $ 0.06  &  14.30   $\pm $ 0.06  &  <14.19$^{\ast}$         & 13.99   $\pm $ 0.03   & <13.04$^{\ast}$       &  <9.43$^{\ast}$            \\
13   & 16.83   $\pm $	  0.17  &  16.09   $\pm $ 0.19	&  15.70   $\pm $0.164  & 15.46   $\pm $  0.04	 & 15.16   $\pm $ 0.07	 & <13.13$^{\ast}$       &   <9.61$^{\ast}$ \\
14   & 13.99   $\pm $     0.03  &  13.40   $\pm $ 0.03  &  13.13   $\pm $ 0.03  &  12.97  $\pm $  0.03   & 12.78   $\pm $ 0.04   & 12.51   $\pm $ 0.27  &  <9.62$^{\ast}$    \\
15   & 15.01   $\pm $     0.05  &  14.27   $\pm $ 0.04  &  14.11   $\pm $ 0.06  &  13.94  $\pm $  0.15   & 13.84   $\pm $ 0.03   & <12.89$^{\ast}$       &  <9.50$^{\ast}$            \\
\hline
\end{tabular}
%\vspace{-2mm}
\tablefoot{``$\ast$'' shows that the upper-limit is reported in the WISE catalogue.}
\end{table*}

\onecolumn

\pagebreak

\section{Image of optical SDSS9 counterparts}
The optical image of the counterpart of the X-ray sources from the SDSS9 survey. The black circles show the $3\sigma$ error of X-ray source positions.
\label{SDSS-image}
\begin{figure*}[!htb]
  \centering 
  \subfloat[Src No.1]{\includegraphics[clip, trim={0.0cm 2.0cm 0.cm 0.0cm},width=0.28\textwidth]{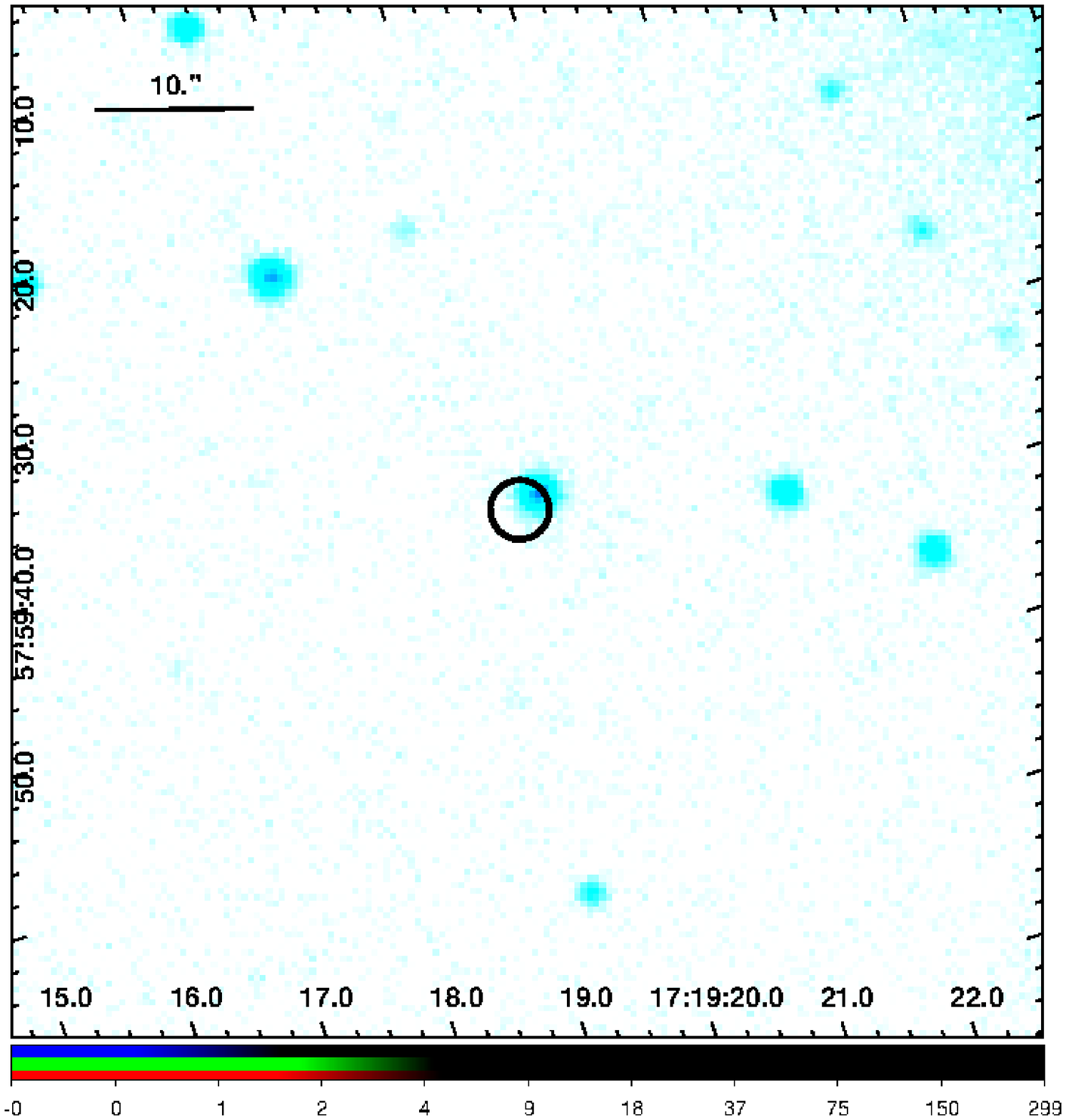}}
\vspace{0.2cm}
  \subfloat[Src No.2]{\includegraphics[clip, trim={0.0cm 2.0cm 0.cm 0.0cm},width=0.28\textwidth]{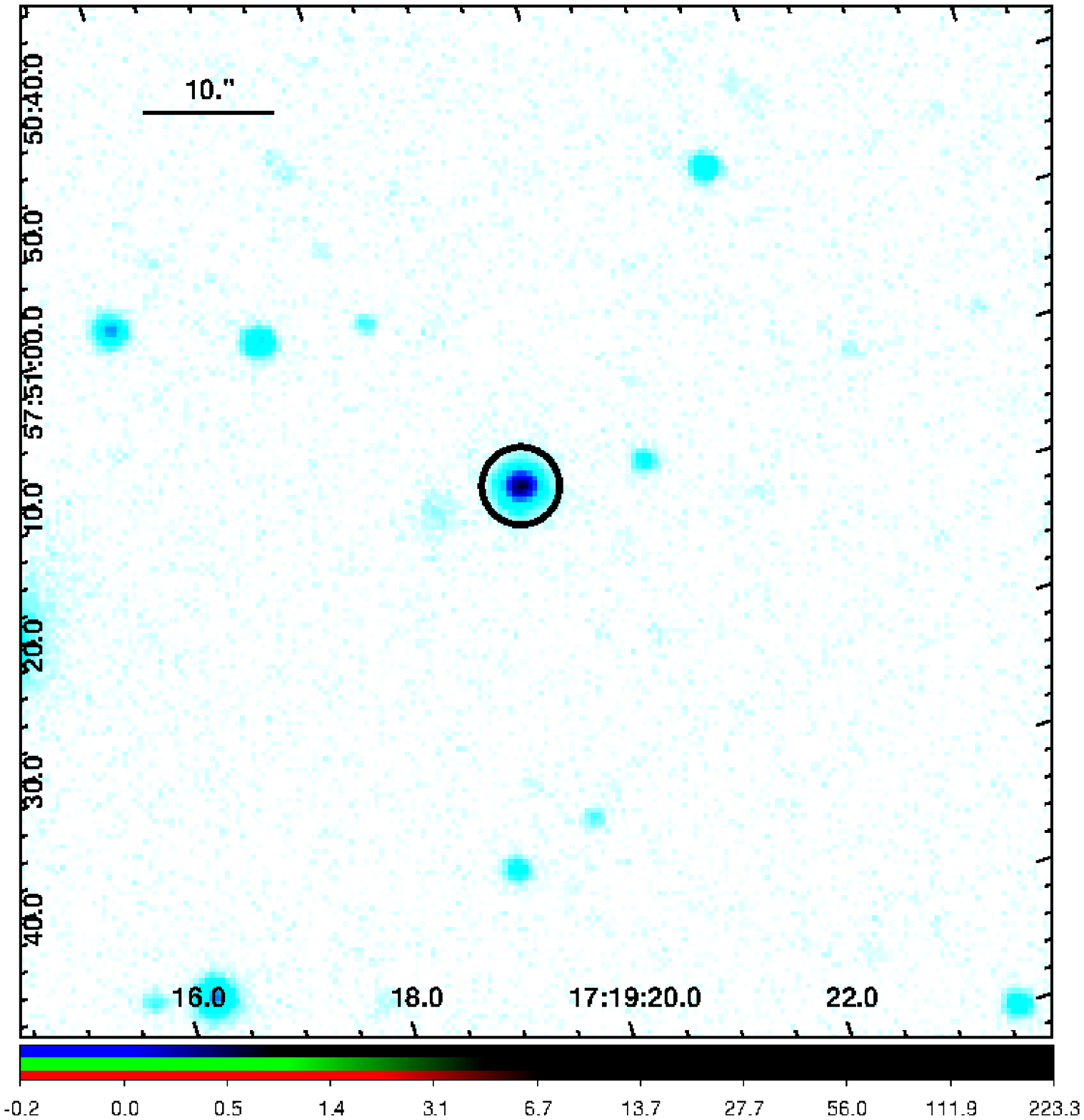}}
\vspace{0.2cm}
  \subfloat[Src No.3]{\includegraphics[clip, trim={0.0cm 2.0cm 0.cm 0.0cm},width=0.28\textwidth]{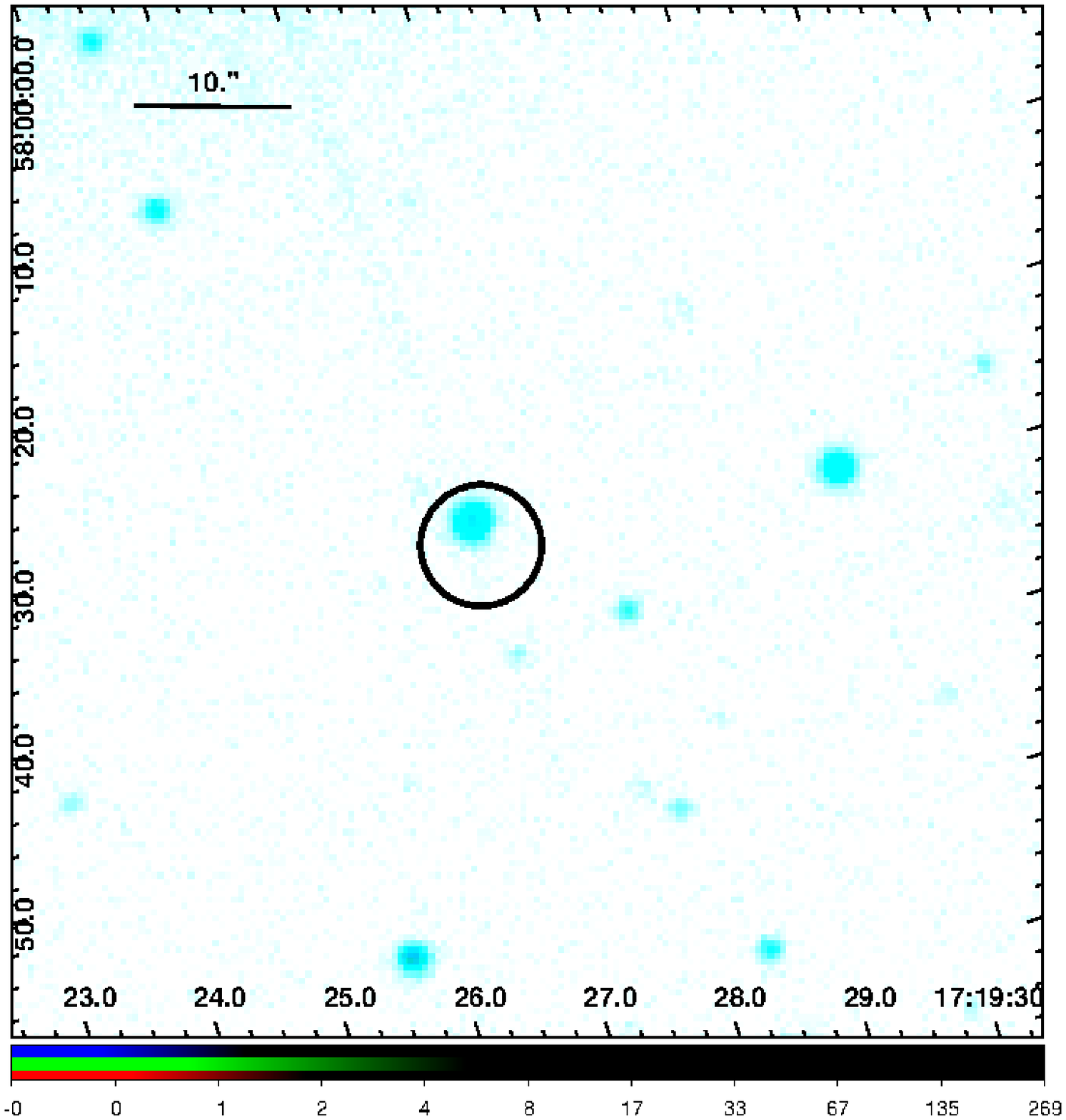}}\\

  \subfloat[Src No.4]{\includegraphics[clip, trim={0.0cm 2.0cm 0.cm 0.0cm},width=0.28\textwidth]{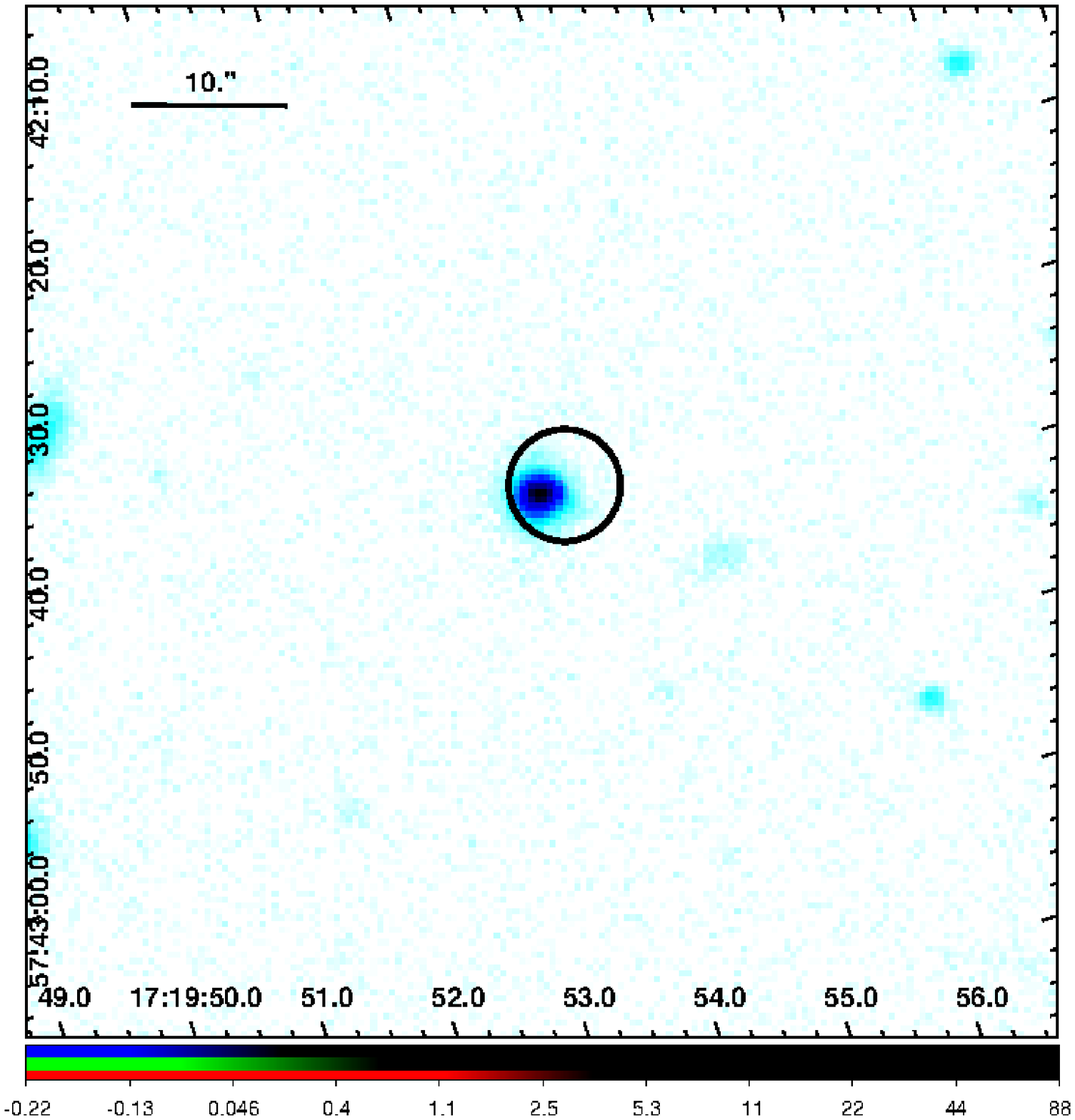}}
\vspace{0.2cm}
  \subfloat[Src No.5]{\includegraphics[clip, trim={0.0cm 2.0cm 0.cm 0.0cm},width=0.28\textwidth]{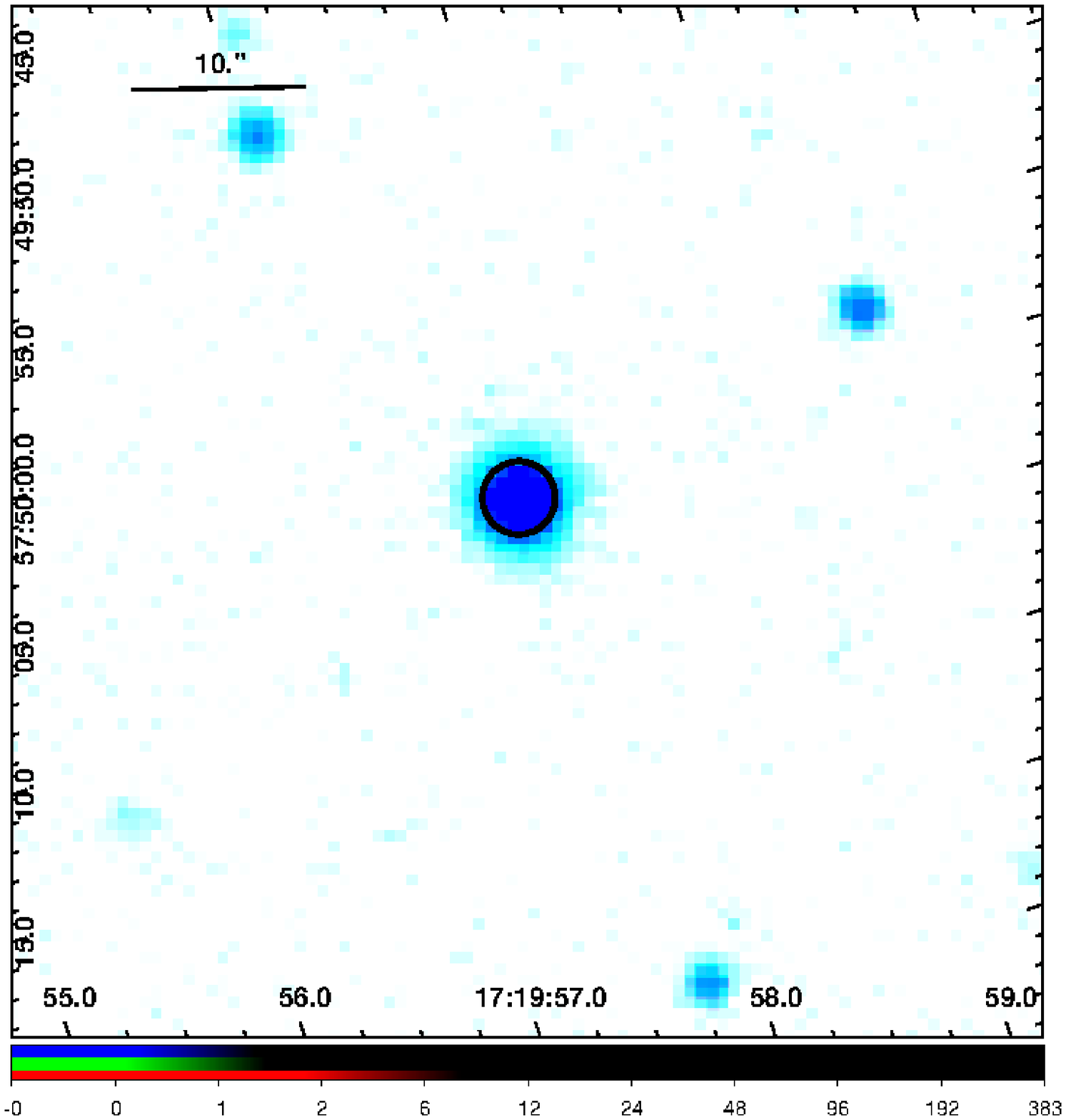}}
\vspace{0.2cm}
  \subfloat[Src No.6]{\includegraphics[clip, trim={0.0cm 2.0cm 0.cm 0.0cm},width=0.28\textwidth]{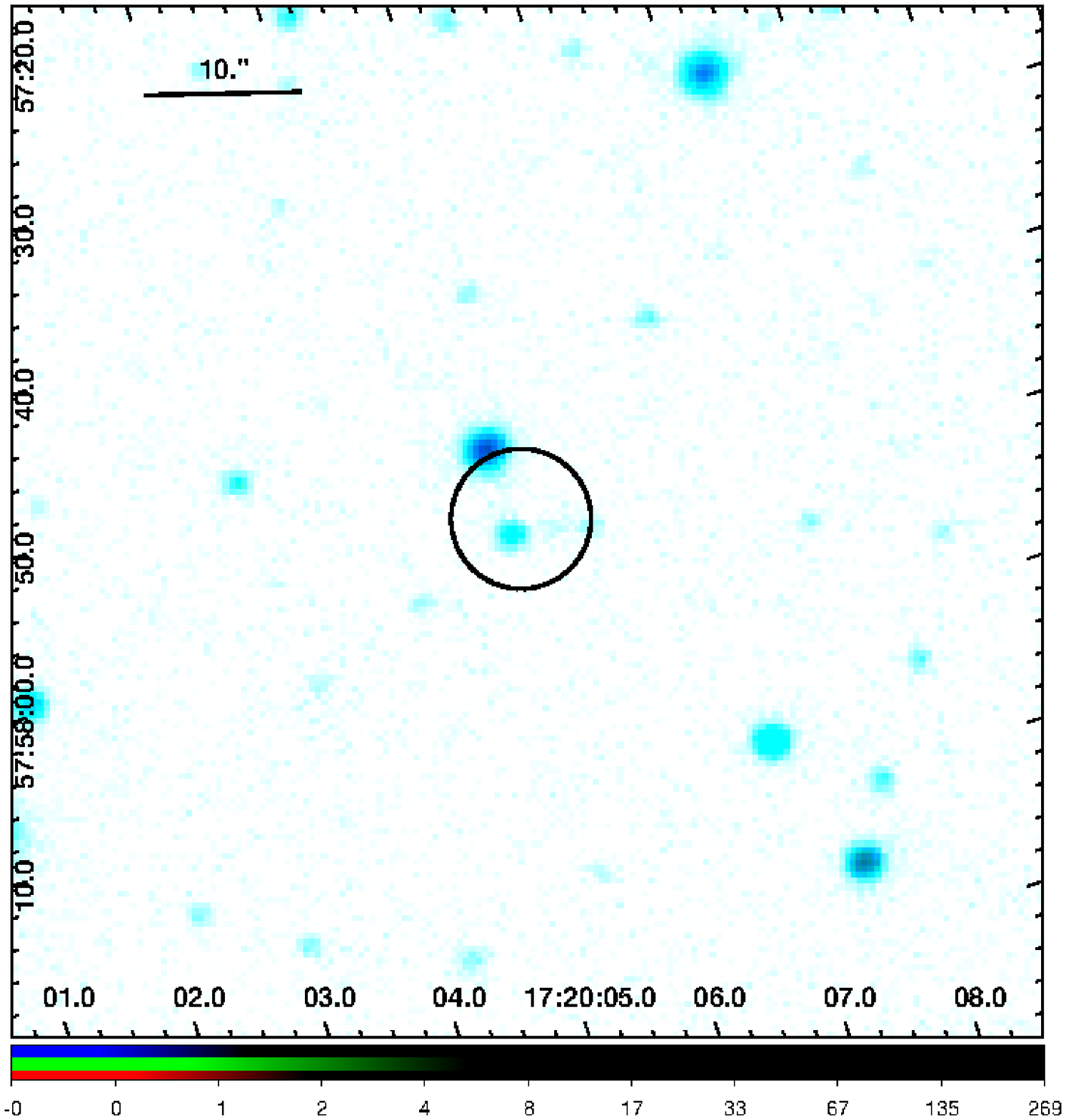}}\\

\subfloat[Src No.7]{\includegraphics[clip, trim={0.0cm 2.0cm 0.cm 0.0cm},width=0.28\textwidth]{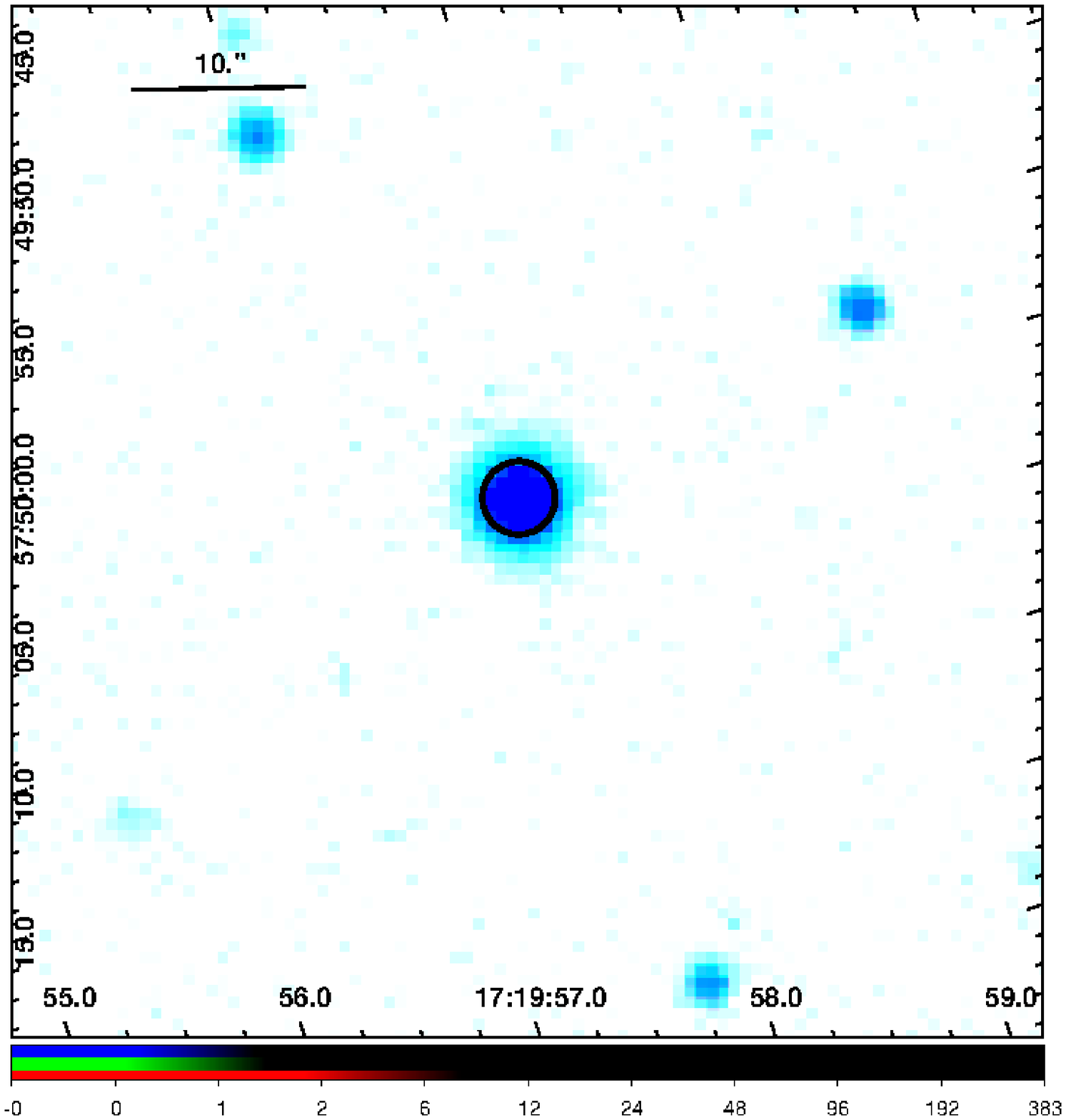}}
\vspace{0.2cm}
\subfloat[Src No.8]{\includegraphics[clip, trim={0.0cm 2.0cm 0.cm 0.0cm},width=0.28\textwidth]{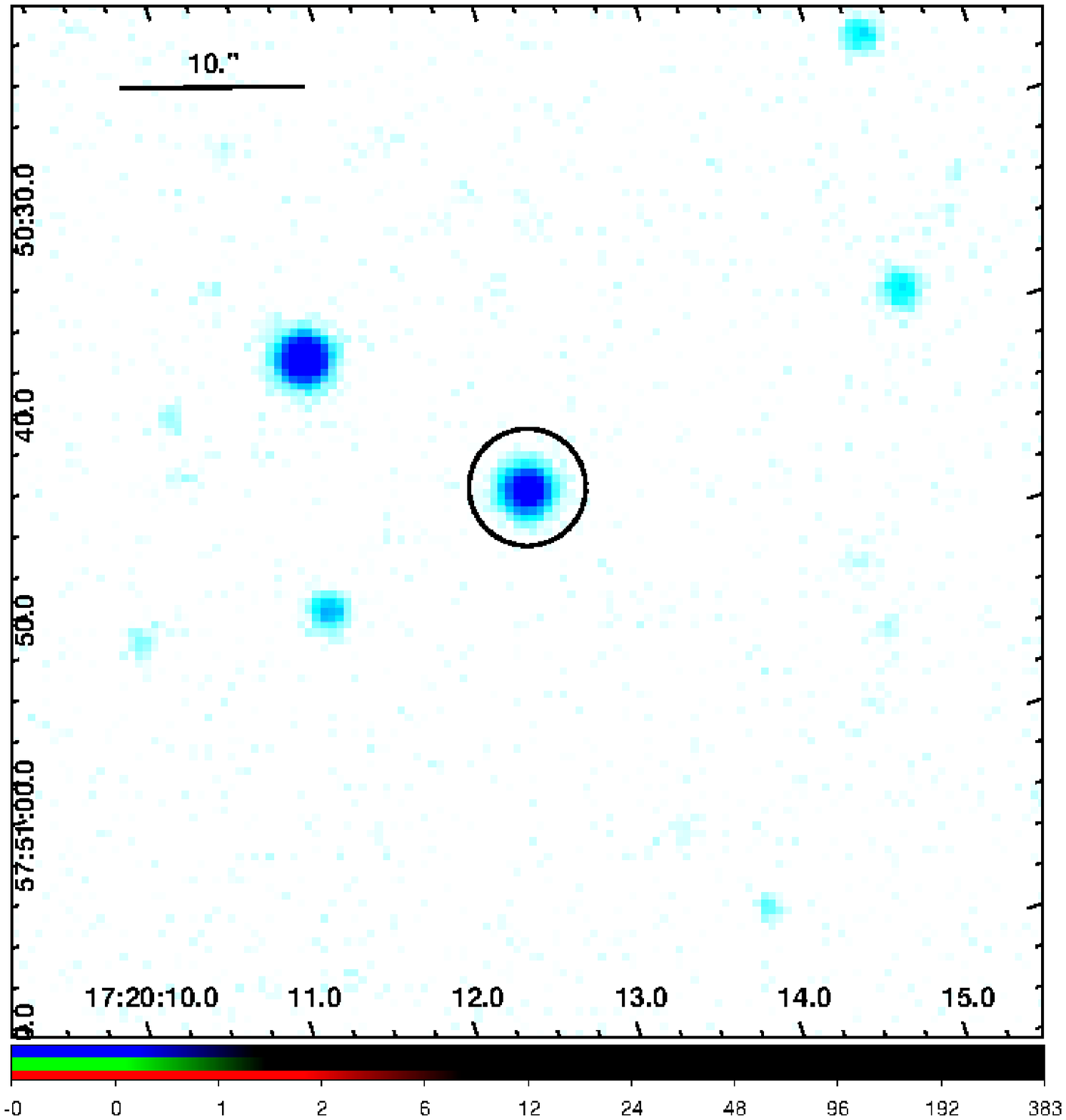}}
\vspace{0.2cm}
\subfloat[Src No.9]{\includegraphics[clip, trim={0.0cm 2.0cm 0.cm 0.0cm},width=0.28\textwidth]{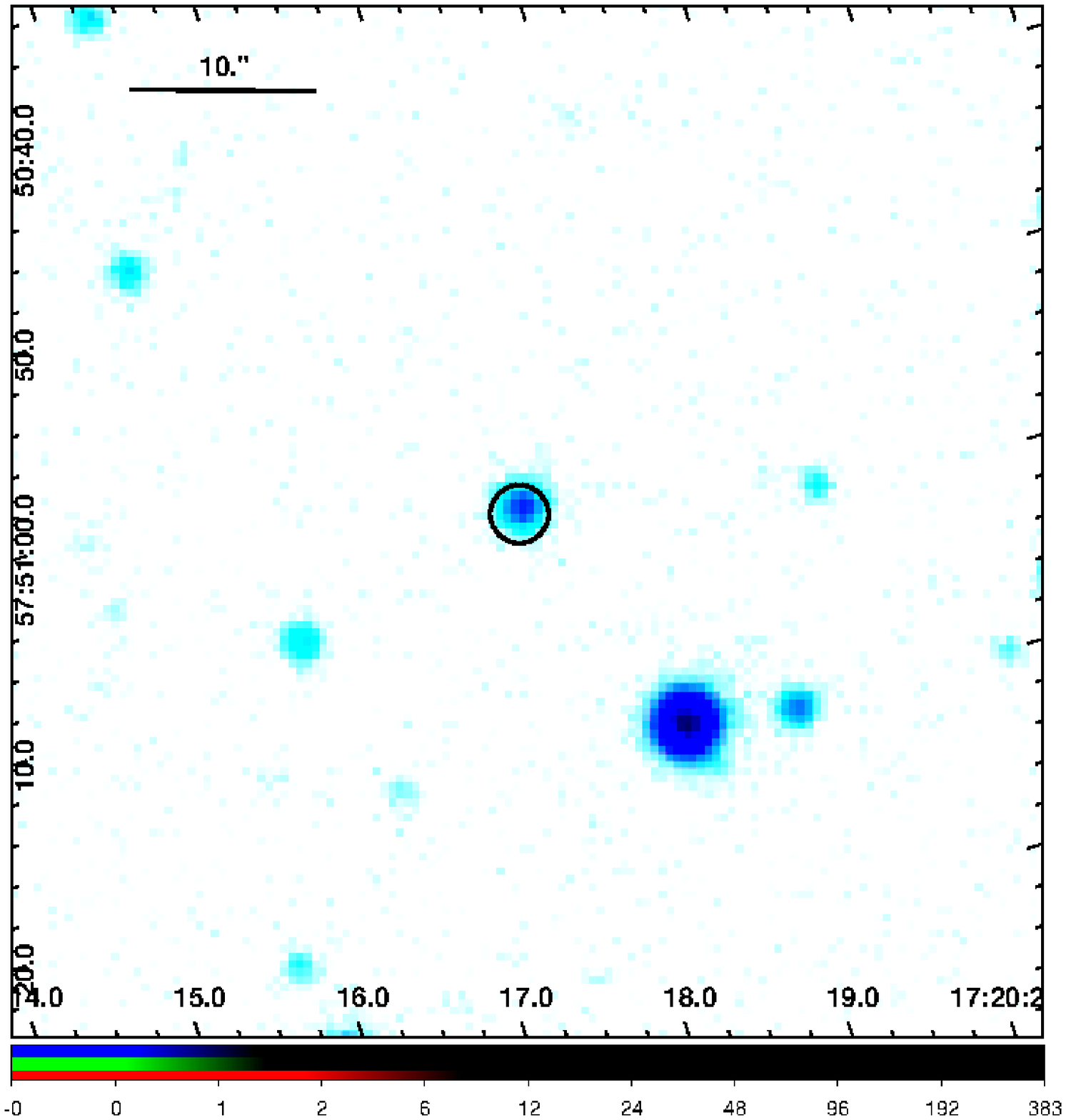}}\\
\end{figure*}

\pagebreak
\clearpage
\hspace{0.3cm}Appendix C continued: Image of optical SDSS9 counterparts
\vspace{-0.4cm}
\begin{figure*}[!htb]
  \centering
  \subfloat[Src No.10]{\includegraphics[clip, trim={0.0cm 2.0cm 0.cm 0.0cm},width=0.28\textwidth]{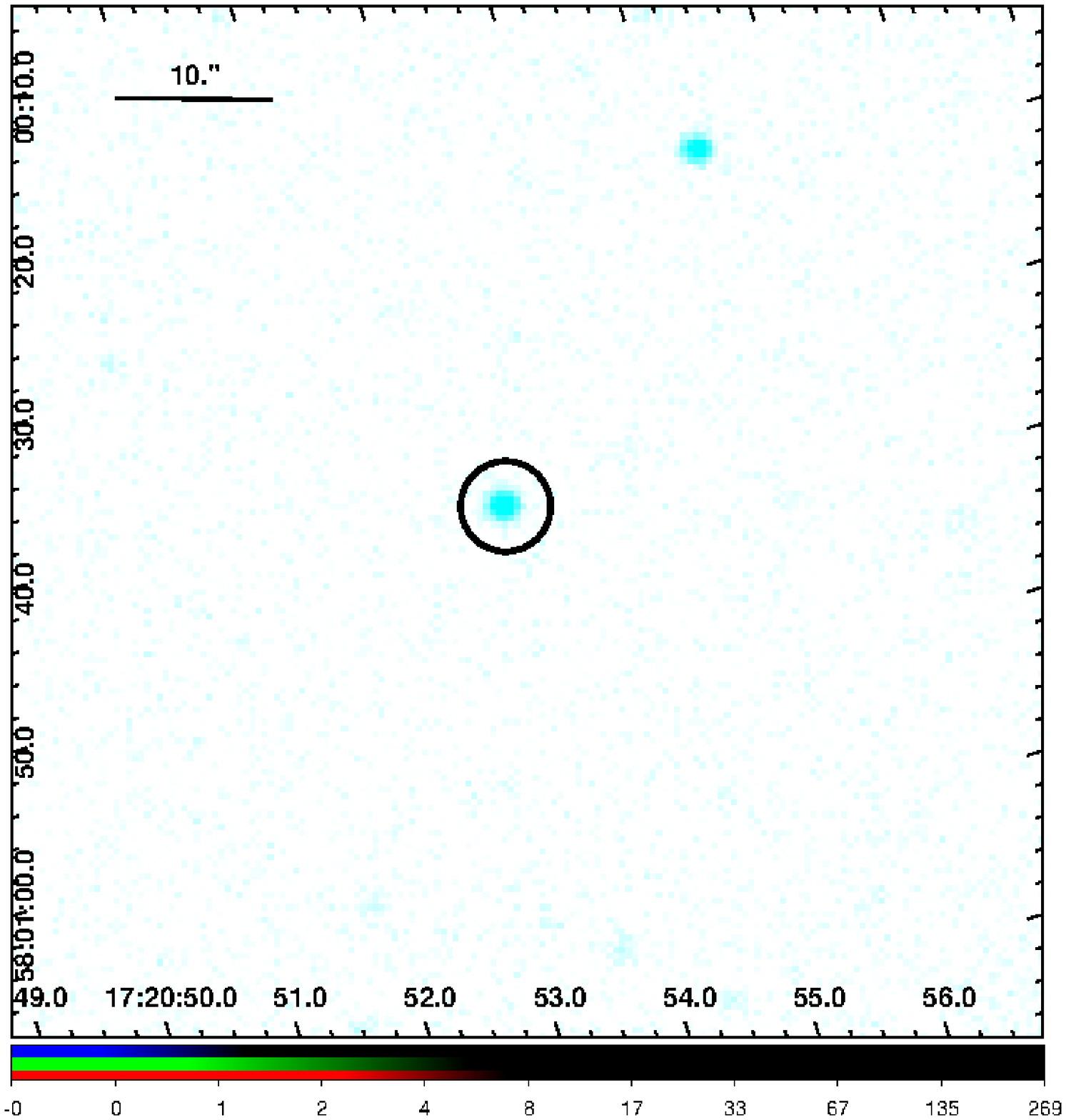}}
\vspace{0.2cm}
  \subfloat[Src No.11]{\includegraphics[clip, trim={0.0cm 2.0cm 0.cm 0.0cm},width=0.28\textwidth]{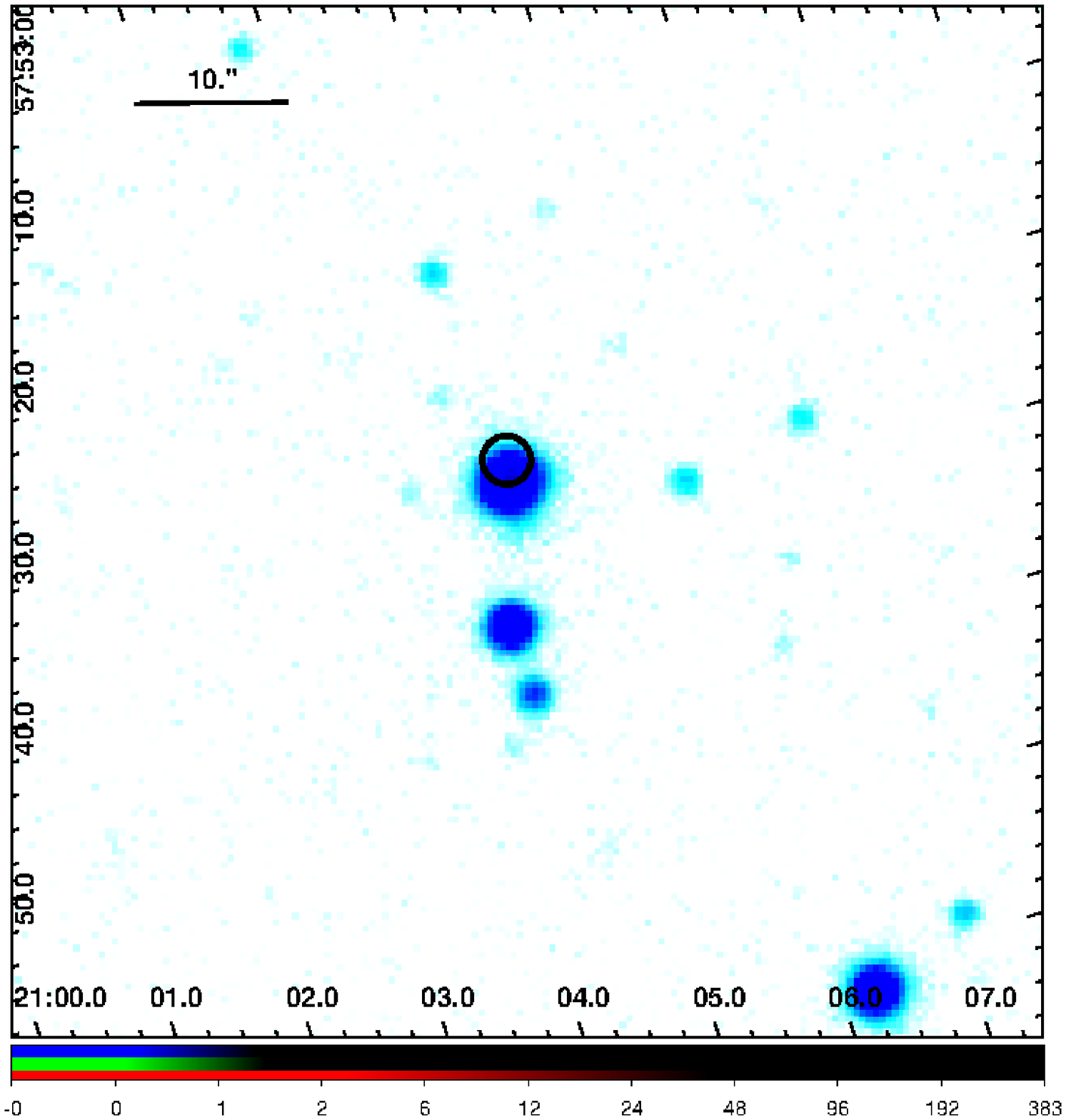}}
\vspace{0.2cm}
\subfloat[Src No.12]{\includegraphics[clip, trim={0.0cm 2.0cm 0.cm 0.0cm},width=0.28\textwidth]{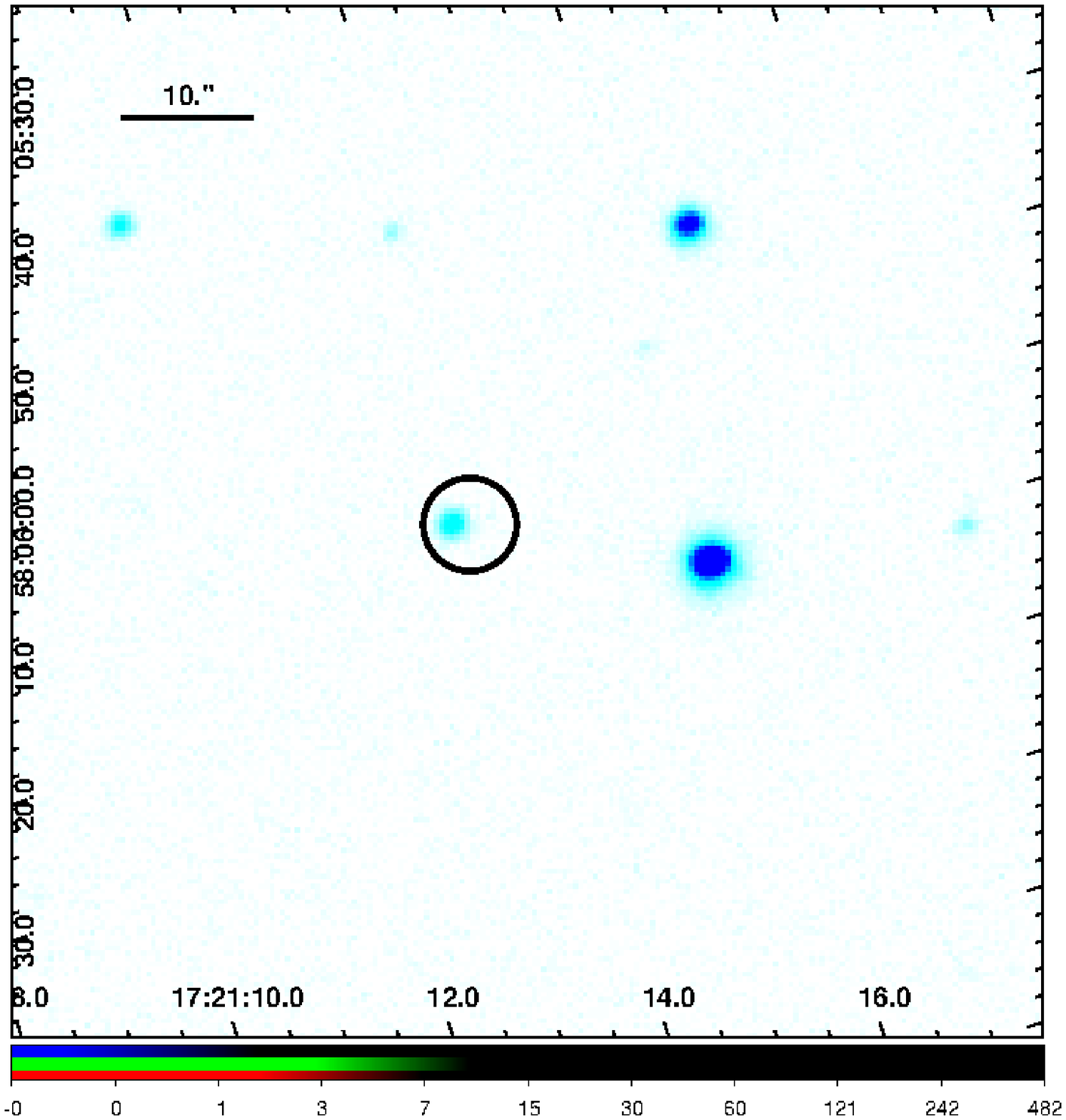}}\\

\subfloat[Src No.13]{\includegraphics[clip, trim={0.0cm 2.0cm 0.cm 0.0cm},width=0.28\textwidth]{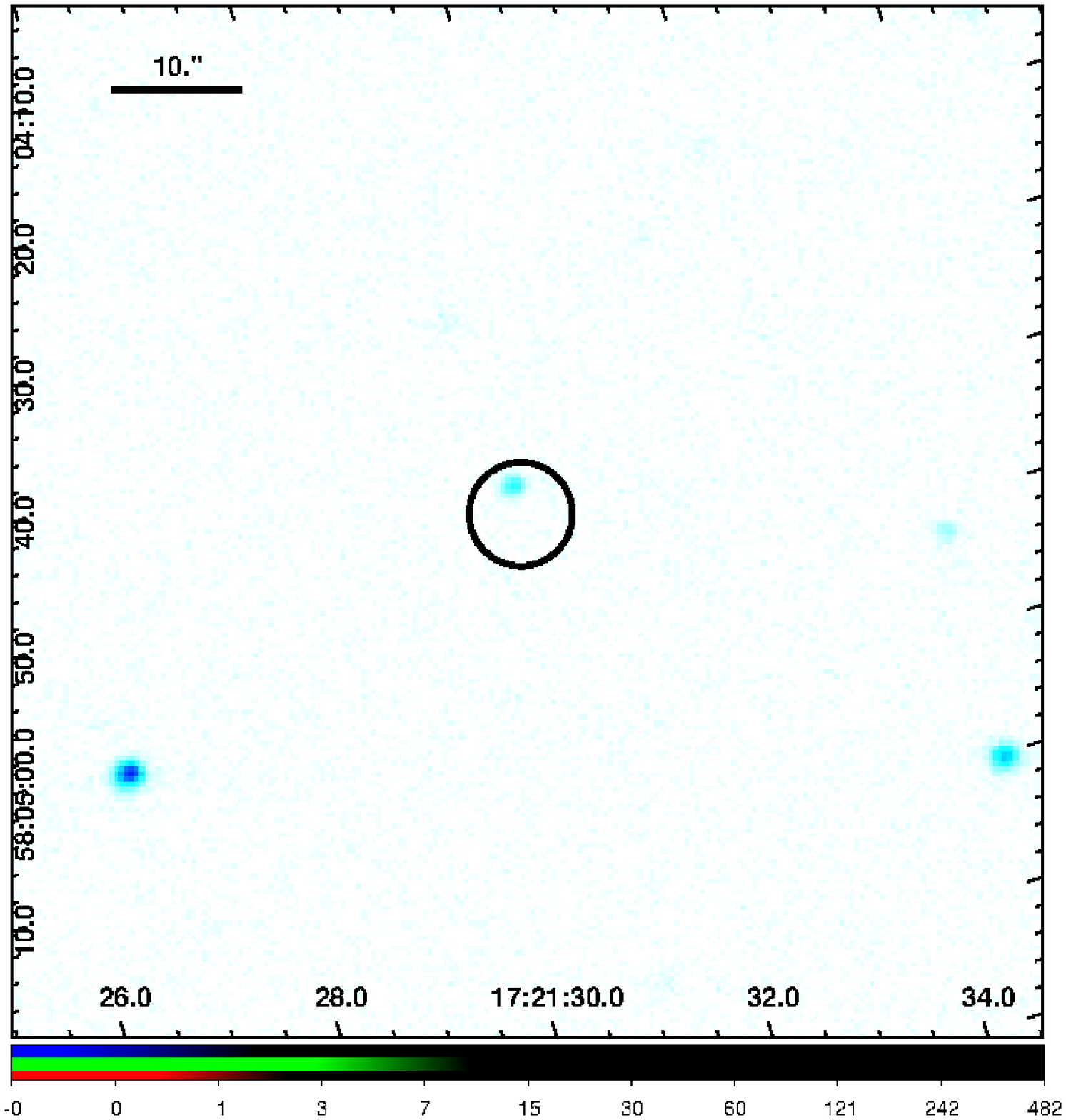}}
\vspace{0.2cm}
\subfloat[Src No.14]{\includegraphics[clip, trim={0.0cm 2.0cm 0.cm 0.0cm},width=0.28\textwidth]{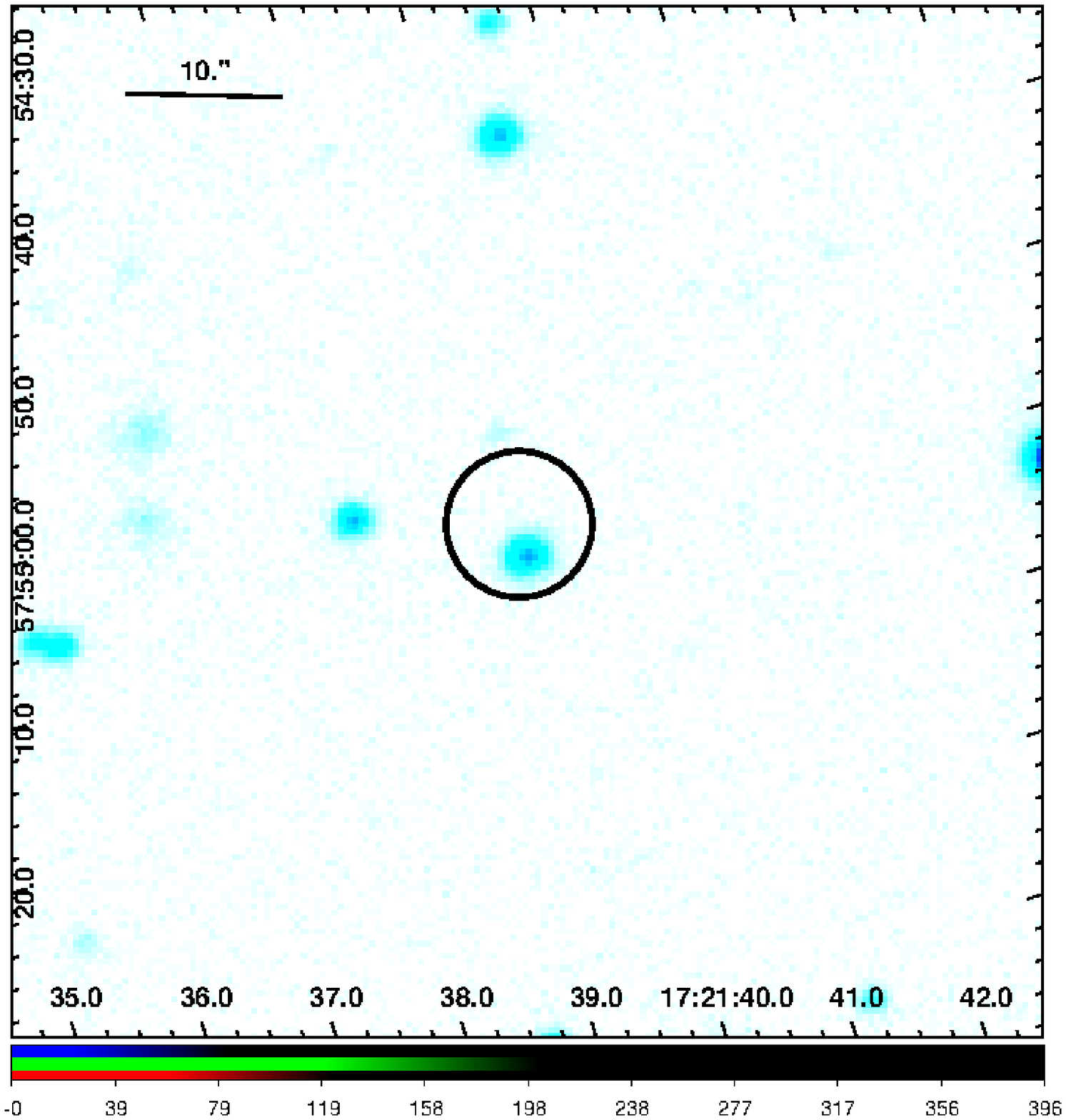}}
\vspace{0.2cm}
\subfloat[Src No.15]{\includegraphics[clip, trim={0.0cm 2.0cm 0.cm 0.0cm},width=0.28\textwidth]{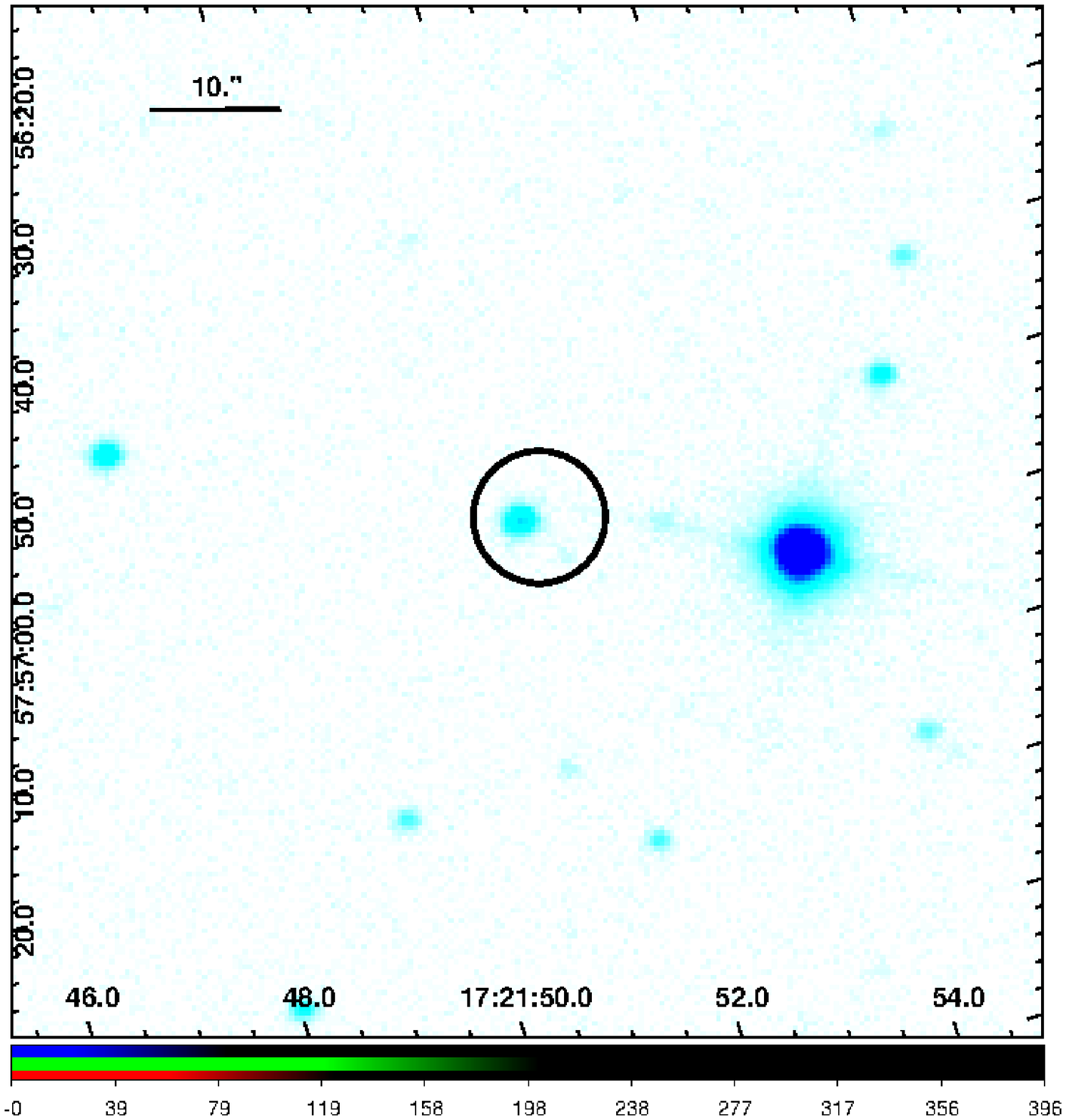}}
\end{figure*}
%\end{landscape}
\end{appendices}

% THIS FILE INCLUDES THE REFERENCES IN BIBTEX FORMAT
%\newpage

%\begin{appendices}
%\label{appen-a}

%\section{Combining the spectra of different EPIC cameras}

\end{document}